\documentclass[fleqn,usenatbib,useAMS]{mnras}

\usepackage{graphicx}	
\usepackage{amsmath}	
\usepackage{multicol}        
\usepackage{bm}		
\usepackage{pdflscape}	

\usepackage[utf8]{inputenc} 

\usepackage{footnote}  
\usepackage{enumitem}  
\usepackage{epstopdf}  

\usepackage{ae,aecompl}

\usepackage{mathptmx}

\title[12P and other Halley-type comets]{Future evolution of 12P and other Halley-type comets in near-polar orbits}

\author[M. Królikowska, P. Kankiewicz, P. Wajer]{Małgorzata Królikowska$^{1}$\thanks{E-mail: mkr@cbk.waw.pl},Paweł Kankiewicz$^{2}$\thanks{E-mail: pawel.kankiewicz@ujk.edu.pl}, Paweł Wajer$^{1}$\thanks{E-mail: wajer@cbk.waw.pl}
\\
$^{1}$Centrum Badań Kosmicznych Polskiej Akademii Nauk (CBK PAN), Bartycka 18A, 00-716 Warszawa, Poland \\
$^{2}$Institute of Physics, Jan Kochanowski University, Uniwersytecka 7, 25-406 Kielce, Poland
}

\date{Accepted 2025 March 20. Received 2025 March 20; in original form 2025 January 14}

\pubyear{2024}

\begin{document}
\label{firstpage}
\pagerange{\pageref{firstpage}--\pageref{lastpage}}
\maketitle

\begin{abstract}
We investigate the future 100\,kyr evolution of six selected HTCs to show their basic commonalities and differences in dynamical behaviour. This includes estimating the probability of sungrazing and flipping. We combined three complementary numerical methods to study the dynamical features: the numerical integrations forwards in time, the Lyapunov time estimations, and the Mean Exponential Growth factor of Nearby Orbits (MEGNO). For each comet, we obtain the osculation orbits from the available observations. We then construct swarms of virtual comets as the basis for all dynamical studies. We show that two comets with $q<1.3$\,au achieve the sungrazing state in the future with high probability: 161P with the likelihood of $\sim$80\% will be a sungrazing object with $q<0.005$\,au in the next 13\,kyr, and 122P with 50\% in the next 100\,kyr. We found that both these HTCs reach the sungrazing states due to Kozai resonances with other planets than Jupiter; for example, Uranus acts as the agent of Kozai resonance for 161P. We indicate that the high sungrazing probability for both comets is connected with a high likelihood of orbit flipping. The other four HTCs have a slight chance to be sungrazers after 100\,kyr ($<2.2$\%); however, three of them can achieve a high flipping probability. We show that the Lyapunov time and MEGNO indicator give a complementary picture of the orbital stability after 10$^4$yr. Results allow us to rank comets from the most to least chaotic, where 161P is a particular case with its high probability to disintegrate due to the Kozai mechanism.
\end{abstract}






\section{Introduction}
\begin{figure*}
	\centering
	\includegraphics[width=8.8cm]{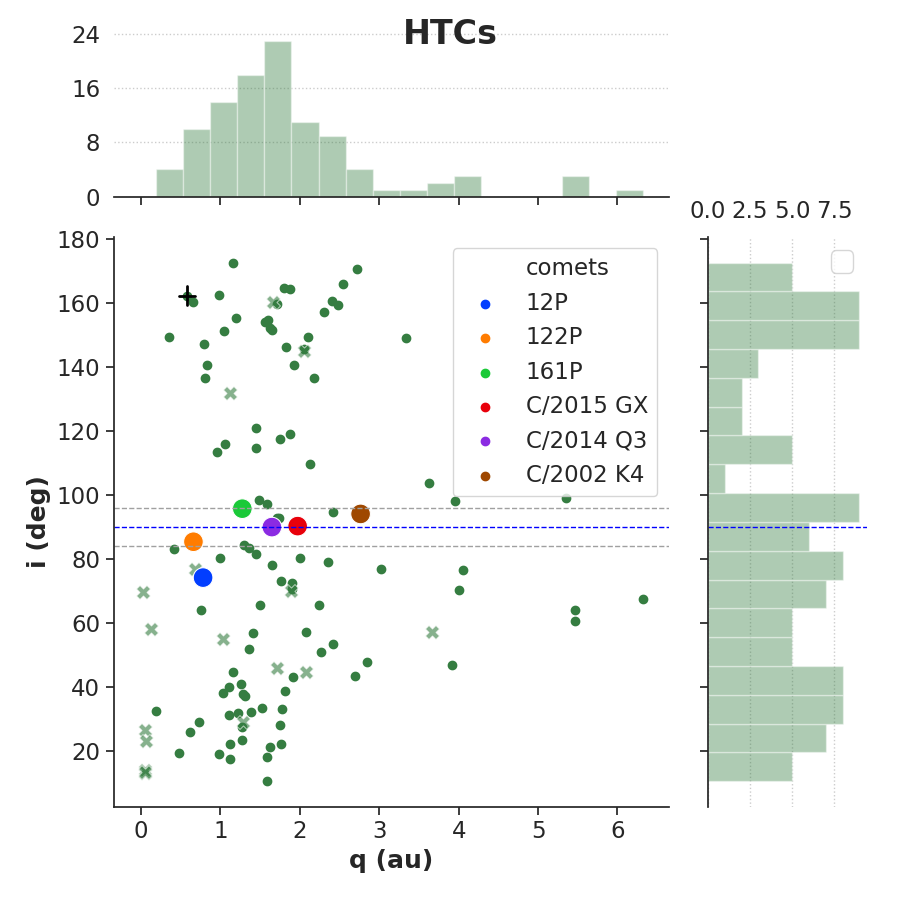}
    \includegraphics[width=8.8cm]{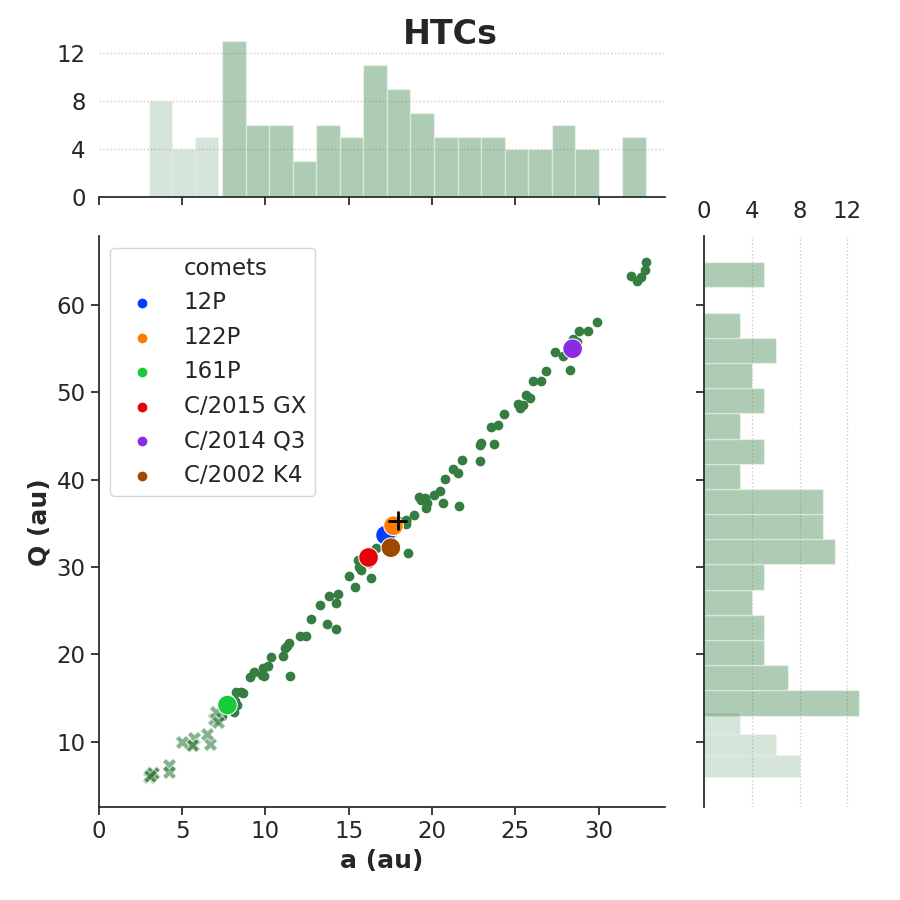}
	\caption{Distribution of known HTCs (defined as having $T_J<2$ and $a<$34.2\,au; as in June 2024) in perihelion distance, $q$, versus inclination, $i$ (left panel), and in semimajor axis, $a$, versus aphelion distance, $Q$ (right panel). Green dots and green marginal distributions represent the sample of HTCs according to classical definition (orbital period from 20\,yr to 200\,yr).  Light-green crosses in both panels and light-green marginal distributions in the right panel show the sample described as 'JFC*' in JPL Database, where six SOHO comets (with 0.028\,au $<q<$0.066\,au) are not excluded. The HTCs studied here are colour-coded and described in the inset.  A black cross indicates the position of 1P/Halley. Dotted grey horizontal lines in the main plot of the left panel indicate the band $i=90$\degr $\pm 6$\degr mentioned in the text.}
 \label{fig:HTCs_q_i} 
\end{figure*}

Named after their famous prototype, Halley-type comets (HTCs) have orbital periods between 20 and 200 years, according to the classical definition used in the Jet Propulsion Laboratory Engine\footnote{{\tt https://ssd.jpl.nasa.gov/tools/}; from JPL, we mainly use two tools: Small-Body Database Query -- for HTCs statistics, and Small-Body Database Lookup -- for an individual object.}(JPL). This range of periods is between Jupiter family comets (JFCs) having shorter orbital periods and long-period comets (LPCs) situated on the opposite side of period values. So alternatively, HTCs are also known as intermediate-period comets. In contrast to JFCs, orbits of HTCs can be highly inclined to the ecliptic (Fig~\ref{fig:HTCs_q_i}). This feature suggests that HTC source can be located in the Oort Cloud, the more or less spherical populations of icy bodies coming to the planetary zone from the outskirt of our solar system due to Galactic tidal force and passing star perturbations \citep[see][and references therein]{wan-bra:2014}. With almost whole orbits inside the planetary zone, HTCs can be notably influenced by the gravity of the giant planets \citep{loh-dvo-fro:1995}.

\cite{chambers:1997} showed that the transition from stable to unstable librations about the mean motion resonances occurs near the orbital period of 200\,yr, the classical definition of the boundary between HTCs and LPCs. He concludes that the fact that HTCs spend a significant part of their lives in resonances and LPCs do not lead to notable differences in the dynamical evolution of these two comet populations and may be an important factor in lengthening the dynamical lifetime of HTCs. This confirms that this once arbitrarily adopted boundary between HTCs and LPCs has its dynamical justification. However, Chambers showed that 200\,yr is an extremely rough assumption because this period boundary increases from about 150\,yr to 270\,yr in function of inclination (crosses in Fig.~5 in Chambers' paper).

Many complex dynamical simulations have been performed to attempt to reproduce observed comet populations, including HTCs \cite[see for example: ][]{bai-eme:1996, lev-don-dun:2001, nurmi-etal:2002, horner-e-b-a:2003, lev-don-dun:2006, wan-bra:2014, fer-gal-you:2016}. However, so far, numerical simulations have not been able to reproduce the observed population of HTCs compared with the observed number of Oort Cloud comets without assuming the existence of cometary fading \citep[][and references therein]{wan-bra:2014, lev-don-dun:2001}.

While the literature on the origin of HTCs is extensive, there is less research on the evolution (especially future evolution) of real comets when we exclude the prototype of the group, comet 1P/Halley. Below, we briefly describe only a few.

In one of the key publications, \cite{carussi-etal:1987} studied dynamics almost a thousand years back and forward in time for a sizeable group of HTCs. Authors found that more or less regular librations with periods of several hundred years around the resonances 1:5, 1:6, or 1:7 with Jupiter is a common phenomenon among the comets with periods around 70\,yr, for example for 12P/Pons-Brooks, 13P/Olbers 23P/Brorsen-Metcalf, and C/1921~H1 (Dubiago). They conclude that resonances can increase the stability of HTC orbit during long-term dynamical evolution.

Almost 10 years later, when computing capabilities have increased significantly, \cite{bai-eme:1996} deal with the dynamic evolution of ten real HTCs over $\pm$1\,Myr. For each comet, they prepared 10 orbital variations. They concluded that the median lifetime of HTCs against dynamical ejection onto a hyperbolic orbit is an order of 1\,Myr. During the HTC evolution, they noticed various secular effects, including Kozai librations, secular resonances, and sungrazing states. They also found that secular perturbation can also cause the evolution of orbit from retrograde to prograde and the evolution of perihelion distance from outside Saturn's orbit to within the orbit of the Earth. However, they considered Jupiter the most important planet controlling the evolution of HTCs. In their sample of 10 comets, only 12P/Pons-Brooks was studied here.

One of the relatively newest is the study of the evolution of 21 HTCs (of which again only 12P is in the group studied here) over 100\,Myr by \cite{hel-jef:2012}. For each comet, they conducted an evolution investigation for five of its clones, with an orbital elements evaluation every 10\,kyr. They obtained 85 ejected clones and 20 sun-colliding or sungrazing in dynamic calculations. For 12P three were ejected and two sungrazing (sun-colliding); however, all five above end-states took place after 500\,kyr. They also found that the comet orbit evolved dramatically over timescales much shorter than the overall dynamical lifetime of HTCs that they found to be of the order of $10^5-10^6$\,yr. Moreover, they noticed the variations in the orbital inclination, $i$, from prograde to retrograde and vice versa.

The study of the evolution of selected HTCs has obvious limitations for long-term numerical integrations. The main reason for this is the presence of chaos in their motion. A good example of such behaviour is the confirmed presence of strong chaos in the motion of comet 1P/Halley, studied by \cite{Munoz:2015}, \cite{Boekholt:2016}, \cite{Perez:2019}, and \cite{Kaplan:2022}. In our HTC sample, chaos is mainly generated by resonances in the mean motion (MMR, RMMR\footnote{retrograde mean motion resonance, first referred to in this way by \cite{Morais2013}}) and regular close approaches to planets. Quantitatively, chaotic properties can be estimated numerically using various indicators based on the expansion of close trajectories and, additionally, using the swarm of virtual comets (hereafter VCs), all representing the orbit obtained using the positional data. We use both approaches together to validate the obtained results. First, we decided to use various research tools: the classical dynamical numerical integrations forwards in time, the Lyapunov time, and the Mean Exponential Growth factor of Nearby Orbits (MEGNO) estimations. Second, all the above-described prompted us to study the future evolution of a few selected HTCs on the timescale of 100\,kyr but for a statistically significant sample of 1000~VCs constructed based on carefully determined orbits for them. To our knowledge, the simultaneous use of three described dynamical methods to swarms of VCs is an innovative approach. We focus on examining the characteristic features of dynamical evolution for chosen HTCs and finding similarities and differences in their dynamical behaviours. In particular, we explore the future orbital evolution of studying HTCs to indicate to what extent the phenomenon of changing the sense of motion from direct to retrograde or vice versa is common for these objects and to extract the characteristic features of this behaviour. Throughout the paper, we call the phenomenon of changing the sense of motion from prograde to retrograde or vice versa as {\it flipping orbit}, as stated first by \cite{Fuente2015}.

Today, the inclination flips appear in literature as a phenomenon occurring in the evolution of some groups of small bodies \citep{Fuente2015, ric_etal:2017, Kankiewicz2020}, including HTCs
\citep{val-ric-mor:2022}. The analytical studies by \cite{val-ric-mor:2022} are particularly relevant to the current research. The authors show that the transition from prograde to retrograde orbit due to a close encounter with Jupiter can take place only if the Tisserand parameter $T_J \leq 2$ (see Sect.~\ref{sec:HTCs-general-statistics} for the definition of $T_J$). This condition is met by all HTCs (Sect.~\ref{sec:HTCs-general-statistics}).
We also aim to investigate the frequency of achieving the sungrazing state in the studied comets and whether inclination flip and evolution towards sungrazers can be connected. The evolution of several HTCs initially moving on highly inclined orbits as possible sources of sungrazing comets was investigated by \cite{bai-cha-hah:1992}, including the evolution of comets 122P/de Vico and 161/Hartley-IRAS studied here.

The organisation of the article is as follows. In Sect.~\ref{sec:HTCs-general-statistics}, we briefly describe current statistics of HTCs and indicate the reasons for selecting particular objects for this study. We base the future evolution study on our osculating orbit determinations for investigated objects from positional observations. We obtained osculating orbits in the purely gravitational (hereafter GR) regime. Additionally, we attempted to obtain the non-gravitational (hereafter NG) orbits, wherever the NG orbits can be determined with reasonable accuracy (Sect.~\ref{sec:orbit-deterination}). This section shows that NG~orbits are uncertain for these objects, except 12P for the assumed data arc; however, the NG~parameters (defined in Sect~\ref{subsec:NG_model}) seem to change from one data arc to another. Since our basic objective is to perform calculations with the fewest possible assumptions, and based on the orbits we obtain directly from positional observations, our further dynamical calculations start with the GR~orbits we have reached in this study, except 12P for which we extend the research to the NG~regime. We consider this stage important because the different data arcs taken for orbit determination can change notably the resulting NG~orbit, and we discuss this issue for 12P. In our comparisons, we also refer to two publicly available and updated all-time orbital minor planet databases: Minor Planet Center\footnote{{\tt https://www.minorplanetcenter.net/db\_search}}(MPC) and JPL. These two databases and our method of orbit determinations mainly differ in their approach to selecting and weighing positional data. We devoted much space here to comet 12P/Pons-Brooks because it is a comet that attracts the attention of researchers with its spectacular behaviour during its current appearance. The comet passed perihelion in April this year, and its motion will most likely be tracked for next year or in the coming years.

In Sects~\ref{sec:evolution-classical}--\ref{sec:evol-flipping} we describe the future evolution of the studied HTCs by long-term numerical integration of their swarms of 1001~orbits in the 100\,kyr period. In particular, we focus on two phenomena common in some of the studying here HTCs: evolution to sungrazing state (Sect.~\ref{sec:evol-to-small-q}) and flipping orbit state (Sect.~\ref{sec:evol-flipping}).

An analysis of the short-period (10\,kyr) stability of our HTC sample, based on the estimation of the Lyapunov time, is presented in Sect \ref{sec:evol-LT}. This section also outlines the method used to estimate this parameter through variational equations.

In Sect.~\ref{sec:evol-MEGNO}, the short-period (10\,kyr) stability of the studied HTCs was analysed for initial conditions near the nominal orbits. This was done using the fast converging indicators MEGNO and their maps in the space of $a-e$ and $a-i$ elements, constrained by the range of orbit determination errors and used 10 times more VCs than above (10,000\,VCs). This allowed us to illustrate the local fluctuations of this indicator and how it can be affected by observational errors. This tool and formalism combined with visualisation was likely used for the first time for this studied sample of comets. We finally report and summarise our conclusions in Sect.~\ref{sec:conclusions}.

\section{Orbital statistics of HTCs and our HTC sample}\label{sec:HTCs-general-statistics}

The JPL database\footnote{see https://ssd.jpl.nasa.gov/tools/sbdb\_query.html} explicitly stated that the classical definition of HTCs is there applied:  20\,yr~$<P<$~200\,yr, and as in June 2024 there were 105 HTCs, where 15 were numbered, 9 had designation P/, and 81 HTCs were marked as C/.  However, dynamically crucial is that all these comets have the Tisserand parameter relative to Jupiter under the restricted circular three-body problem:  

\begin{equation}
T_J = \frac{a_J}{a} + 2 \left[(1 - e^2)\frac{a}{a_J}\right]^{1/2} \cos(i)
\end{equation}

\noindent smaller than 2 \citep[see][]{kresak:1972}, where $a_J$ is semimajor axis of Jupier and $a,e,i$ are semimajor axis, eccentricity, and orbital inclination of a comet. Value of $T_J =2$ as additional condition to $P<200$\,yr (separation between HTCs and LPCs) was first proposed by \cite{carussi-etal:1987} to distinguish between HTCs ($T_J <2$) and JFCs  ($T_J >2$);  however, they pointed out that three exceptional comets with $T_J<2$ are classified as JFCs, namely 8P/Tuttle, 96P/Macholz~1, and 126P/IRAS. Peculiar dynamics of the last two comets were discussed in literature widely, see also \cite{car-val:1992}. Next, \cite{levison:1996} used this $T_J$-condition in his classification  based on the dynamical characteristics of comets (see Fig.~7 therein). Still  in 1996, he noted only above mentioned three comets classified as JFCs which have $T_J <2$.  Now, 17 comets with $T_J <2$ and $P<20$\,yr have been discovered and were separated into the 'JFC*' group of orbit class in the JPL Database. This means that the situation of these 17 objects is unclear and they could just as well be classified as HTCs, which is what \cite{wan-bra:2014} did. They used the upper limit for the semimajor axis as $a<34.2$\,au (Levison set 40\,au as the upper value for $a$) and added an assumption of $q > 0.01$\,au to exclude the sungrazing Kreutz comets (see their list of HTCs). However, we noticed that they in practice used the criterion  $q > 0.07$\,au excluding from the HTC list four comets discovered by SOHO (P/1999~J6, C/2002~R5, P/2002~S78, and P/2008 Y12). Today, in this group of JFCs with $T_J <2$ are six comets with $q<0.07$\,au (two more are: 342P and C/2015~D1).

Here, we present the statistics of HTCs in the classical picture (comets with orbital periods in the range of 20--200\,yr). However, we display this small sample of 'JFC*' in Fig.~\ref{fig:HTCs_q_i} (light green crosses) to show some similarity in the orbital inclination distribution to the group of HTCs. 

HTCs signed by 'P/' have orbital periods from 20\,yr to 31\,yr and semimajor axes in the range 7.4--10\,au. Each of remaining groups (C/ and numbered) have periods between 20--200\,yr and semimajor axes between 7.4\,au and 34.2\,au.

Table~\ref{tab:HTCs-numbers} shows a very uneven rate of HTC discovery during the last twenty-five years. In the period 2005--2009, only 5 new HTCs were discovered, and in 2015--2019 as many as 33. However, in the last five years, the number of discovered comets seems to be notably smaller than in the previous period.

\begin{table}
	\caption{\label{tab:HTCs-numbers} Number of HTC discoveries in function of time intervals (as in October 2024).}  \setlength{\tabcolsep}{12.0pt} \centering
	\setlength{\tabcolsep}{13.0pt}
	\begin{tabular}{rrrrr}
		\hline  
		Years         &    all    &  numbered   &      P/      &         C/   \\
		\hline 		\hline
		before 2000   &    29   &    14     &       0      &       15   \\
		2000-2004     &    12   &     1     &       0      &       11   \\
		2005-2009     &     5   &     -     &       2      &        3   \\
		2010-2014     &    14   &     -     &       2      &       12   \\
		2015-2019     &    33   &     -     &       5      &       28   \\
		2020-2024     &    12   &     -     &       0      &       12   \\
		\hline
	\end{tabular}
\end{table}

\begin{table}
	\caption{\label{tab:HTCs-stat-i} HTC statistics as a function of inclination $i$. HTCs studied here are listed in columns 3 and 6.  }  \setlength{\tabcolsep}{2.0pt} \centering
	\setlength{\tabcolsep}{1.5pt}
	\begin{tabular}{rlccrlcc}
		\hline  
		\multicolumn{2}{c}{Range}	&  Number  & HTCs         & \multicolumn{2}{c}{Range}	&  Number  & HTCs  \\
		\multicolumn{2}{c}{of $i$}	&  of HTCs & here         & \multicolumn{2}{c}{of $i$}	&  of HTCs & here  \\
		\hline 		\hline
		0\degr -- & 10\degr	&       0  &                     	&      170\degr -- & 180\degr &    2 & \\
		10\degr -- & 20\degr	&       5  &                      	&      160\degr -- & 170\degr &    7 & \\
		20\degr -- & 30\degr	&       8  &                      	&      150\degr -- & 160\degr &    9 & \\
		30\degr -- & 40\degr	&      10  &                     	&      140\degr -- & 150\degr &    8 & \\
		40\degr -- & 50\degr	&       8  &                     	&      130\degr -- & 140\degr &    2 & \\      
		50\degr -- & 60\degr	&       5  &                     	&      120\degr -- & 130\degr &    1 & \\
		60\degr -- & 70\degr	&       6  &                     	&      110\degr -- & 120\degr &    5 & \\
		70\degr -- & 80\degr	&       9  &    12P              	&      100\degr -- & 110\degr &    2 & \\
		80\degr -- & 90\degr	&       8  &   122P                	&       90\degr -- & 100\degr &   10 & 161P/ \\
		&     &          &   C/2014 Q3            &             &     &      & C/2002 K4 \\
		&     &          &                        &             &     &      & C/2015 GX \\ 
		0\degr -- & 90\degr  &      59  &                        &       90\degr -- & 180\degr &   46 &           \\                      
		\hline
	\end{tabular}
\end{table}

\begin{table}
	\caption{\label{tab:HTCs-stat-q} HTC statistics as a function of perihelion distance $q$. HTCs studied here are listed in the last column.}  \setlength{\tabcolsep}{2.0pt} \centering
	\setlength{\tabcolsep}{6.0pt}
	\begin{tabular}{rlcl}
		\hline  
		\multicolumn{2}{c}{Range}	    &  Number  & HTCs           \\
		\multicolumn{2}{c}{of $q$ [au]}	&  of HTCs & studied here   \\ 
		\hline  \hline  
		0.0 -- & 0.5 &    5 & \\
		0.5 -- & 1.0 &   14 &  12P ,        122P \\
		1.0 -- & 1.5 &   26 &  161P \\
		1.5 -- & 2.0 &   31 &  C/2014 Q3 ,  C/2015 GX \\
		2.0 -- & 2.5 &   13 & \\
		2.5 -- & 3.0 &    5 &  C/2002 K4 \\
		3.0 -- & 3.5 &    2 & \\
		3.5 -- & 4.0 &    3 & \\
		4.0 -- & 4.5 &    2 & \\
		4.5 -- & 5.0 &    0 & \\
		5.0 -- & 5.5 &    3 & \\
		5.5 -- & 6.0 &    0 & \\
		6.0 -- & 6.5 &    1 & \\
		\hline
	\end{tabular}
\end{table}

Statistics of HTCs as a function of orbital inclination $i$ is given in the left panel of Fig.~\ref{fig:HTCs_q_i} and Table~\ref{tab:HTCs-stat-i}. More generally, the number of HTCs on prograde orbits is greater than those moving on retrograde orbits and this issue has been widely discussed in the literature \citep{fer-gal:1994, lev-don-dun:2001, wan-bra:2014}. Additionally; however, when comets are counted in four bins with a width of 45 deg then we get 29 (0\degr--45\degr) and 30 (45\degr--90\degr) comets moving in prograde orbits, and 18 (90\degr--135\degr) and 28 (135\degr--180\degr) objects moving retrograde, respectively. More precisely, we have a clear deficiency of HTCs with orbital inclinations between 90\degr and 135\degr. It can be seen that an even greater deficiency of HTCs relative to random numbers is observed in the range 100\degr--140\degr -- there are only 10 comets. This heterogeneous $i$-histogram looks intriguing, but confirming the statistical significance of this feature requires a larger sample of HTCs. Moreover, it is striking that all 9~HTCs known today with $q>3.5$\,au have 40\degr ~$ < i < 110$\degr; see left panel in Fig.~\ref{fig:HTCs_q_i}. This is however all the more an insufficiently large sample to conclude about the significance of the difference in this distribution compared to the $i$-distribution for HTCs with $q <$~3.5\,au ($\sim10$\degr ~$ < i < \>\sim170$\degr). A similar $i$-distribution signature was shown by  \cite{gallardo:2019} in Fig.~3 for the sample of asteroids and comets with $a<40$\,au, including about known then 70~HTCs (as in 2019). \cite{gallardo:2019} pointed out the excess of objects on orbits with inclinations from 130\degr ~to 170\degr ~and argued that it is connected with the existence of the stability band around 150\degr ~for all small bodies along the Solar system. He also showed that retrograde orbits with inclinations below this band are in the unstable region.

Table~\ref{tab:HTCs-stat-q} shows that 89 of 105 known HTCs have perihelion distances from 0.5\,au to 3.0\,au (76\%); only 4 have $q$~>~5.0\,au (all four discovered in the last decade). Fig.~\ref{fig:HTCs_q_i} reveals a drop in $q$-distribution for HTCs starting from $q=2$\,au (or even for slightly smaller $q$). This drop was discussed in detail by \cite{wan-bra:2014}. They attributed this decline in the number of HTCs to observational biases and showed that the similar drop beyond 2\,au was then observed for all comets with Tisserand parameter $T_{\rm J} < 2$ (Fig.~2 therein). In the last decade, after the publication of their work, the number of HTCs has almost doubled \citep[Table~1 in][]{wan-bra:2014}. We additionally found that, they included to HTCs seven comets with periods less than 20\,yr from the mentioned above sample 'JFC*' available at the JPL Database, and one LPC with a period of 500\,yr (C/1906~V1). So, the current HTC sample is notably richer. We found that 17 of 49 HTCs discovered after 2014 have $q>2.0$\,au (35\%), including all today known HTCs with $q>5.0$\,au whereas before 2014 was known 12 of 55~HTCs with $q>2.0$\,au (22\%).  

The right panel of Fig.~\ref{fig:HTCs_q_i} shows the position of chosen comets among HTCs in the $a$--$Q$ plane. One can see that the four of them are very close to each other and, additionally, close to 1P/Halley. They are inside the middle peak of HTCs in this 2D~plane of orbital elements. Among these four there is 12P which was in resonance 1:6 with Jupiter about a thousand years around the observed perihelion passage according to \cite{carussi-etal:1987}. As they pointed out, some concentrations in $a$-distribution (visible around the 12P position in the right panel of  Fig.~\ref{fig:HTCs_q_i}) can be associated with resonances with Jupiter commonly occurring in dynamical evolution of HTC in the past. In the proximity of 12P are two other comets, C/1921~H1 (Dubiago) and C/1942~EA (V\"ais\"al\"a), and \cite{carussi-etal:1987} obtained that both experience resonances with Jupiter during their past evolution, 1:5 and 1:7, respectively.

Great hopes for discovering more HTCs, especially those with large perihelion distances, can be placed in the {\it Vera C. Rubin} Observatory, which will most likely be operational in 2025. Estimates say that this telescope has a chance to discover objects with $q>5$\,au on a mass scale \citep{schwamb-etal:2023, ivezic-etal:2019}.

\begin{table*}
	\caption{ General characteristics of positional material used in this study for the analysed Halley-type comets; observations were taken from the MPC database in 2024 March for five of them, and 12P -- data were taken in 2024 July~8; for more see text.  Comets are ordered according to decreasing semi-major axis (and decreasing orbital period). }\label{tab:objects} 
	\centering
	\setlength{\tabcolsep}{2.0pt} 
	\begin{tabular}{cccccccccccc}
		\hline 
		Name &  a    & q      & e        & P    & i      & Q      & data arc              & T & Epoch  & type of      & T$_{\rm J}$\\
		& [au]  & [au]   &          & [yr] & \degr  & [au]   &\multicolumn{3}{c}{[yyyy mm dd]}    & data arc     & \\
		$[1]$    & $[2]$  & $[3]$  & $[4]$    &$[5]$ & $[6]$  & $[7]$  & $[8]$    & $[9]$        &$[10]$    &  $[11]$ & \\
		\hline\hline
	C/2014 Q3 (Borisov)           & 28.44 & 1.65 & 0.942 & 151.7 & 89.9 & 54.97 & 2014 08 22 -- 2015 02 27& 2014 11 19&2014 12 09& 1opp & 0.184\\ 
		122P/de Vico                  & 17.69 & 0.659& 0.963 &  74.39& 85.4 & 34.71 & 1995 09 17 -- 1996 06 25& 1995 10 06&1995 10 10& 1opp & 0.375\\
		C/2002 K4 (NEAT)              & 17.53 & 2.76 & 0.842 &  73.10& 94.1 & 32.20 & 2002 05 27 -- 2002 10 01& 2002 07 12& 2002 07 25& 1opp & 0.157\\ 
		12P/Pons-Brooks               & 17.21 & 0.782& 0.954 &  71.37& 74.1 & 33.63 & 1883 10 01 -- 2024 07 05& 2024 04 21&2022 01 22& 3opp & 0.598\\ 
		C/2015 GX (PanSTARRS)         & 16.19 & 1.97 & 0.878 &  67.23& 90.3 & 31.08 & 2015 04 08 -- 2016 02 28& 2015 08 26& 2015 09 15& 1opp & 0.314\\ 
		161P/Hartley-IRAS              & 7.718 & 1.28 & 0.834 &  21.44& 95.7 & 14.19 & 1983 11 04 -- 2005 10 12& 1984 01 08&1983 11 02& 2opp & 0.540\\
		\hline
	\end{tabular}
\end{table*}

\subsection{The choice of HTC sample; initial consideration}\label{sub:HTC-sample}

The first comet we started to study was 12P/Pons-Brooks because of its unique behaviour in the current apparition; see also Sect.~\ref{subsec:orbit-12p}.
Initially, we wanted to check to what extent the choice of data arc affects the prediction of the orbital future of this comet. This issue is important because there are at least two ways to obtain the starting orbit of an HTC to study its future orbital evolution. The first is to determine the NG~orbit using the longest possible data arc covering all observed appearances of a given comet. This approach considers the physical nature of a comet, which is a great advantage over GR~models that ignore it. However, it forces the use of some simplifying assumptions regarding the character of sublimation from the comet nucleus. In Sects.~\ref{subsec:NG_model}--\ref{subsec:orbit-5comets}, we show how accurately these NG~orbits and NG~parameters are obtained for the studied HTCs and compare their GR and NG orbits.

A completely different approach may be to choose the last apparition of the comet and to obtain the starting orbit in the GR~regime. In this strategy, one should limit to a GR~orbit because it is rarely possible to accurately determine the NG~parameters. Then, however, the starting orbits for orbital evolution studies usually have larger uncertainties of the orbital elements even though it is a GR~orbit (for the same data arc, the NG~orbit always has larger uncertainties than the GR~orbit because of more parameters to determine using the same set of positional observations). Nevertheless, this can be considered an advantage for examining orbital evolution based on a sufficiently numerous cloud of clones (what we mean by numerous cloud of clones is explained in Sect.~\ref{sec:evolution-classical}) because then the starting swarm of clones covers a broader range of values in each orbital element. This, in turn, allows us to hope that important behaviours will not be missed when studying future orbital evolution. Generally, in the present research, we choose the latter approach and focus on the future dynamics of HTCs in the GR~regime, except 12P for which the preliminary NG~results are also presented.

During the initial inspection of the future orbital evolution of 12P, we noticed several cases where its inclination evolved between $i\,<\,90$\degr and $i\,>\,90$\degr. Next, we decided to find similar occurrences in other HTCs and try to estimate the probability of this phenomenon. Therefore, we selected the remaining comets so that their orbital inclinations were close to 90\degr, in the range 90\degr $\pm$ 6\degr. These five comets make up about half of all known comets in the range of inclinations around 90\degr; see left panel of Fig.~\ref{fig:HTCs_q_i}.

\section{Orbit determination from the positional data}\label{sec:orbit-deterination}

Table~\ref{tab:objects} shows the general characteristics of orbital elements and observational material used by us for orbit determination of analysed HTCs; observations we retrieved from the MPC database in 2024~March for five of them, and for 12P -- in 2024 July~8. We always used full sets of observational materials available in the MPC to determine the most suitable osculating orbit for further dynamic studies, except in the case of comet 122P/de~Vico for which we used only data from the last apparition in the orbit determination process; the reason is explained in Sect.~\ref{subsec:orbit-5comets}.

\subsection{Pure gravitational orbits for investigated HTCs}

In our numerical calculations leading to orbit determinations from positional data set, the equations of motion were integrated using the recurrent power series method \cite{sitarski:1989}, taking into account perturbations by all planets, Pluto and including relativistic effects (it means a Newtonian N-body problem with the eight planets, Pluto, and the Sun modeled as point masses and the comet considered as a massless point particle where relativistic effects are included); the compatible ephemeris obtained by \cite{sitarski:2002} was used. We applied the selection and weighting procedure simultaneously with the orbit determination from the data; for more see \citet{kroli_dyb:2017,kroli-dyb:2010}. The orbits we have determined are discussed in Sects.~\ref{subsec:orbit-12p}--\ref{subsec:orbit-5comets} and are listed in Tables~\ref{tab:12P-starting-orbits}--\ref{tab:5HTCs-starting-orbits}, where GR~solution used as a starting orbit for studying the future evolution of each comet is highlighted in bold in the first column indicating the orbit code used.

\subsection{Non-Gravitational orbits}\label{subsec:NG_model}

In order to obtain a cometary NG orbit for HTCs (in the region well inside the planetary zone), as in \cite{kroli_dyb:2017}, we used a standard formalism proposed
by \citet[][MSY]{marsden-sek-ye:1973}, where three orbital components of the NG~acceleration acting on a comet are proportional to the
$g(r)$-function, which is symmetric relative to perihelion,

\begin{eqnarray}
F_{i}=A_{\rm i} \> g(r),& A_{\rm i}={\rm ~constant~~for}\quad{\rm i}=1,2,3,\nonumber\\
& \quad g(r)=\alpha(r/r_{0})^{-m}[1+(r/r_{0})^{n}]^{-k},\label{eq:g_r}
\end{eqnarray}

\noindent where $F_{1},\, F_{2},\, F_{3}$ represent radial, transverse, and normal components of the NG~acceleration, respectively, and the radial acceleration is defined as positive outwards along the Sun-comet line.  Here, we use two types of the $g(r)$-function defined above; their parameters are given in Table~\ref{tab:gr-like_functions}. 

The coefficient $\alpha$ is determined by the condition  $g(1\,{\rm au}) = 1$. The values of $r_0$ for isothermal water-ice sublimation or more volatile ices sublimation have a simple interpretation, as they represent the approximate distance at which gas production rates drop to about $10^{-3}$ of the rates at a distance of 1~au from the Sun \citep{kro-don:2023}. 

Table~\ref{tab:gr-like_functions} shows how these $g(r)$-like functions differ in the values of the exponents $m$, $n$, and $k$, but most importantly, on the heliocentric distance range of the effective action of the NG~acceleration, which is mainly controlled by the $r_0$ parameter \citep{sekanina:2021}.  

\begin{table}
	\caption{\label{tab:gr-like_functions}Parameters used for the $g(r)$-like formula introduced in Eq.~\ref{eq:g_r}.}  
	\centering
	\setlength{\tabcolsep}{8.0pt}
	\begin{tabular}{ccccc}
		\hline 		 
		$\alpha$    & $r_0$ & $m$      & $n$   & $k$           \\
		\hline 	\hline  
		\multicolumn{5}{c}{Standard $g(r)$ function} \\ 
		0.1113      & 2.808 & 2.15  & 5.093 & 4.6142     \\
		\\
		\multicolumn{5}{c}{More volatile ices than water ice (CO$_2$ or CO sublimation)} \\
		0.01003     & 10.0  & 2.0   & 3.0   & 2.6        \\
		\hline
	\end{tabular}
\end{table}

All the comets studied here have $q<$2.8\,au and we should expect that the sublimation of water ice will dominate at perihelion. However, these comets were also observed at greater distances, some beyond the orbit of Jupiter and even Saturn (as 12P). Therefore, we also checked whether the formula representing the more volatile ices  gives a better fit. However, in each case studied, the standard $g(r)$ formula gave the same or better fit than that based on CO sublimation with r$_0$=10\,au. Therefore, we next discuss only NG~orbits based on the standard $g(r)$ formula. 

\subsection{Osculating orbits for 12P/Pons-Brooks}\label{subsec:orbit-12p}


In the MPC, observations from three appearances are available: 1883-84 (120 observations), 1953-54 (21 observations), and the current one from 2020 06 10 -- 2024 07 05 (over 8 thousand measurements as of 2024 July 8). The comet passed the perihelion ($q\,=$ 0.78\,au) in 2024 on April 21. There were two orbits given in the MPC, both GR, the first one based on the data arc: 2024 Jan. 01 -- 2024 July 05. and the second one with data arc: 2023 Feb. 27 -- 2024 Jan. 30. On the same day, in the JPL the NG orbit was listed with parameters A$_1 = 0.5561 \pm 0.0207$ A$_2 = 0.1196 \pm 0.0122$ in units $10^{-8}$\,au\,day$^{-2}$ (see also Table~\ref{tab:12P-starting-orbits}, orbit JPL/NG/1953-2024; data arc: 1953 June 20 -- 2024 June 26).

To know to what extent the assumption of constant~NG parameters can reflect reality, we attempted to determine the NG~orbit from the first two appearances available in MPC, that is from the years 1883--1954. The MPC has 120~observations from the data arc 1883 Oct 1 -- 1884 April 23. They were made in 6 observatories: Paris, Kuoba, Płońsk, Palermo, Bogenhausen (M\"unchen), and the Royal Observatory from the Cape of Good Hope.  It turned out that the full set of 13 observations made by the Polish amateur astronomer, Jan Jędrzejewicz in his observatory in Płońsk lies perfectly in orbit determined by all data with an $rms$ (rms=3\farcs 00) less than $rms$ for all observations  ($rms$=4\farcs 15). This was also the case with observations made in Bogenhausen ($rms$=2\farcs 43, 9 observations)  and Paris ($rms$=3\farcs 77, 26 observations), although several measurements were rejected as outliers in the last case.  Orbit based on this single appearance is also given in Table~\ref{tab:12P-starting-orbits} (orbit: GR/1883-1884).

For the next appearance, 1953--54, only 21 observations are available in the MPC. They include observations made at the Lick Observatory and the Argentine National Observatory. We found in the literature 54 observations made at the Yerkes and McDonald Observatories by \cite{Bies:1955}. The 1953-54 orbit obtained from this enriched data set (75 positional observations) is given in Table~\ref{tab:12P-starting-orbits} (orbit: GR/1953-1954).

The NG~orbit based on above two appearances is included in the same table, where, for comparison, a GR orbit based on the same set of positional data is also given (orbits NG/1883-1954 and GR/1883-1954, respectively). One can see, that from the data arc 1883-1954, we obtained A$_1<0$, which means that the radial component of the recoil vector is directed towards the Sun. In addition, the normal component dominates over radial and transverse components. Neglecting the normal component, we also get a radial component with a negative sign and the same $rms$ as for GR orbit.

Next, we obtained NG~orbit using 75 observations from 1953--54 and nearly 8,000 observations made during the comet's current appearance; data arc: 2020 June 10 -- 2024 July 5 (data we took from MPC on July~8). Previously, we stated that the assumption of a non-zero normal component does not improve the $rms$ and the value of the A$_3$ parameter itself is small (of the order of $10^{-10}$\,au\,day$^{-2}$). Therefore, we assumed it to be zero, similar to JPL's orbit (see JPL's orbit in Table based on the same data arc). The NG orbit along with the obtained two NG parameters can be found in Table~\ref{tab:12P-starting-orbits}, where, for comparison, we also provided the GR orbit obtained using the same data set. One can see that both NG~orbits (JPL and ours) are in excelent agreement.

It turned out to be possible to determine the NG~orbit with two well-defined NG~acceleration components from the current apparition, 2020--24, where the radial component of the acceleration is larger than that based on the years 1953--2024, and A$_2$ has an opposite sign -- the orbit and NG parameters are given in Table~\ref{tab:12P-starting-orbits} (orbit NG/2020-2024). Considering that the uncertainties of the NG parameters are small compared to their values, it appears that the assumption about the constancy of NG~parameters from 1953 to 2024 is very vague. Therefore, we conclude it is better to base our dynamic study on this comet future on the orbit determined only from the last appearance (orbits: GR/2020-2024 and NG/2020-2024).

\subsection{Osculating orbits for the remaining five comets}\label{subsec:orbit-5comets}

We used exactly the same data arcs for osculating orbit determinations as were done in the MPC and JPL Databases for four comets: 161P, C/2014~Q3,  C/2015~GX, and C/2002~K4. These data arcs are listed in column [8] of Table~\ref{tab:objects}. This table also shows that only 161P was observed in two perihelion passages in this group, data arcs are shorter than one year for three others. For these four comets only GR orbits are given in both cometary databases mentioned above. Our orbital elements with their uncertainties are given in Table~\ref{tab:5HTCs-starting-orbits}. Additionally, we attempted to obtain the NG~orbits for these objects, despite the short data arcs for three of them. Only for comet C/2014~Q3, values of NG~parameters were reasonable enough to include this NG solution in Table~\ref{tab:5HTCs-starting-orbits}. However, the uncertainties of these parameters are comparable to their values, moreover, the $rms$ decrease is less than 0\farcs 01. We present this NG solution in Table~\ref{tab:12P-starting-orbits} only for clear evidence that it is impossible to determine the NG orbits for these four HTCs with reasonable accuracy. 

The case of comet 122P/de Vico is more complex because MPC offers data starting from the 1846 return to perihelion  (two months data arc from February 27 to April 29). There is no data from the return to perihelion around 1920, and next, there are observations from the last perihelion return in 1995-1996 (data arc: 1995 Sept. 18 -- 1996 June 25). However, both databases provide only GR orbits for this comet.  Due to the lack of NG solutions in both databases, it can be concluded that determining the NG~orbits, assuming constant NG parameters, was impossible despite the period covering three perihelion passages. We confirm this conclusion. Therefore, we decided to obtain the GR~orbit only for the  1995-1996 return to perihelion and attempt to obtain the NG~orbit from the same data arc, see Table~\ref{tab:5HTCs-starting-orbits}. It turned out that the NG~parameters for this comet can be determined with reasonable accuracy, however, with a dominant value of the normal component of the NG acceleration and a slight decrease of the $rms$ by 0\farcs 04 relative to the GR~orbit. Compared to the accuracy of the NG parameters achieved for the comet 12P from its current apparition (solution NG/2020-2024 in Table~\ref{tab:12P-starting-orbits}); however, the values of the NG~parameters are notably less accurate for 122P.

\subsection{Nominal orbits for studying future evolution}\label{subsec:starting-osc-orbits}

In practice, the above results show that if we want to include NG~effects in the long-term motion of real HTCs, we would have to make many additional rough assumptions about the level of NG effects and their changes from one perihelion to another. This goes beyond our basic objective, to base on the starting orbits we obtain directly from positional observations and perform calculations with the fewest necessary assumptions. 

Therefore, we finally decided in this investigation to limit all future evolution studies to GR cases, except for comet 12P. We plan similar studies on the some HTCs based on time-consuming calculations on the NG~model grid in the future. 

From the orbital element uncertainties given in Table~\ref{tab:12P-starting-orbits}, it can be seen that the osculating orbits of comets 122P, 161P, and C/2015~GX are of much better accuracy than the orbits of C/2014~Q3 and C/2002~K4. However, the most accurate orbit of the six studied is the GR~orbit of comet 12P, which is a natural consequence of the relatively long data arc used for orbit determination and a rich number of positional observations (more than 8,000 compared to several hundred for the remaining five comets). Further in the text, we use the term {\it nominal orbits} for these osculating orbits obtained directly from observations.

\section{'Classical' evolution}\label{sec:evolution-classical}

\begin{figure*}
	\centering
	\includegraphics[width=8.4cm]{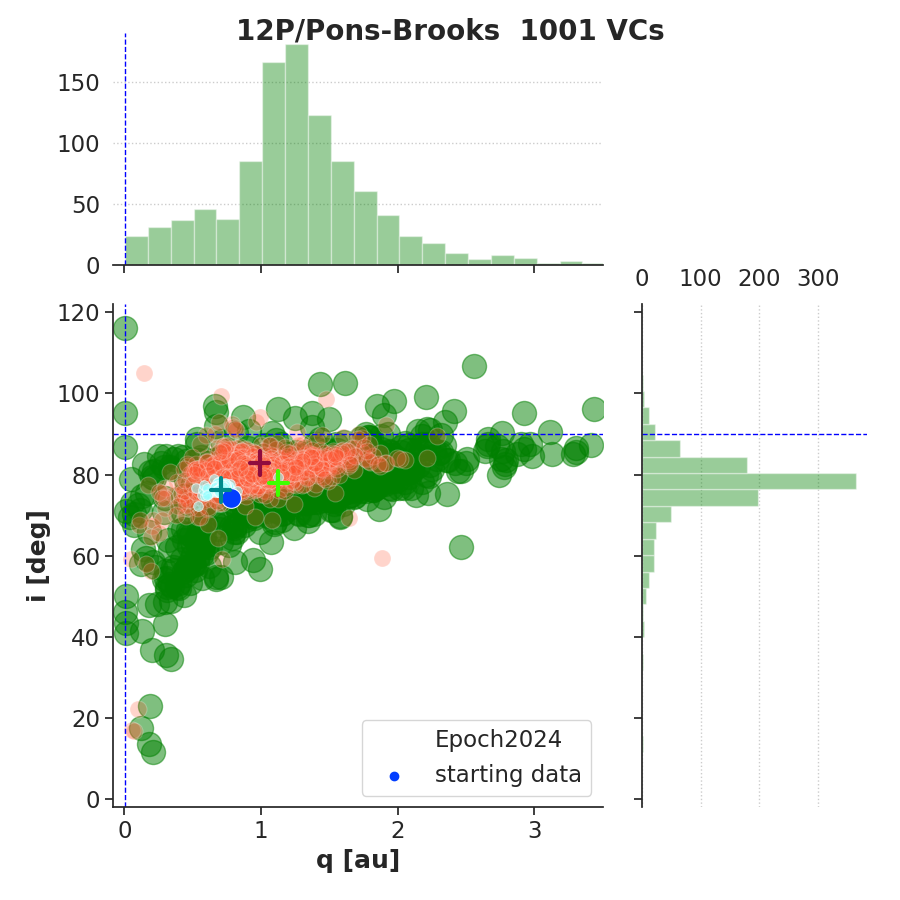}
    \includegraphics[width=8.4cm]{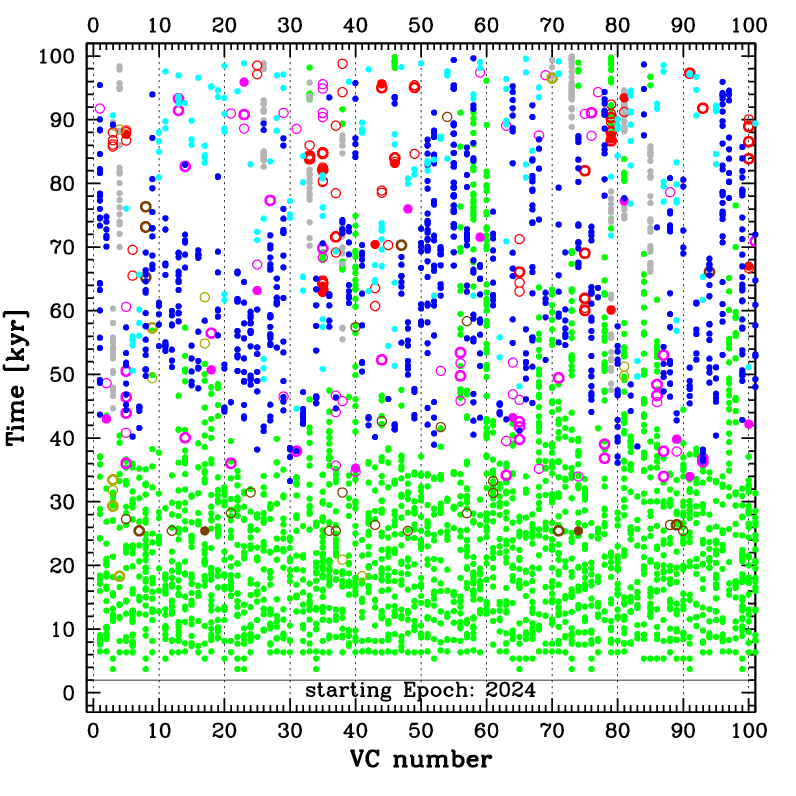}
	\caption{GR evolution of 12P/Pons-Brooks.  Left panel: Orbital elements distribution of 1001\,VCs in 2D plot of $q$ and $i$. Green swarm of big dots in the background described the distribution of VCs at the end of evolution (after 100\,kyr) where the histograms of $q$ and $i$ are visible as marginal plots; only two VC with $i>$110\degr are outside the vertical axis scale. The nominal orbit position is shown by light-green cross. Against this background, are shown the distributions that VC-swarm achieved after 60\,kyr  (orange scatter plot with nominal orbit position as dark red cross within this orange cloud of points) and after 20\,kyr (cyan scatter plot with nominal orbit evolution as blue cross). A few VCs having 3.5\,au~$<q<$5.02\,au are outside the right border of the image box. Vertical dashed line marks $q=0.005$\,au, horizontal: $i=90$\degr. Right panel: Time distribution of approaches to eight planets at a distance below 0.1\,au (solid coloured points) and giant planets in the range of 0.1--0.2\,au (rings) and 0.2--0.3 au (thin rings) for the first 100 clones from the VC~swarm; first VC is the nominal orbit. Colour-coded marks represent: grey -- approaches to Mercury, green -- to Venus, blue -- to the Earth, cyan -- to Mars, red -- to Jupiter, magenta -- to Saturn, brown -- to Uranus, and darkgold -- to Neptun.}	\label{fig:12p-b5-q-i-planets}
\end{figure*}

\begin{figure*}
	\centering
\includegraphics[width=5.9cm]{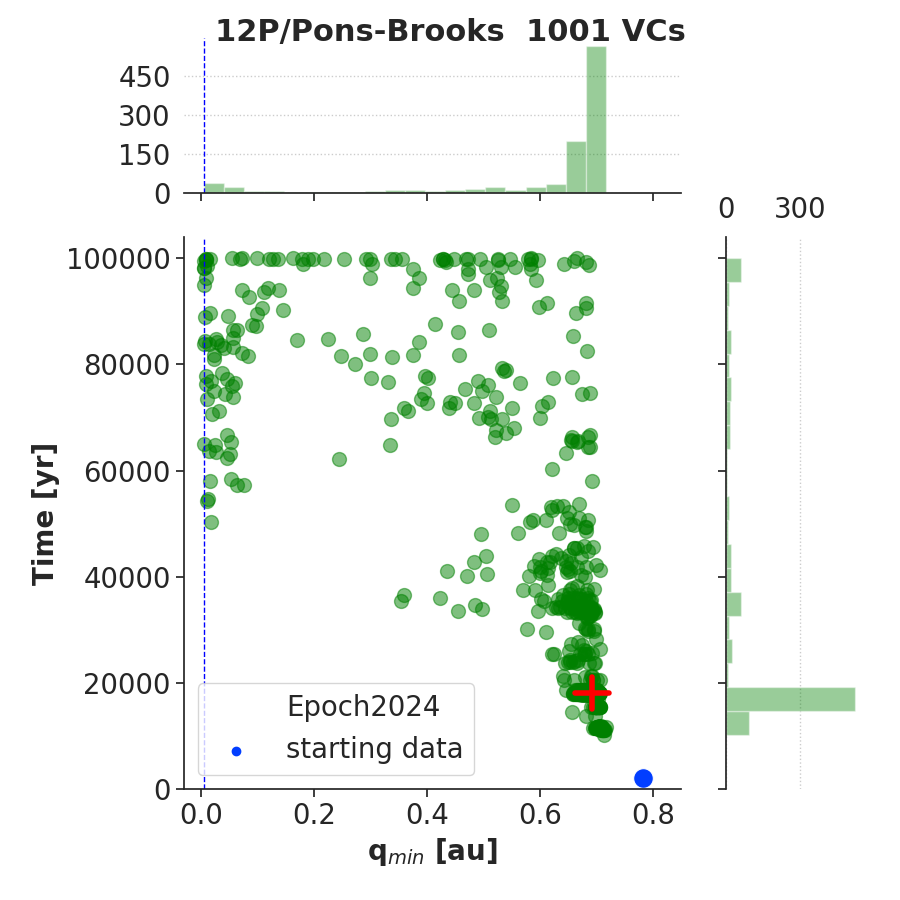}
\includegraphics[width=5.9cm]{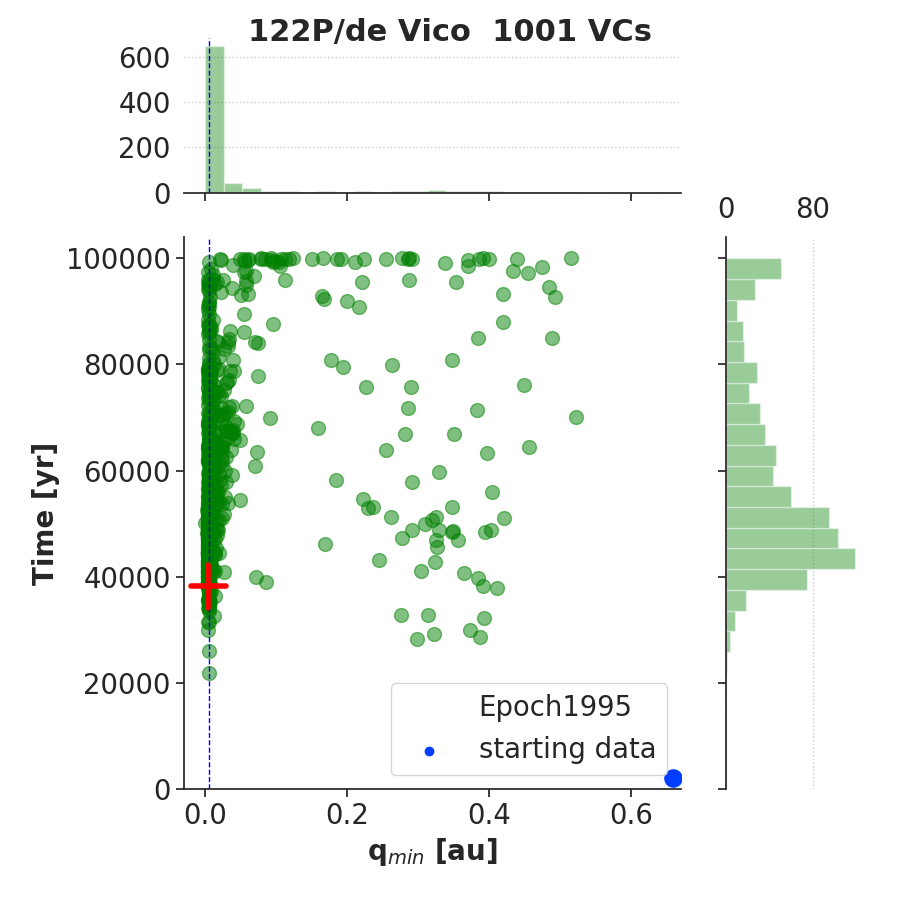}
\includegraphics[width=5.9cm]{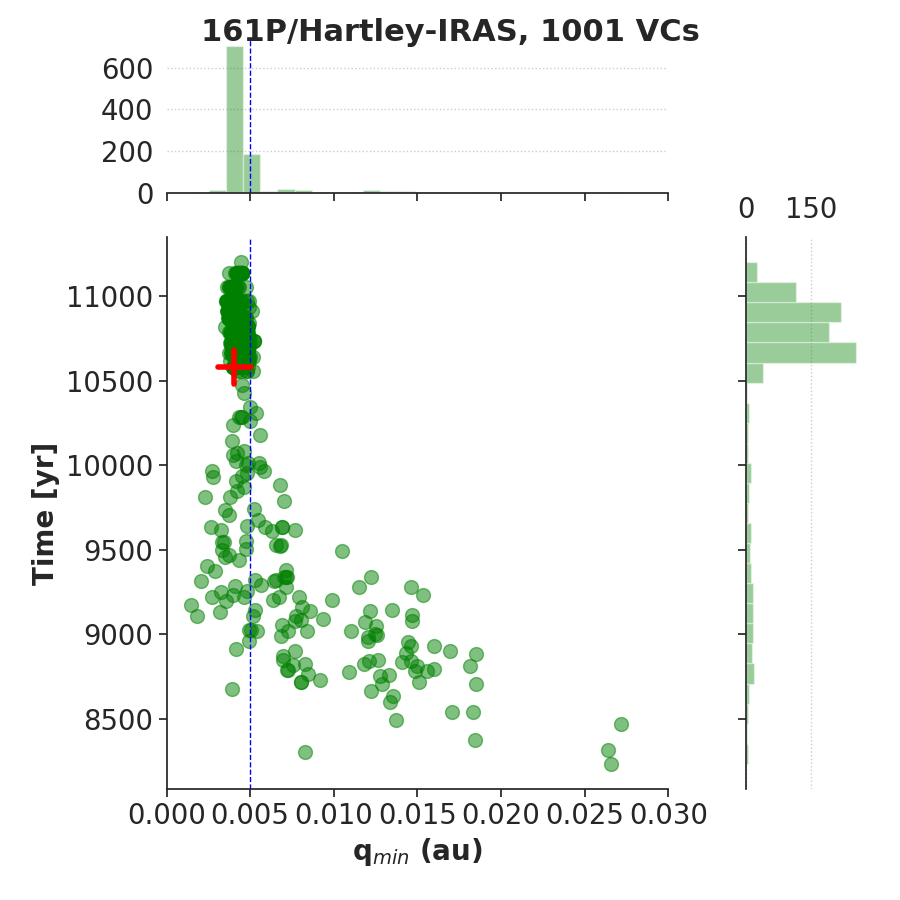}
\caption{Distribution of $q_{\rm min}$ and the moment of approaching this minimum for the swarm of 1001\,VCs for GR~evolution of 12P/Pons-Brooks ($q=0.782$\,au), 122P/de Vico ($q=0.659$\,au), and 161P/Hartley-IRAS ($q=1.28$\,au); the state of $q_{\rm min}$ for nominal orbit is shown by red cross in each plot. Blue dashed vertical line represents the $q=0.005$\,au, for more see text; all VCs are in the image box for 12P and 122P.} \label{fig:12p122p161p_qmin_time}
\end{figure*}

\begin{table}
	\caption{Statistics of orbital elements at the end of dynamical evolution for 1001\,VCs for 12P/Pons-Brooks in the GR and NG model of motion. In the last column, we show maximal values for orbits with $e<1$. }\label{tab:12P-stat_at-the-end} 
	\centering
	\setlength{\tabcolsep}{3.0pt}
	\begin{tabular}{ccccccc}
		\hline 
		Orbital	  &  minimal & 10\%    & 50\%    & 90\%    & maximal  & nominal \\
		element	  &  value   &         & median  &         & value    & value \\
		$[1]$     & $[2]$    & $[3]$   & $[4]$   &$[5]$    & $[6]$    & $[7]$  \\
			\hline\hline \\
		\multicolumn{7}{c}{GR evolution 1001 VCs}		\\ 
		$q$ [au]  & 0.00470  &  0.540  & 1.23    &   1.93  &      5.02 & 1.126  \\
		$e$~~~~~~~~~& 0.5667   &  0.8902 & 0.9318  &   0.9742&     0.9997& 0.9326 \\
		$i$ [deg] & 11.6     &   68.4  &  77.9   &   84.8  &    133.0  & 77.9   \\
		$a$ [au]  & 5.90     &   15.0  &  17.6   &   24.1  &    3767.  & 16.7   \\
		$Q$ [au]  &  9.24    &   28.3  &  34.0   &   47.2  &    7532.  & 32.3   \\
		$P$ [yr]  & 14.3     &   58.1  &  73.7   &  118.   &  231\,174.& 68.3   \\
		\\
		\multicolumn{7}{c}{NG evolution 250 VCs}		\\ 
		$q$ [au]  & 0.00471  &  0.211  & 1.587   &   2.68  &      4.25 & 0.959  \\
		$e$~~~~~~~~~& 0.2695   &  0.7119 & 0.8534  &   0.9350&     0.9981& 0.8558 \\
		$i$ [deg] & 4.03     &   48.0  &  71.2   &   86.3  &    157.3  & 53.3   \\
		$a$ [au]  & 0.10     &   2.42~~&  10.3   &   18.2  &    76.8~~ & 6.65~~ \\
		$Q$ [au]  & 0.14     &   4.30~~&  18.8   &   34.4  &   151.    & 12.3   \\
		$P$ [yr]  & 0.01     &   3.77~~&  32.8   &   77.6  &   674.   .& 17.2   \\
		\hline
	\end{tabular}
\end{table}

\begin{figure*}
	\centering
\includegraphics[width=8.8cm]{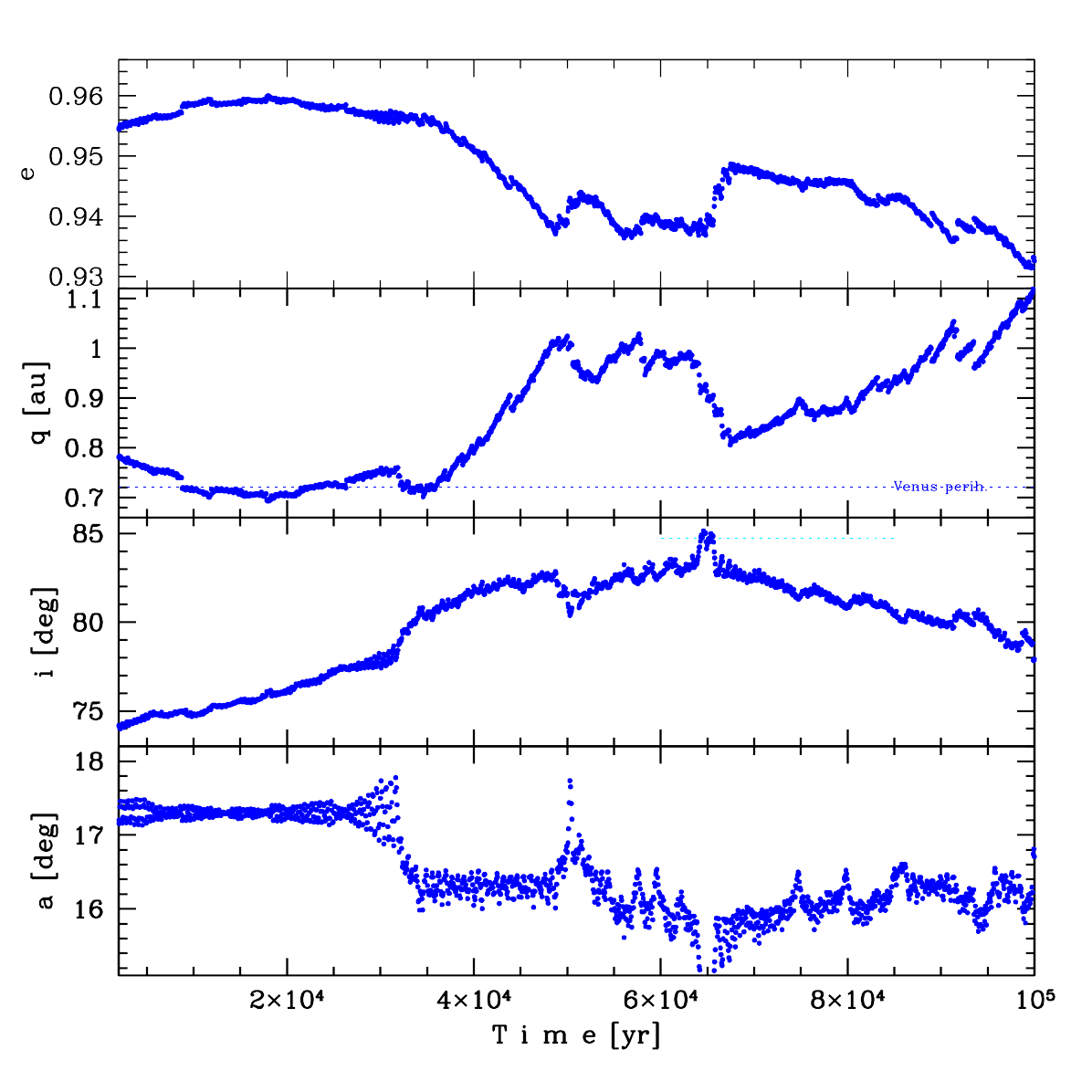}
\includegraphics[width=8.8cm]{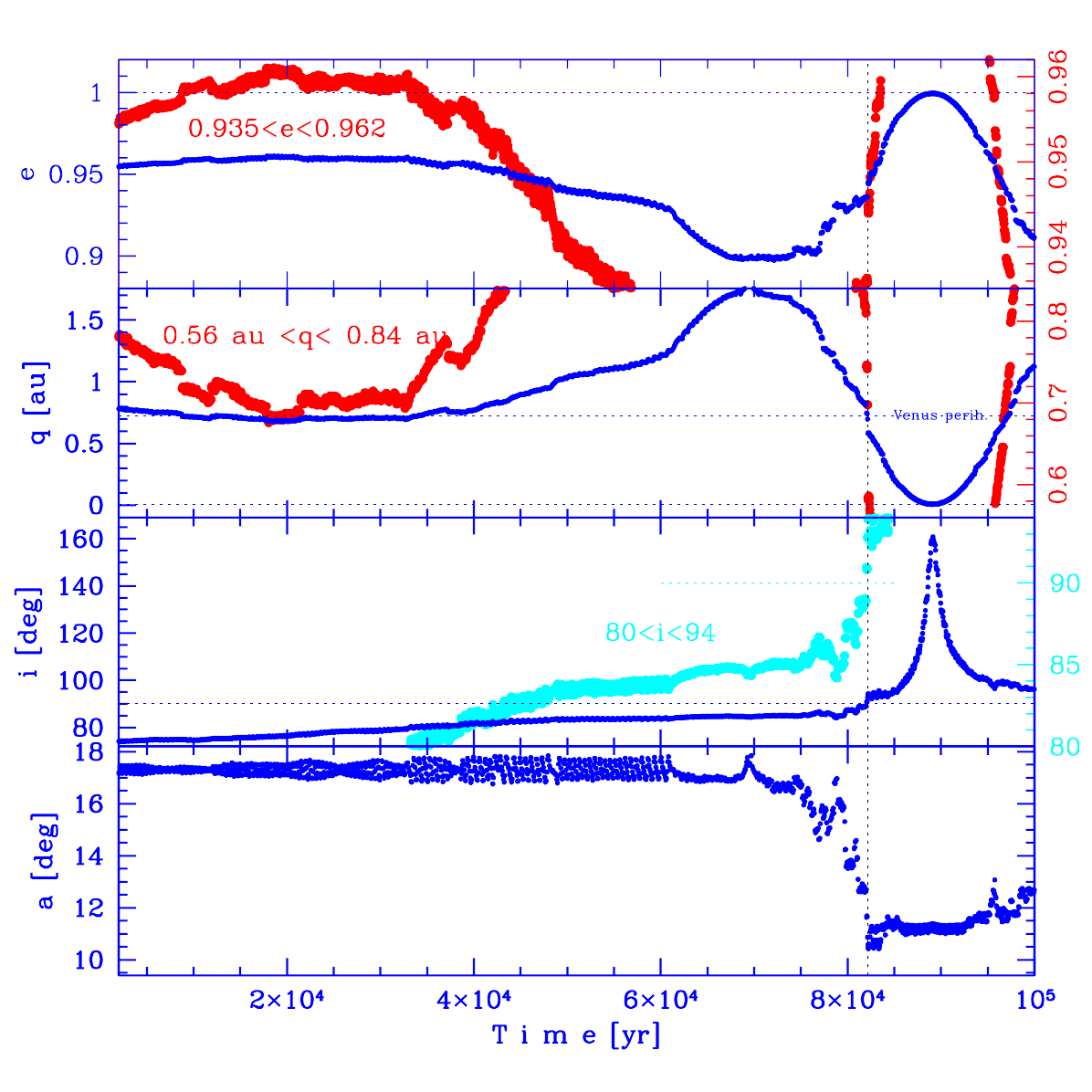}
	\caption{Future GR evolution of orbital elements $e$, $q$, $i$, and $a$ for nominal orbit (left panel) and VC number 211 of 12P/Pons-Brooks (right panel) where values of orbital elements in each perihelion passages are plotted. Right panel: Left blue vertical scales correspond to the blue curves of evolution. The red and cyan vertical scales describe red and cyan curves, respectively, showing some more details of evolution in $e$, $q$, and $i$, the moment of flipping orbit is pointed by vertical dotted thin line.}\label{fig:12p-b5-nominal-211} 
\end{figure*}

\begin{figure}
	\centering
\includegraphics[width=8.8cm]{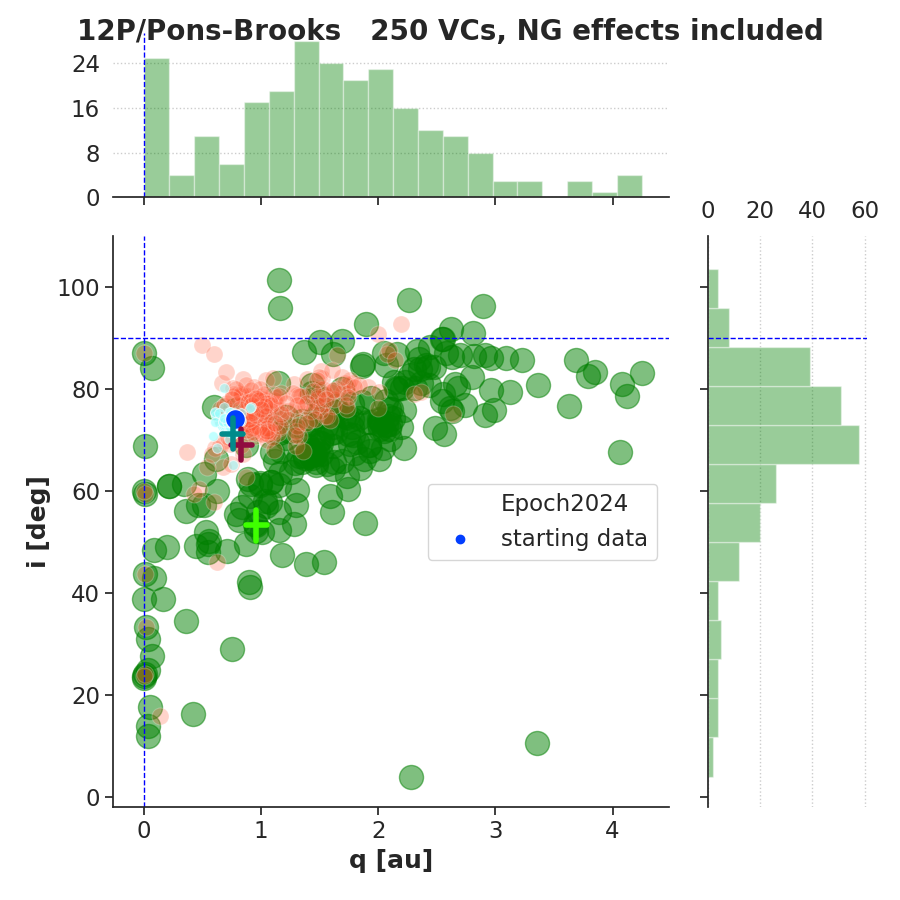}
	\caption{NG evolution of 12P/Pons-Brooks. Orbital elements distribution of 250\,VCs in 2D plot of $q$ and $i$; scales on both axes as in Fig.~\ref{fig:12p-b5-q-i-planets}.}	\label{fig:12p_q_i_ng}
\end{figure}

For studying the future evolution and dynamical behaviour described in Sects.~\ref{sec:evolution-classical}--\ref{sec:evol-LT}, we constructed a swarm of randomly selected 1001\,VCs (including the nominal orbit), in the 6D~space of orbital elements for a given comet where each of VC is well-consistent with observations. The method we used here was described by \cite{sitarski:1998}, and next applied for many dynamical studies of Oort Cloud comets, in particular, to obtain the uncertainties of original semimajor axes of LPCs \citep{krolikowska:2001} and to investigate uncertainties in predicting their past orbital motion in the Oort Cloud; see for example \cite{kroli-dyb:2010,dyb-etal:2024}. 
In the considerations presented below, we have intended to describe the orbital element distribution with uncertainty at a level not worse than 10\% at each stage of the dynamical evolution. Thus, we decided to use the statistics based on 1000\,VCs (plus nominal orbit) which allowed us to obtain the number in each wing of the distribution containing 10\% of VCs with an accuracy of about 10\% ($100 \pm 100^{0.5}$). Consequently, when describing the dispersion of distributions during evolution we use the first decile, the median, and the 9th decile.

In this section, each VC from the swarm is dynamically evolved forwards for a time of 100\,kyr. Analysing the full swarm dynamics and its increasing dispersion in orbital elements with time, we can obtain the full possible spectrum of the future dynamical evolution of a given HTC caused by different dynamical interactions with eight planets. All figures in the current section show orbital elements' evolution at perihelia in the heliocentric ecliptic coordinates. 

\subsection{12P/Pons-Brooks}\label{subsec:evo-12p}

Comet 12P/Pons-Brooks is classified as NEO in JPL with Earth MOID of 0.1807\,au and Jupiter MOID of 2.0177\,au.  

\vspace{0.1cm}

First, we describe the evolution in the GR~regime where the starting swarm of 1001\,VCs is based on the solution GR/2020-2024 listed in Table~\ref{tab:12P-starting-orbits}. Using this swarm, we obtained that the comet makes frequent approaches to Earth in the near future, but all of them are above 0.1\,au, for example in December 2310 it will pass the Earth a bit closer than 0.2\,au. The first close approach to Earth (below 0.1\,au) may not happen until 30\,kyr, see blue dots in the right panel of Fig.~\ref{fig:12p-b5-q-i-planets}. One can see in the same plot that during the future 30\,kyr this comet will experience many close encounters with Venus and close approaches to Uranus may occur at the end of this period of evolution. 

The cloud of cyan dots in the left panel of Fig.~\ref{fig:12p-b5-q-i-planets} shows the evolution of the comet from its starting position (blue point) in the $q-i$ plane after about 20\,kyr. At this stage of evolution, the cloud is still relatively little dispersed with $q$: 0.52\,au~ -- 0.71\,au (median) -- 1.12\,au, $e$: 0.938--0.959--0.970, $i$: 72.2\degr -- 76.2\degr -- 79.4\degr, $a$: 15.2\,au -- 17.3\,au -- 20.5\,au, $Q$: 29.6\,au -- 34.0\,au -- 40.4\,au, and $P$: 59.2\,yr -- 72.0\,yr --93.1\,yr, and in a typical evolution $q$ decreases, $i$ increases, and $e$, $a$, $Q$ a bit increase. It is worth emphasizing here that in the first few thousand years a large part of the VC cloud, including the nominal orbit, may fall into 1:6 resonance with Jupiter (semimajor axis around 17.17\,au), as the semimajor axis evolves from the initial value of 17.155\,au (starting data) to 17.24\,au (after about 3\,kyr from now); see below. 

After this period, close approaches to the giant planets also began, mainly Uranus, Saturn, and Jupiter (right panel of Fig.~\ref{fig:12p-b5-q-i-planets}). Close approaches to Neptune are very rare in this 1001\,VC cloud. After 60\,kyr, the swarm of VCs is notably more dispersed, see orange dots in the left panel of Fig.~\ref{fig:12p-b5-q-i-planets}. Although the typical $q$ and $i$ have increased compared to the previous state, the first VCs with $q$ below 0.05\,au appear. Left plot in Fig.~\ref{fig:12p122p161p_qmin_time} shows the distribution of minimum values of $q$ reached during the evolution up to 100\,kyr. One can see that after 60\,kyr a few VCs reached so low $q$ values that they became sungrazers, more below.

The swarm of green dots on the background in Fig.~\ref{fig:12p-b5-q-i-planets} shows the situation at the end of evolution where the marginal distributions in $q$ and $i$ are dispalyed. It can be seen that although the VC cloud is scattered, the position of the maximum of the distribution in $q-i$ plane is clearly visible and its position does not differ much from the initial state. The obtained ranges of orbital parameters after 100\,kyr are given in Table~\ref{tab:12P-stat_at-the-end}. We found, that all six orbital elements at the end of the evolution of nominal orbit are in the main part of peaks of their distributions and all are close to the median values of VC~swarm.  One can see that we deal here  with a flipping orbit phenomenon in a small percent of VCs, see below for more.

Fig~\ref{fig:12p-b5-nominal-211} shows details of orbital element evolution for nominal orbit (left panel) and chosen VC (number 211). In the last VC, the flipping orbit phenomenon and the evolution to the sungrazing state with $q_{\rm min}=0.0065$\,au  occur.

In nominal orbit evolution we have no encounter with Jupiter and Neptune below 0.9\,au, two encounters with Uranus below 0.9\,au, after about 4\,kyr ($\sim 0.4$\,au) and 24.4\,kyr ($\sim 0.8$\,au) from now, and a numerous series of Saturn encounter below 0.9\,au after about 30\,kyr of evolution, the deepest is 0.29\,au. During $\sim$20\,kyr from the beginning, this VC may be in the resonance 1:6 with Jupiter (semimajor axis about 17.17\,au, see lowest panel of the left plot in the figure). 

In the evolution of VC with sungrazing state, we noticed one encounter with Uranus below 0.9\,au, after about 4\,kyr ($\sim 0.8$\,au), two encounters below 0.9\,au with Saturn (between 37-39\,kyr), and a numerous series with Jupiter after about 77\,kyr. Here the similar as above resonance signature seems to last $\sim$60\,kyr from the beginning (lowest panel of the right plot in the same figure). One of closest encounters with Jupiter to 0.197\,au after about 82\,kyr causes flipping from prograde to retrograde orbit ($i$ change from 88.9\degr to 90.9\degr whereas $q$ decreases from 0.810\,au to 0.745\,au). Two next encounters with Jupiter within the 200\,yr from flipping to 0.272\,au and 0.129\,au gives increase to $i>94$\degr and decrease to $q<0.59$\,au (see red and cyan points at about 82\,kyr in the right panel of Fig~\ref{fig:12p-b5-nominal-211}). In both cases, similarly as in evolution of all VCs, there are numerous encounter with four Earth-like planets. Possible scenarios of the comet falling into an 1:6 J MMR, located around 17.17 au, are also noted. While the comet currently is only just near this resonance, some VCs easily fall into this region after longer numerical integration as mentioned above, as a potential possibility. This is evident in the example shown in Fig.~\ref{fig:12p-b5-nominal-211}. 

The decrease in $q$ from 0.2\,au to the minimal value of 0.0065\,au occurs in 104 orbital revolutions in this case and lasts about 4\,kyr; see the evolution path after 80\,kyr in the right panel in Fig.~\ref{fig:12p-b5-nominal-211}. At the same time, as $q$ decreases, $e$ and $i$ systematically increase to a value very close to 1 and $\sim$160\degr, respectively. Next we observe an increase in $q$ with a decrease in $i$ and $e$. This correlation between $q$ and $i$ ($e$) during the evolution to the sungrazing state could point to the Kozai mechanism in this case (and some others in the VC cloud), see sec.~\ref{sub:evol-Kozai}.
The possibility of this object being in this type of resonance may also be supported by the  arguments presented in \cite{bai-cha-hah:1992}. 
They noticed an episode of small perihelion distance during the future evolution until 1\,Myr for their test particle PB04 having the starting elements similar to the orbit of 12P; this event occurred after 150\,kyr from the present (see their Fig.~8). However, they only examined a single orbit corresponding to 12P. We noticed here several similar cases in the swarm of 1001\,VCs during 100\,kyr of future evolution; all took place always after 80\,kyr. So, to be sure that the Kozai mechanism took place in all detected cases, this would require dynamical study over longer periods than 100\,kyr. 

The decrease to $q$ below 0.01\,au occurs in the evolution of 14~VCs (1.4\%), however only for 4~VCs are below 0.005\,au (0.4\%). In three of the last four, we observe the interaction of inclination flipping with the evolution to the sungrazing state with q<0.005\,au, similar as is shown in the right panel of Fig.4. However, the VC is moving in a direct orbit in nine of the remaining ten cases during the evolution to sungrazers with $q>0.005$\,au.

{{\bf} The flipping orbit phenomenon (mainly from prograde to retrograde type of motion) occurs in about 3\% of VCs (see also Sect.~\ref{sub:evol-below-limit}). }

\subsubsection{NG evolution of 12P. Initial approach. }\label{sub:12P-NG-evo}

Below, we present preliminary NG evolution results for 250~VCs of 12P, where the nominal orbit is the NG/2020-2024 solution given in Table~\ref{tab:12P-starting-orbits}. In our dynamical calculations, we keep constant NG~parameters for each VC; however, each VC has a unique pair of values $A_1, A_2$, randomly selected simultaneously with six orbital elements (where each VC is well-consistent with observations of 12P).

During the study of NG~evolution it turned out that $q$ drops to values below 0.1\,au much more often than in the GR~case; when comparing marginal distributions in Fig.~\ref{fig:12p_q_i_ng} with those in Fig~\ref{fig:12p-b5-q-i-planets} it should be remembered that in the NG case we have 4 times less numerous swarm of VCs compared to the GR case.  It seems that one should scale the NG parameter values so that the NG acceleration never increases dramatically when evolution leads to large decreases in $q$ (see eq.~\ref{eq:g_r}). This is not implemented in the current calculations, and therefore one can suspect that the number of instances of reaching the sungrazer state is overestimated.

The second problem is the number of revolutions (numbers of returns to perihelion) the comet makes during its future evolution of 100\,kyr. It is estimated that JFCs are in the active stage during about 1000 orbits around the Sun, about 10\,kyr \citep{lev-dun:1997}. Next, they typically become dormant comets, in which the sublimation-driven activity is at most below the level that can be detected.
In our calculations, typical VC makes from 1400 to 3000 orbits around the Sun in the NG~regime, but there are cases of such a large decrease in the orbital period that there are a much larger number of revolutions. For comparison, there were from 1300 to 1500 revolutions during 100\,kyr of GR evolution. Therefore, to be closer to reality, it is probably worth introducing the extinction of the cometary activity (comet ageing) by reducing the NG parameters over time. This can be done in many ways, finding the most fitted to-reality method requires many tests with different parameters.

The above suggests that the results presented here should rather be considered as hints of how NG effects may operate in the dynamical evolution of this comet.  Therefore, we present only some general remarks on the results obtained in the approach described above. To our knowledge, this is, however, the first such attempt to study NG~evolution based on the swarm of NG~orbits fitted to positional data of this HT comet.

\vspace{0.1cm}
The main results of NG~evolution of 12P are as follows.

\vspace{0.1cm}
(i)~~Fig.~\ref{fig:12p_q_i_ng} compared with the left panel of Fig.~\ref{fig:12p-b5-q-i-planets} (their marginal distributions) and Table~\ref{tab:12P-stat_at-the-end} confirm that NG~effects lead to a larger dispersing of the VCs cloud than in the case of GR~evolution. In particular, relatively more VCs evolve towards orbits with lower inclination and to sungrazer states. The flipping orbit from prograde to retrograde type of motion is similarly rare as in the GR regime of motion.

\vspace{0.1cm}
(ii)~~The nominal orbital elements are notably different from the median values; however, it should be remembered that the swarm of VCs is four times less numerous than for GR orbital evolution presented here.

\vspace{0.1cm}
(iii)~~It follows from Table~\ref{tab:12P-stat_at-the-end} that when we take NG effects into account, comet 12P evolves much faster towards decreasing eccentricity and semi-major axis.

\vspace{0.1cm}
(iv)~~In the GR model, all VCs did not substantially change values of $T_j$ during the evolution, and all clones remained HTCs (always with $T_J <0.75$) on a timescale of 100\,kyr studied here. In the NG model, the changes of $T_j$ are faster, and 33 of VCs (13\%) reach $T_j>2.0$ during the studied period (the starting value was $T_j=0.5993$). The first case of crossing the threshold between HTCs and JFCs in the swarm of 250~VCs occurs after about 50\,kyr. The cases of reaching the JFC state can be divided into three groups:
\begin{enumerate}
		\item Nine VC ($\sim 1$\%) evolved to JFC orbits ($2 < T_j <3$) at the end of the evolution (the earliest after 71 kyr) and stayed in this state until the end of integration. As JFC their $q$ and $i$ evolve in the ranges 0.07\,au\,$ < q < $\,3.5\,au and 4\degr $ < i < $\,65\degr.
		\item In 21 cases ($\sim$8\%) the VC orbit became a JFC orbit and next Encke type comets (ETCs). Typically, this happened after a few thousand orbits, the earliest after 51 kyr (after 1060 orbits). The JFC stage usually lasted for several hundred to several thousand orbits when the orbit turns to Encke type and orbit is this type until the end of calculations (6 VCs, for JFCs+ETCs states: 0.07\,au\,$ < q < $\,2.1\,au and 14\degr $ < i < $\,64\degr) or evolved to sungrazing state and ended in the Sun (15 cases, for JFCs+ETCs states: 0.12\,au\,$ < q < $\,1.0\,au and 12\degr $ < i < $\,73\degr ~and one VC on retrograde orbit in both states). A fairly typical example is VC number 6 which, after about 84\,kyr, moved in an orbit with orbital elements consistent with JFCs and, after another about 10\,kyr, evolved to an Encke-type orbit. Having ET~orbit the VC stays for about 3\,kyr and ends in the Sun. However, the number of revolutions is worrying because the JFC state was typically reached after more than 2000 revolutions. The Encke state started after about 4000 orbits from the beginning of evolution.
	\item Three VCs ended as sungrazers with  2~$< \> T_J \> <$~3 and $q\><\>$0.005\,au after 39\,kyr, 48\,kyr, and 55\,kyr, respectively.
\end{enumerate}
(v)~~Additionally, we noticed 8~VCs with $q\><$\,0.1\,au and $T_J<2$ at the end of the studied evolution.

\vspace{0.1cm}
These results and high number of revolutions obtained suggest that the long-term NG dynamical evolution scenarios of an actual comet should include the recipes on limiting the level and lasting of NG~effects during the 100\,kyr of dynamical evolution. 

\begin{figure*}
	\centering
\includegraphics[width=8.4cm]{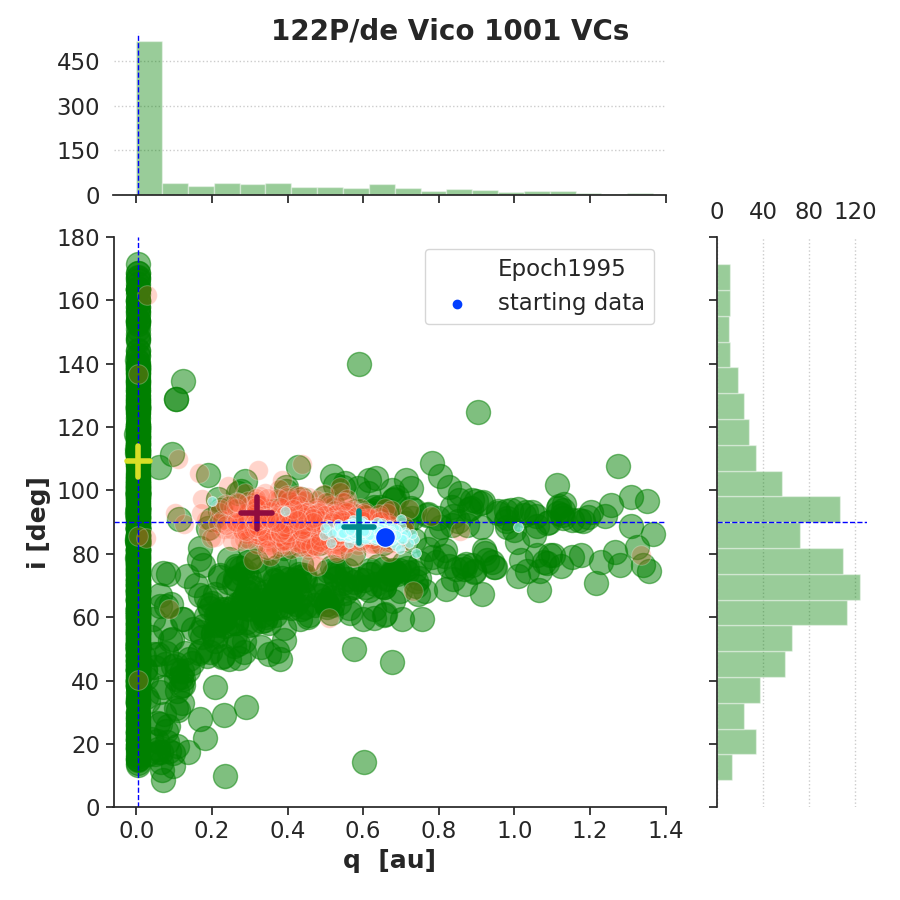}
\includegraphics[width=8.4cm]{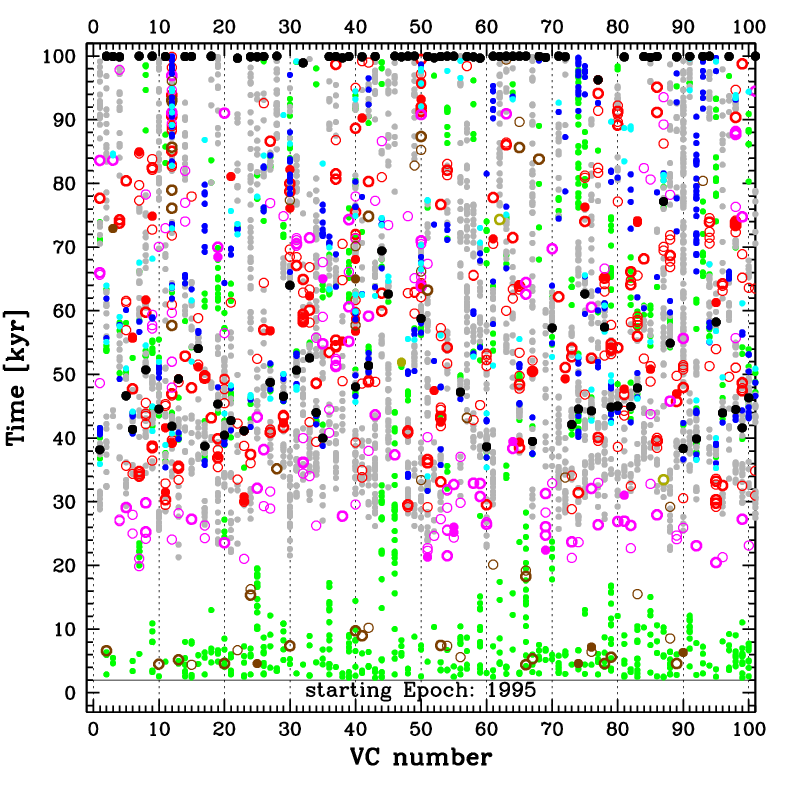}
	\caption{GR evolution of 122P/de Vico.  Left panel: Orbital elements distribution of 1001\,VCs in 2D plot of orbital elements in $q$ and $i$ in an analogous convention as in Fig.~\ref{fig:12p-b5-q-i-planets}; however here cyan, orange, and green  swarms of dots show the distributions that VC-swarm achieved after $\sim$10\,kyr,  $\sim$30\,kyr and  $\sim$100\,kyr (yellow cross represents the nominal orbit) of future evolution. Right panel: Time distribution of close approaches to planets; symbols are colour-coded as in Fig.~\ref{fig:12p-b5-q-i-planets}. The black dot indicates when each clone reached its minimal $q$ value during the evolution.} \label{fig:122p-b5-q-i-planets}
\end{figure*}

\begin{figure}
	\centering
\includegraphics[width=8.8cm]{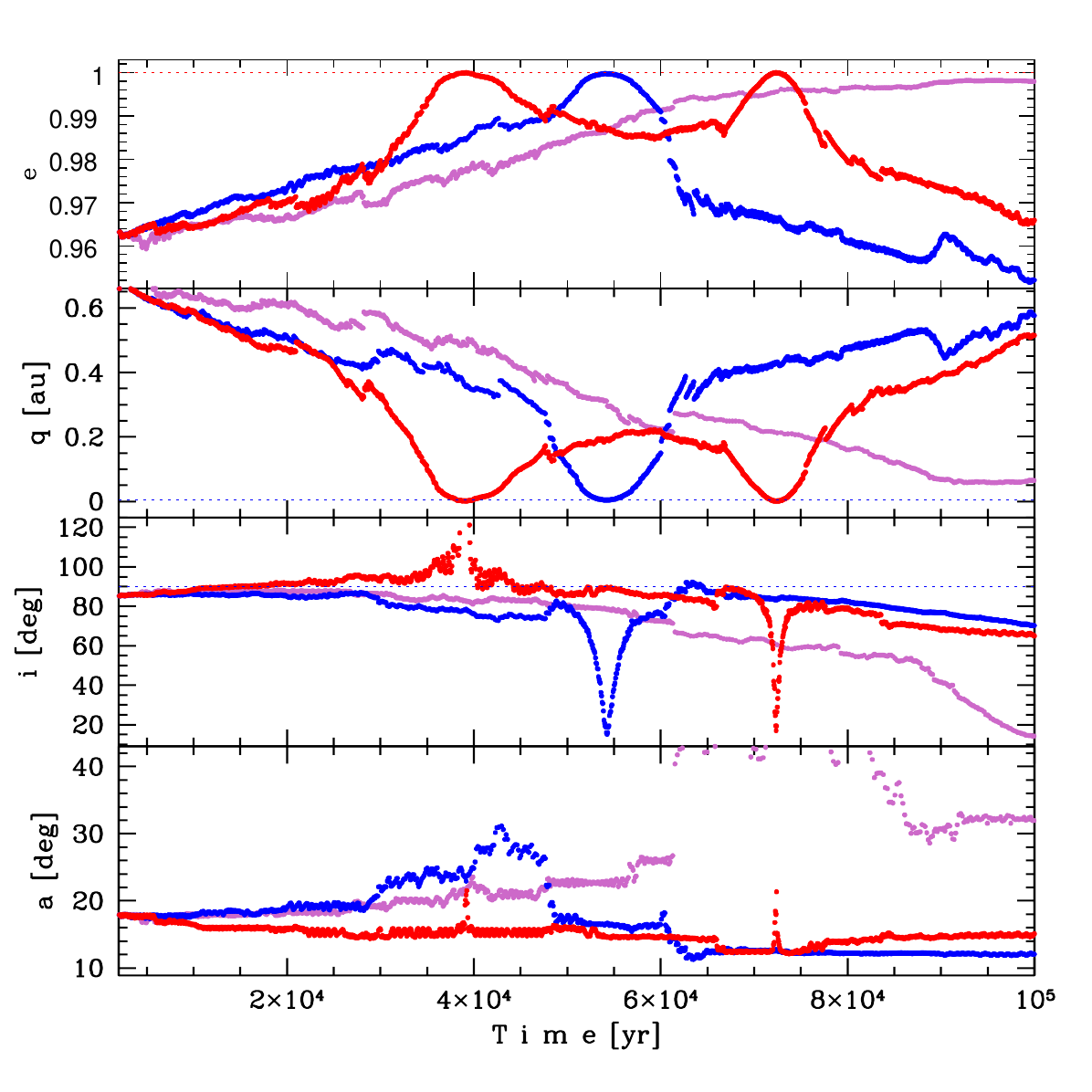}
	\caption{Three different path of future GR evolution towards small $q$ for comet 122P/de Vico where values of four orbital elements in each perihelion passages are plotted. 
		Nominal orbit are given in red ($q_{min}<0.005$\,au), and VC numbers 927  ($q_{min}<0.005$\,au), 310 ($q_{min}=0.06$\,au) in blue and violet, respectively; for more see text. In this picture, we did not freeze the evolution after achieving the $q<0.005$\,au.} \label{fig:122p_ewo_smallq_flip} 
\end{figure}

\begin{figure*}
	\centering
\includegraphics[width=8.8cm]{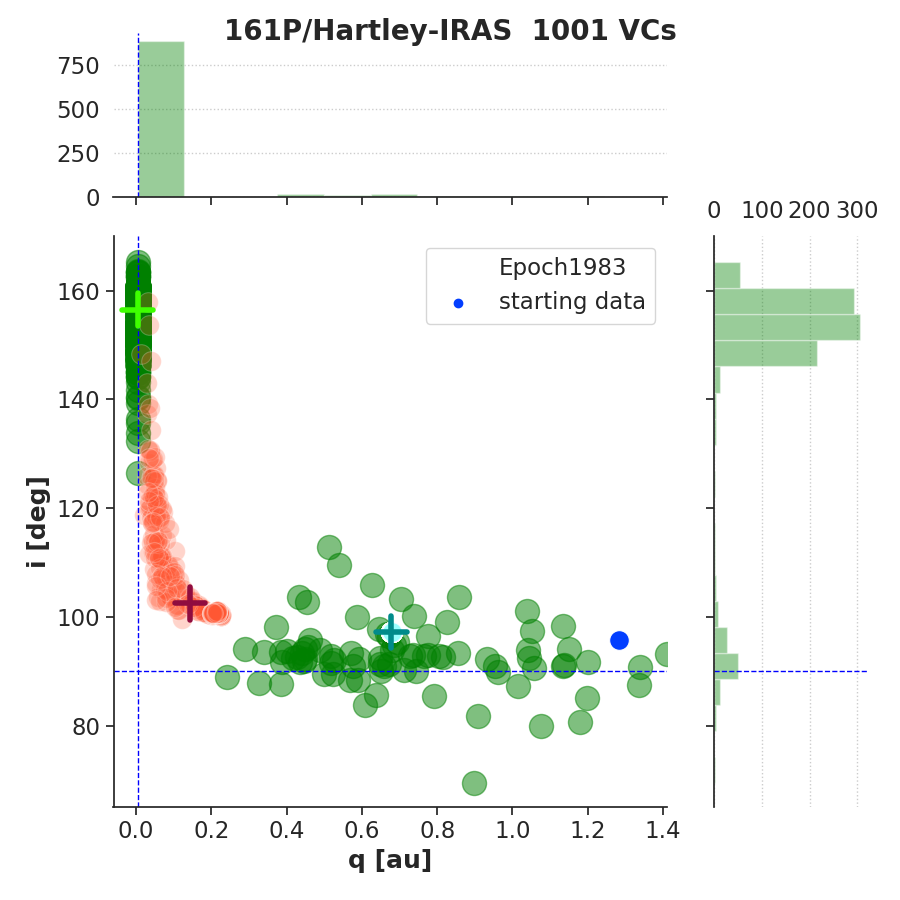}
\includegraphics[width=8.8cm]{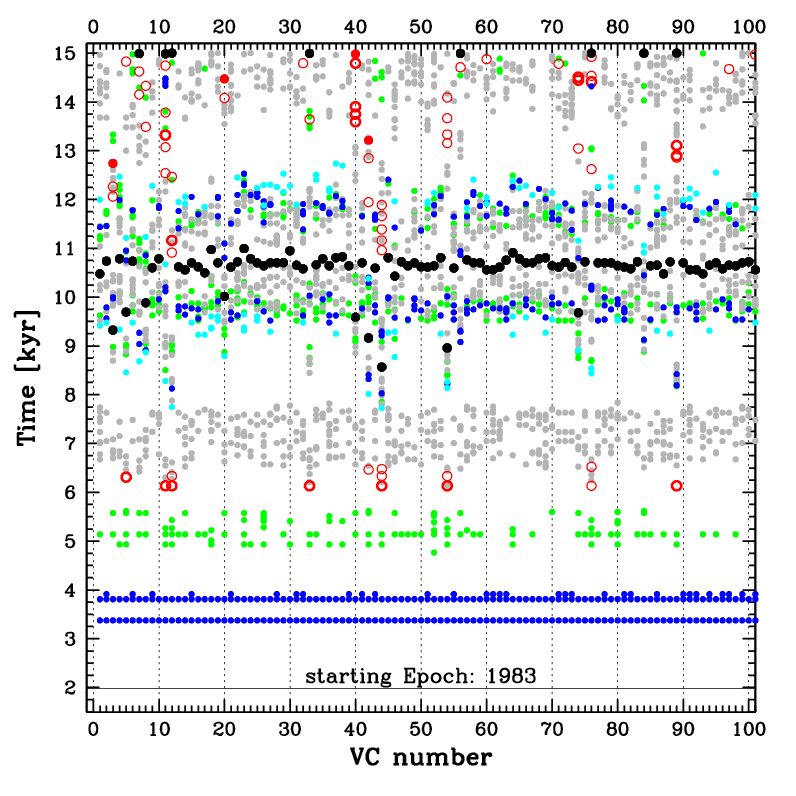}
	\caption{GR evolution of 161P/Hartley-IRAS in the time interval 2000--15000~AD. Left panel: Orbital elements distribution of 1001\,VCs in 2D plot of orbital elements in $q$ and $i$ in similar convention as in Fig.~\ref{fig:12p-b5-q-i-planets}, where cyan  orange, and green dots represent swarms after $\sim 3$\,kyr, $\sim 6.5$\,kyr, and  $\sim 13$\,kyr from now, respectively. Right panel: Time distribution of approaches to planets at a distance below 0.1\,au (solid coloured points) and giant planets in the range of 0.1--0.2\,au (rings) and 0.2--0.3\,au (thin rings) for the first 100 clones from the VC~swarm. Colour-coded marks represent: grey -- approaches to Mercury, green -- to Venus, blue -- to the Earth, cyan -- to Mars, red -- to Jupiter.  There were no close approaches to Uranus or Neptune in the entire swarm of 1001\,VCs. The black dots indicate moments when each clone reached its minimal $q$ value during this 13\,kyr of future evolution.}	\label{fig:161p-b5-q-i-planets}
\end{figure*}

\subsection{122P/de Vico}\label{subsec:evo-122p}

Similarly as 12P, this comet also is classified as NEO in JPL with Earth MOID of 0.3168\,au.  The evolution of 122P is more dynamic as the swarm of VCs experiences numerous approaches to giant planets. They happen in greater numbers than in the case of 12P, and a close approach with Uranus may occur just a few thousand years after the beginning of evolution. The right panel of Fig.~\ref{fig:122p-b5-q-i-planets} shows the picture for the first 100~VCs where only deep close-ups of the planets are marked. It can be seen that regardless of the VC, at the beginning of evolution (0--20\,kyr), we are most often dealing with close approaches to Venus (green dots) and, rarely occurring, to Uranus (2-3\% of VCs have then close approaches before 20\,kyr). After 20\,kyr, the semimajor axis of VCs stays within 15--21\,au, with a few VCs outside this range (starting value: 17.7\,au).
During this period, the nominal orbit shown by coloured crosses in the left panel in Fig.~\ref{fig:122p-b5-q-i-planets}, represents the typical evolution of VC and is as follows: that $q$ from its starting value (blue big point) slowly decreases in $q$ whereas $i$ slowly increases, and the nature of the motion begins to change from the prograde to retrograde; for the nominal orbit it is occured after about 20\,kyr.

After 10\,kyr we have $q$ in the range 0.199\,au -- 0.564\,au (10\%) -- 0.602\,au (median) -- 0.650\,au (90\%) -- 1.008\,au, and 
$i$ : 80.4\degr -- 86.0\degr --  87.1\degr -- 88.3\degr -- 96.6\degr (cyan points in $q$ vs $i$ planet in Fig.~\ref{fig:122p-b5-q-i-planets}). After 30\,kyr these ranges are notable wider with $q$: 0.0037\,au -- 0.326\,au -- 0.455\,au -- 0.577\,au -- 1.33\,au  and $i$: 40.1\degr --  84.79\degr --  89.58\degr --  93.05\degr -- 161.58\degr (orange points at the same plot). It is also visible in the scatter of these orange points that the VCs have typically evolved to slightly larger value of $i$ than its starting value and the perihelia are more closer to the Sun. Therefore, in this period, the phenomenon of changing motion from direct to retrograde is already widely visible (for more details see Sect.~\ref{sec:evol-flipping}), and the first of VCs become sungrazing comet. 
 
From the moment the first VCs with perihelion below 0.005\,au appear (see also middle plot of Fig.~\ref{fig:12p122p161p_qmin_time}), we continue to study the evolution also of these orbits; however, in the left panel of Fig.~\ref{fig:122p-b5-q-i-planets}, we proceed as we had finished the evolution when $q$ drops below 0.005\,au, that is we assume that VC did not survive this close encounter with the Sun (for more see Sec.~\ref{sub:evol-below-limit}). As a result of this procedure, it can be seen that many of the 1001\,VCs became sungrazing comets with 10\degr$<i<$170\degr in the period between 30--100\,kyr. On the right panel, the moment the VC reaches the minimum distance from the Sun during the entire period up to 100\,yr is marked with a black point. Typically, the decrease in $q$ from 0.2\,au to 0.005\,au occurs in less than 100 orbital revolutions. At the same time, as $q$ decreases, $e$ systematically increases to a value very close to 1. This is a similar correlation between $q$ and $i$ pointing to Kozai mechanism as we mentioned for 12P. Fig.~\ref{fig:122p_ewo_smallq_flip} shows the nominal orbit evolution and one of the chosen VC (number 927); both with a decrease to $q<0.005$\,au and such a signature. For nominal orbit the sungrazing state is achieved after 38.2\,kyr (see first deep minimum in $q$ for red curve), and for VC number 927 -- after 53.8\,kyr ($q$-minimum in blue). In both cases, the inclination flip occurs. The nominal orbit moves on a retrograde orbit for a long time, whereas the inclination flips to retrograde motion for VC number 927 is a short event in its dynamics.  
The evolution of VC number 310 shown in Fig.~\ref{fig:122p-b5-q-i-planets} in violet is also worth mentioning. 
Here we are dealing with a relatively slow orbital changes to the state of $q\simeq 0.06$\,au and $i\simeq 14$\degr ~after 100\,kyr with simultaneously $e$ increasing; flipping not occur in this case.

After 40\,kyr we have a group of about 13\% of all VCs with an episode of $q$ below 0.005\,au, including the nominal orbit (Fig.~\ref{fig:12p122p161p_qmin_time}). The remaining VCs have their main concentration between 0.2\,au~$<q<$~0.8\,au, and the semimajor axes of these VCs stay within 10--40 au, with a few VCs outside this range.

After 50\,kyr, about 30\% of the VCs experienced an episode of $q$ below 0.005\,au, and most of the remaining have 0.005\,au~$<q<$~0.6\,au with the maximum of distribution located on the side of retrograde orbits, but the number of prograde orbits dominating in general. 

At the end of evolution (after 100\,kyr), two notable groups we have under our assumptions of cometary decay. The first group contains about 50\% of all VCs with a $q<0.005$\,au episode. In all these cases we noticed a remarkable correlation between $q$ and $i$ (Kozai mechanism, see Sect.~\ref{sub:evol-Kozai}) which seems to effectively operating during evolution to sungrazing state in the case of 122P. Consequently, the VCs experienced episodes of $q<0.005$\,au are highly dispersed in $i$. 
The rest of VCs form a roughly equal group in number in which $q$ did not fall below 0.005\,au during the 100\,kyr future evolution, and 10\% of entire swarm of VCs have $q>0.94$\,au. There is a clear preference for VCs in prograde orbits in both groups. The general statistics of orbital elements after 100\,yr are shown in the Table~\ref{tab:122P-stat_at-the-end} where these statistics include the frozen VCs when $q$ drops below 0.005\,au. This last group of $q$ below 0.005\,au includes the nominal orbit. Values of orbital elements of nominal orbit differ from the median values of full VC~swarm, the most in $i$.

In five VCs (0.5\% of all) we got hyperbolic orbits due to a close approach to Jupiter, below 0.01\,au. All these events occurred after more than 49\,kyr of evolution and always after an episode of VC reaching a minimum $q$ in the range from 0.008\,au to 0.12\,au.  

The evolution of 122P in the context of migration to the sungrazing state was studied by \cite{bai-cha-hah:1992}. Starting from the orbit given in Marsden's {\it Catalogue of Cometary Orbits} of 1989, they obtained for 122P a decrease in $q$ to the sungrazing level together with a simultaneous reduction in $i$ to about 20\degr ~(their Fig.~2), with a minimum of $q$ reached after about 50\,kyr during the future evolution. From the case illustrated by \cite{bai-cha-hah:1992} it can be concluded that no inclination flip occurred during the period studied (100\,kyr). In terms of reaching the sungrazing state, the cited result corresponds well with our findings (our orbit is based on the observations that took place after the publication by \cite{bai-cha-hah:1992}), the sungrazing states start to occur in our swarm of 1001\,VCs after 30\,kyr, as was pointed above. Until 60\,kyr, every second VC already reached this state, including the nominal. However, in the evolution of the swarm of VCs we study, flipping occurs on a  common scale (see Fig.~\ref{fig:122p-b5-q-i-planets} and Sect.~\ref{sec:evol-flipping}), including the nominal (Fig.~\ref{fig:122p_ewo_smallq_flip}). 

\begin{table}
	\caption{Statistics of orbital elements at the end of GR~dynamical evolution for 1001\,VCs for two comets with large probability of destruction in the scenario where the evolution of individual VC is stopped when $q$ decreases below 0.005\,au. In the case of 122P, 5 VCs with $e>1$ are excluded from this statistics.}\label{tab:122P-stat_at-the-end} 
	\centering
	\setlength{\tabcolsep}{2.0pt} 
	\begin{tabular}{ccccccc}
		\hline 
Orbital	  &  minimal & 10\%    & 50\%    & 90\%    & maximal  & nominal \\
element	  &  value   &         & median  &         & value    & orbit   \\
$[1]$     & $[2]$    & $[3]$   & $[4]$   &$[5]$    & $[6]$    & $[7]$   \\
			\hline\hline
	\\
\multicolumn{7}{c}{122P/de Vico (after 100\,kyr) }		\\ \\
 $q$ [au]  & 0.00013  &  0.0043 & 0.0387  &   0.956 &      7.17  & 0.00446\\
 $e$~~~~~~~~~& 0.6208   &   0.9477&  0.99942 &  0.99976 &  0.99989   & 0.99971 \\
 $i$ [deg] & 5.79     &   40.1  &  75.3   &  117.~~~   &    171.    & 109.~~~ \\
 $a$ [au]  & 5.42     &   13.6  &  17.4   &   39.3  &    956.    & 15.6 \\
 $Q$ [au]  & 9.83     &   27.1  &  34.5   &   77.8  &   1911.~~~~& 31.2 \\
 $P$ [yr]  & 12.6     &   50.1  &  72.8   &  246.   &  29544.~~~~~& 61.6 \\
		\\
\multicolumn{7}{c}{161P/Hartley-IRAS (after 15\,kyr)}		\\ \\
$q$ [au]    & 0.00386  &  0.00456& 0.00483 &   0.407 &      2.48  &  0.00466 \\
$e$~~~~~~~~~& 0.5684   &   0.9453&  0.99936 &  0.99941 &  0.99951   & 0.99939 \\
$i$ [deg]   & 69.5~~~~ &  101.~~~~~~& 153.~~~~~~  &  160.~~~~~~ &  165.~~~~ &  ~~~156.~~~~ \\
$a$ [au]    & 5.30     &  ~~7.60 & ~~7.65  &  ~~7.69 &     18.2   &  ~~7.68   \\
$Q$ [au]    & 9.02     &   15.2~ &  15.3~  &   15.4  &     35.5~~ &  15.4 \\
$P$ [yr]    & 12.2~~~  &   20.9~ &  21.2~  &   21.3  &     77.6~~ &  21.3 \\
		\hline
	\end{tabular}
\end{table}

\subsection{161P/Hartley-IRAS}\label{subsec:evo-161p}

\begin{table}
	\caption{Deepest approaches of 161P to planets over the next 1,500\,yr}\label{tab:161P-planet-approaches} 
	\centering
	\setlength{\tabcolsep}{3.0pt} 
	\begin{tabular}{ccc }
		\hline 
Planet	  &  Date [yyyymmdd]  & distance [au] \\
			\hline\hline
Jupiter & 2028\,06\,02 & 0.396 \\
Jupiter & 2171\,02\,03 & 0.668 \\
Mars    & 2584\,12\,31 & 0.22 \\
Earth   & 2979\,10\,08 & 0.16 \\
Mars    & 3062\,09\,19 & 0.20 \\
Earth   & 3374\,10\,16 & 0.095$\pm$0.003 \\
\hline
\end{tabular}
\end{table}

\begin{figure*}
	\centering
\includegraphics[width=5.9cm]{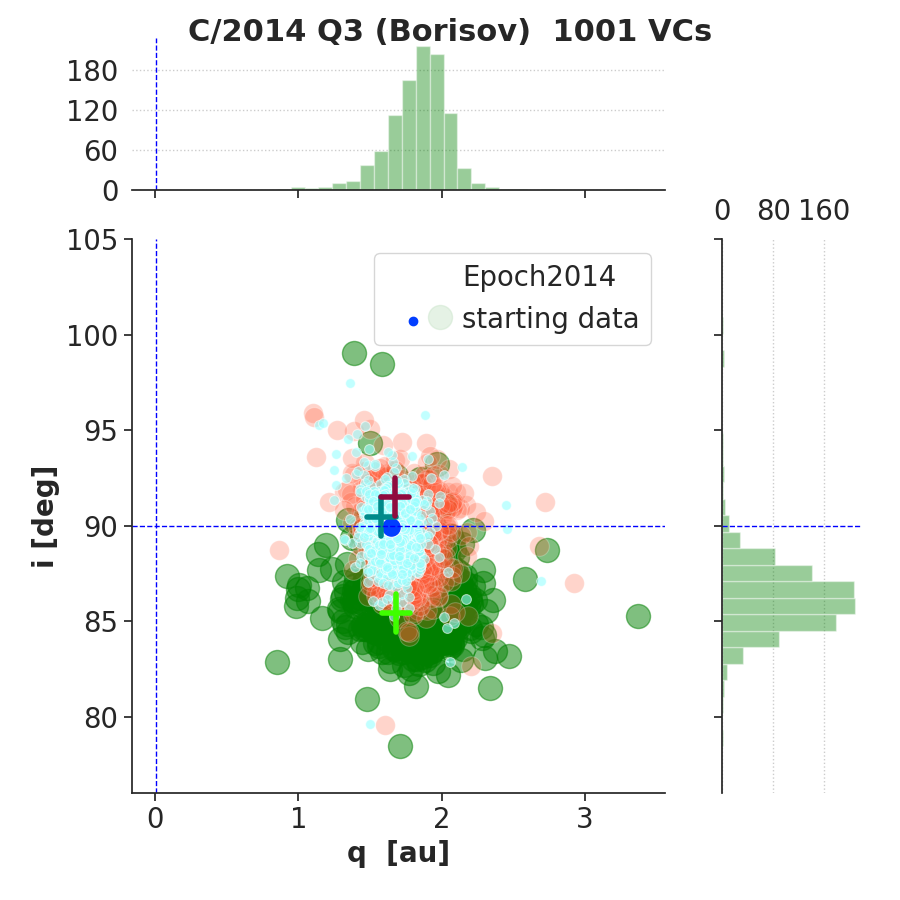}
\includegraphics[width=5.9cm]{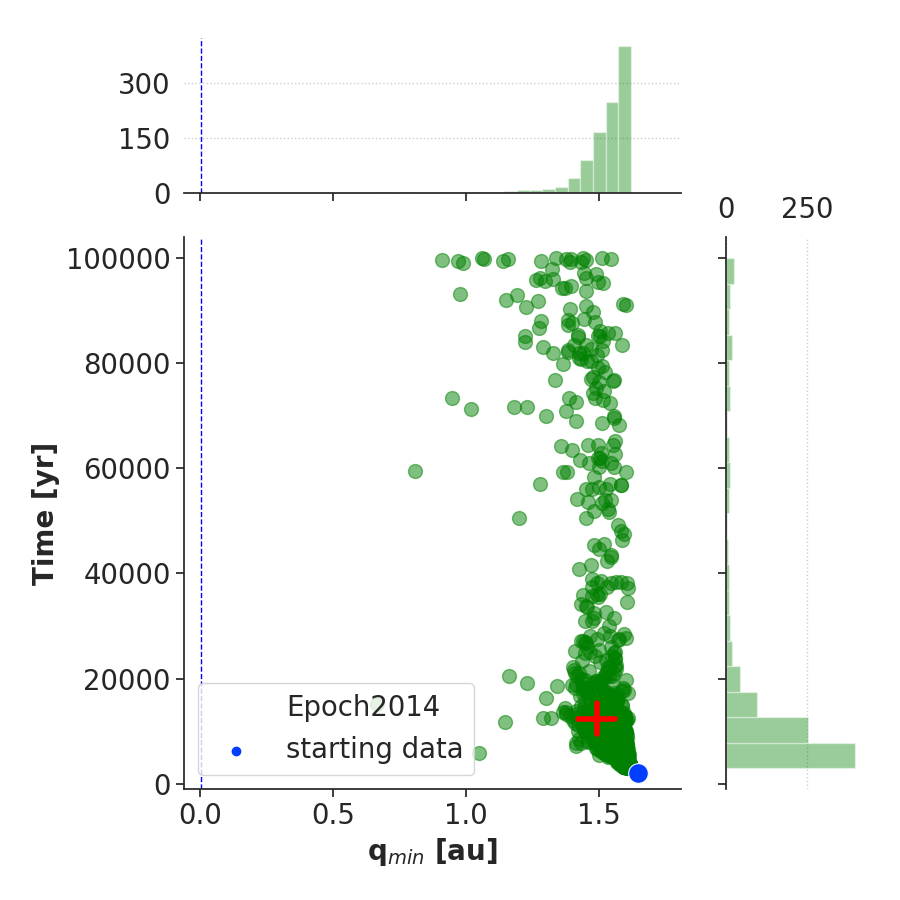}
\includegraphics[width=5.9cm]{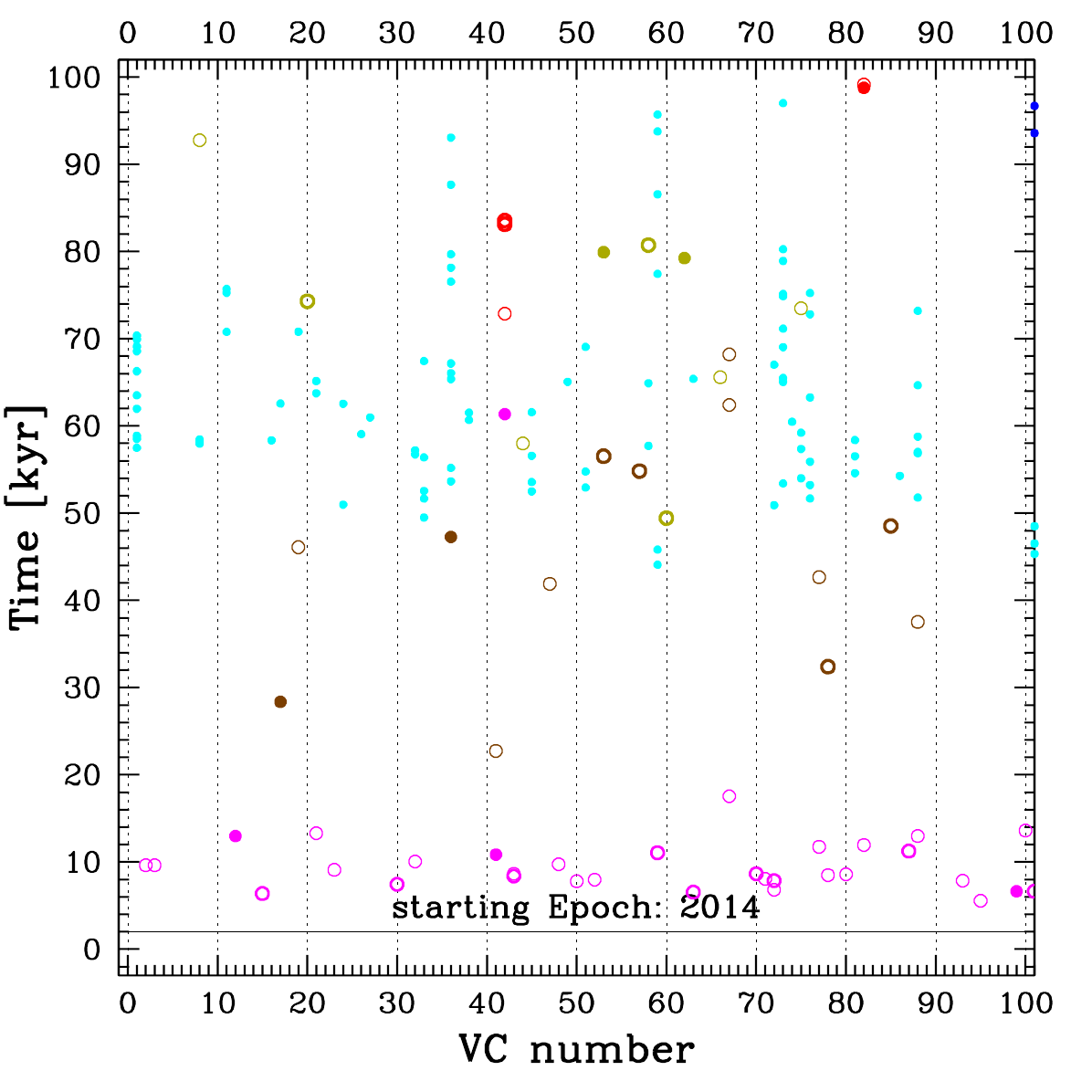}
\includegraphics[width=5.9cm]{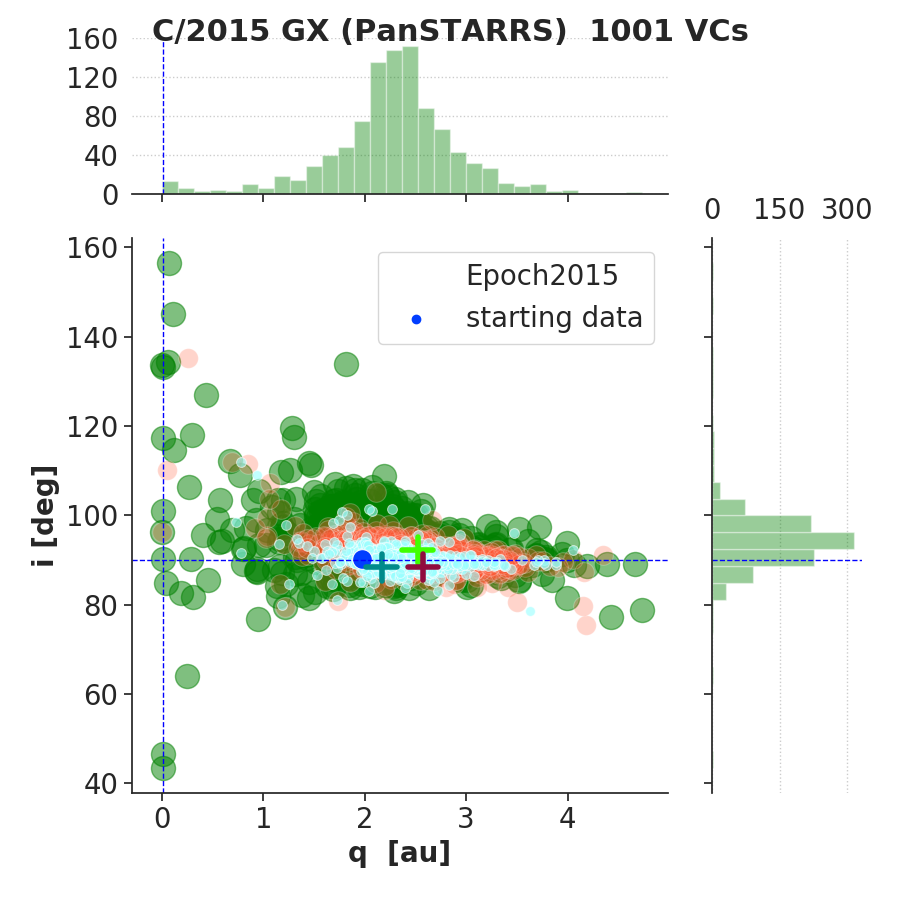}
\includegraphics[width=5.9cm]{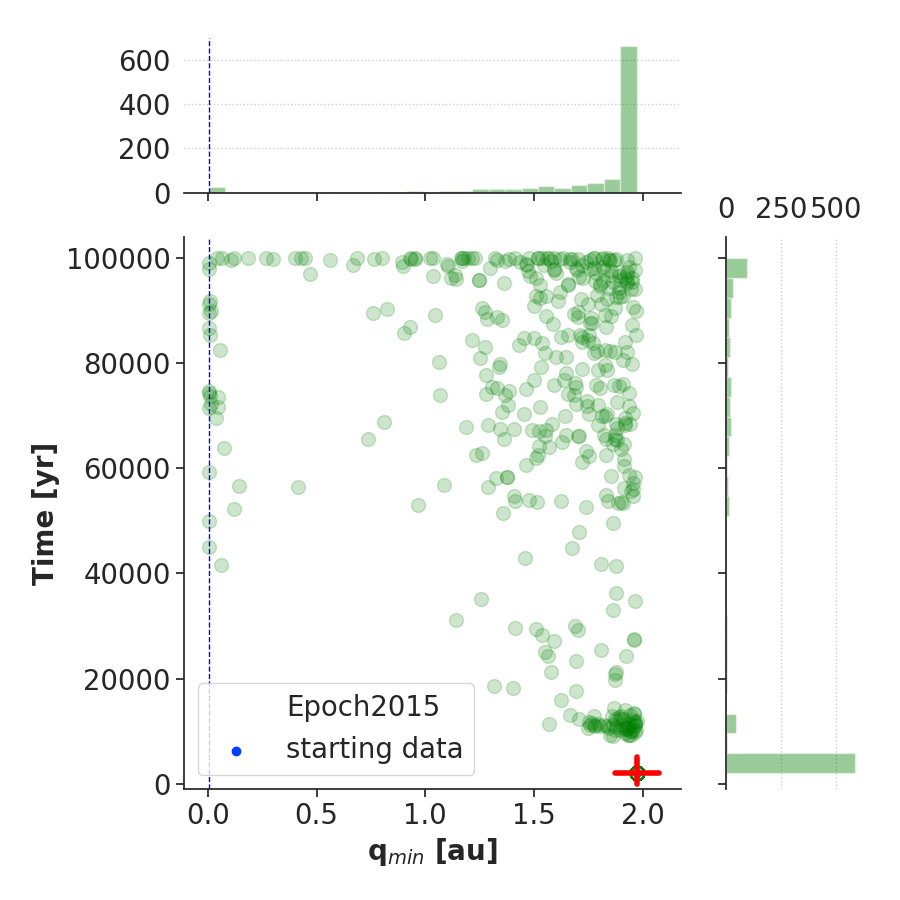}
\includegraphics[width=5.9cm]{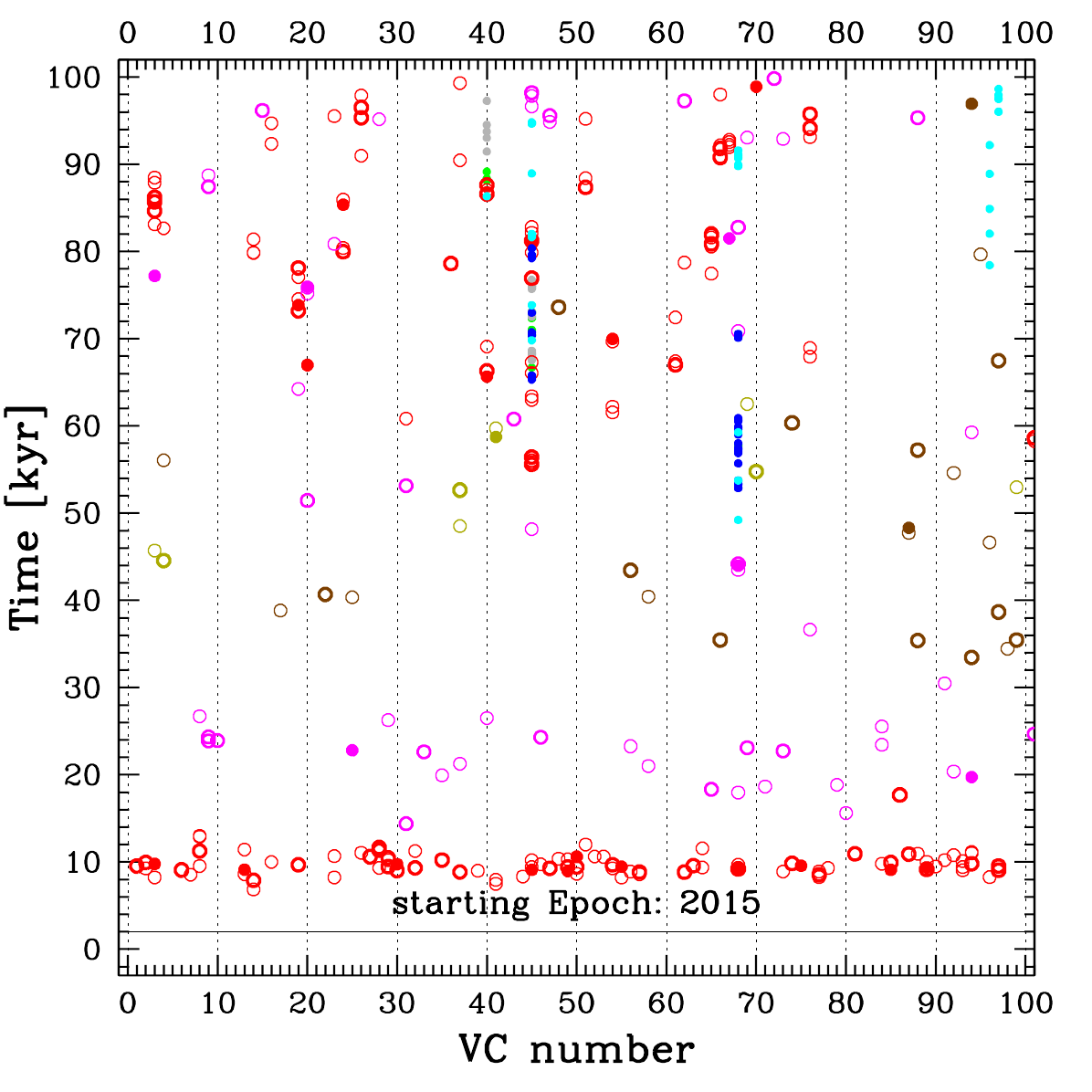}
\includegraphics[width=5.9cm]{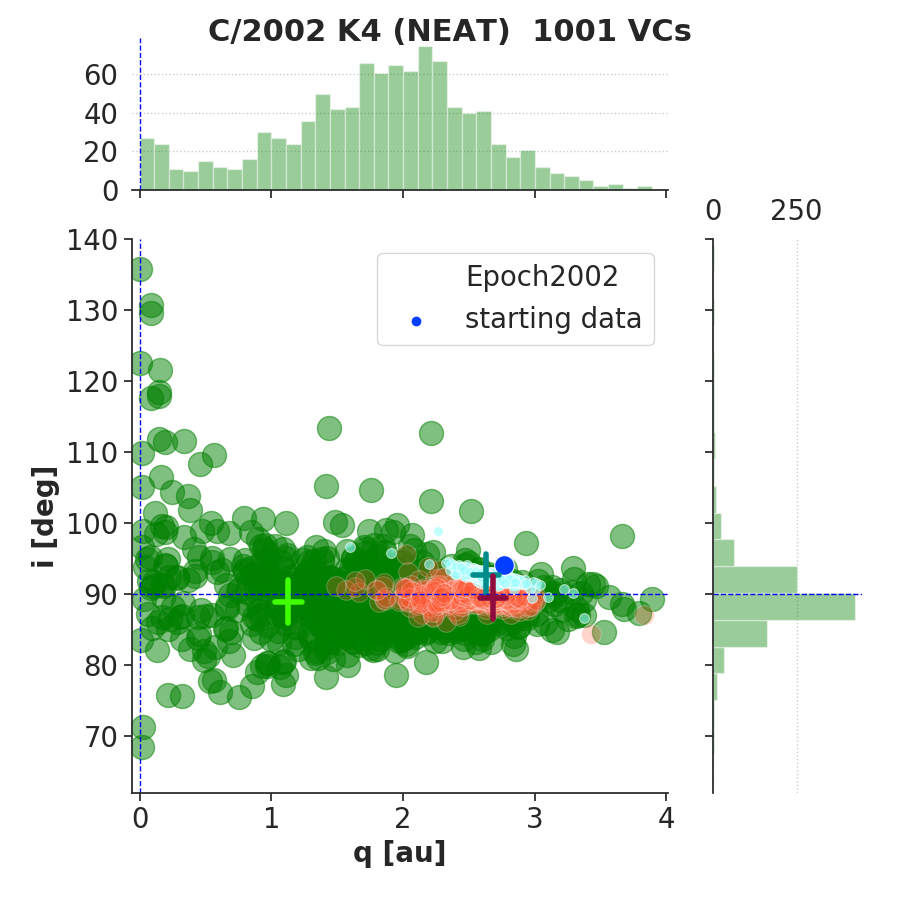}
\includegraphics[width=5.9cm]{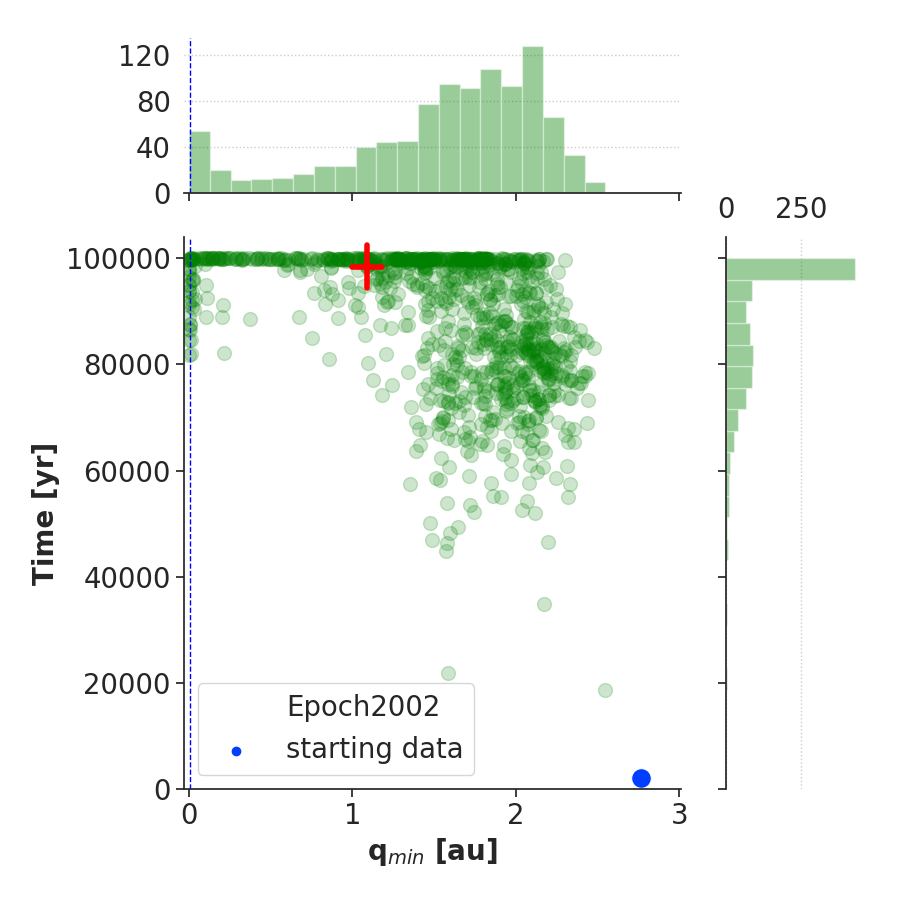}
\includegraphics[width=5.9cm]{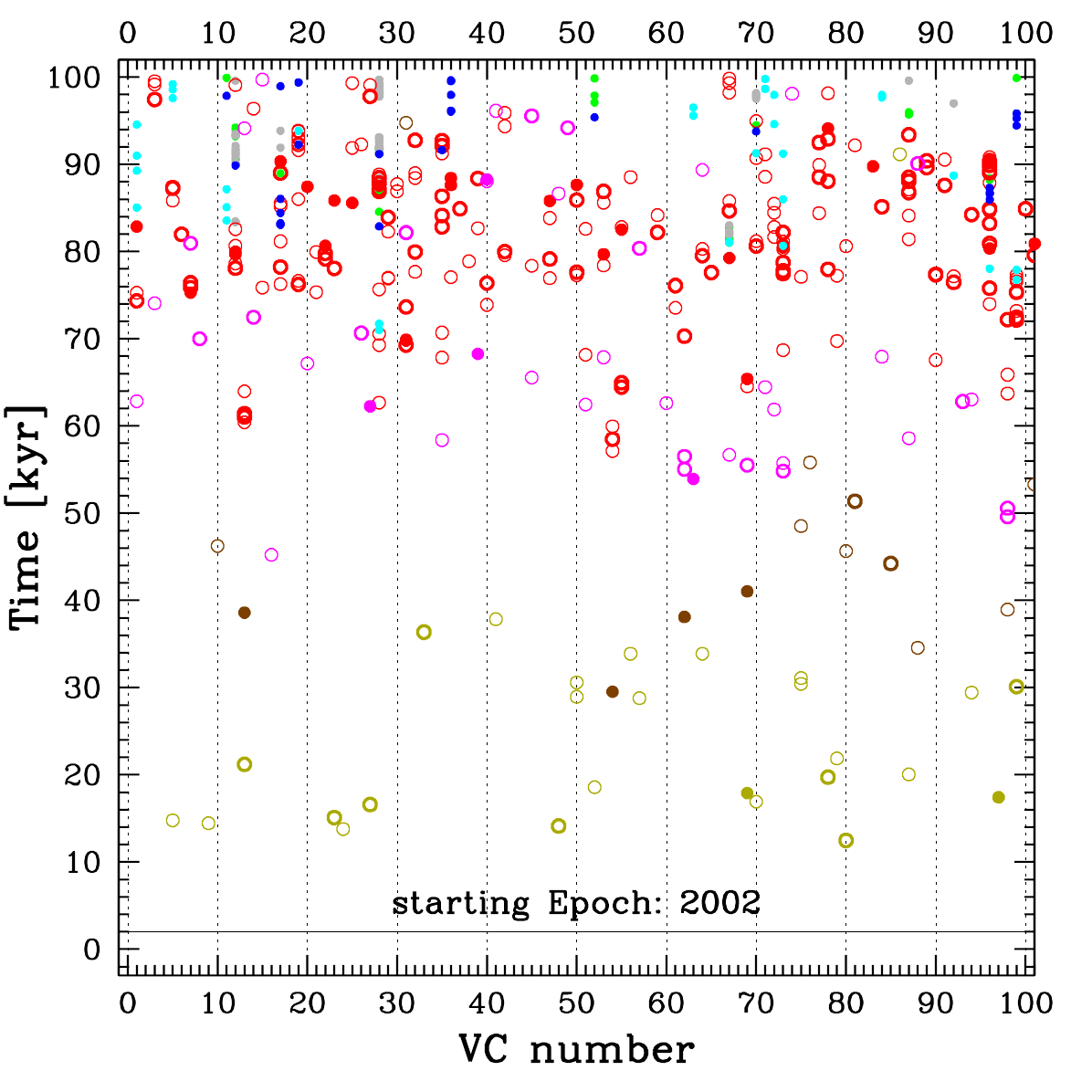}
	\caption{GR evolution of C/2014~Q3 (Borisov), C/2015~GX (PanSTARRS), and  C/2002~K4 (NEAT) in the time interval 2--100\,kyr. Left panel:  Orbital elements distributions of 1001\,VCs in 2D plot of orbital elements in $q$ and $i$ in a similar convention as in Fig.~\ref{fig:12p-b5-q-i-planets}, where cyan, orange, and green dots represent swarms after $\sim 20$\,kyr, $\sim 50$\,kyr, and  $\sim 100$\,kyr, respectively. Middle panel: Distributions of $q_{\rm min}$ and the moment of approaching this minimum for the whole swarm. The evolution of nominal orbits to the same state are shown by red crosses. Blue dashed vertical lines represent the $q$-limit for survival:  $q_{\rm lim}=0.005$\,au. Right panel: Time distributions of approaches to planets at a distance below 0.1\,au (solid coloured points) and giant planets in the range of 0.1--0.2\,au (rings) and 02--0.3\,au (thin rings) for the first 100 VCs from the VC~swarm; Colour-coded symbols as in Fig.~\ref{fig:12p-b5-q-i-planets}; for more see text.} \label{fig:three_HTCs-evo} 
\end{figure*}

\begin{figure*}
	\centering
\includegraphics[width=8.8cm]{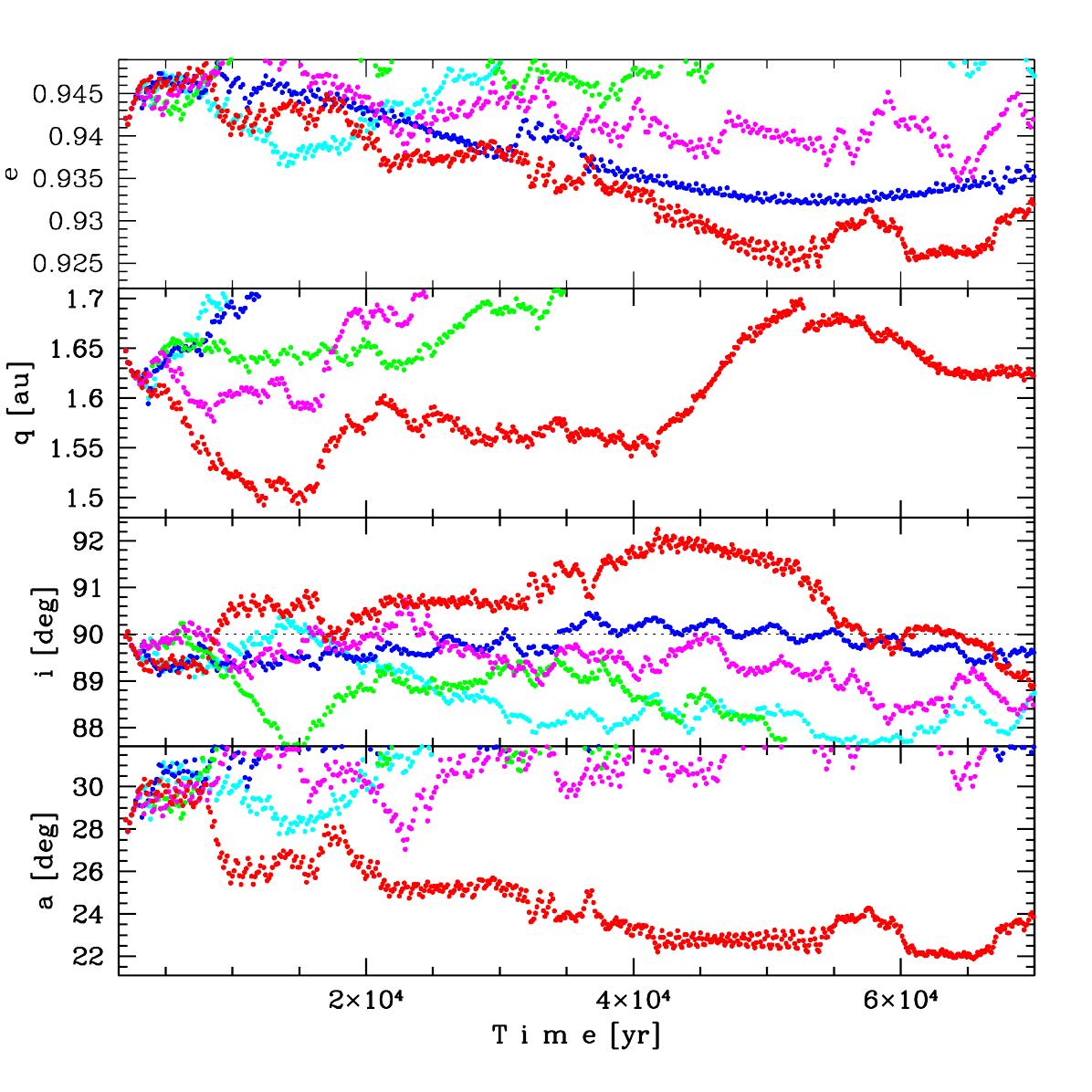}
\includegraphics[width=8.8cm]{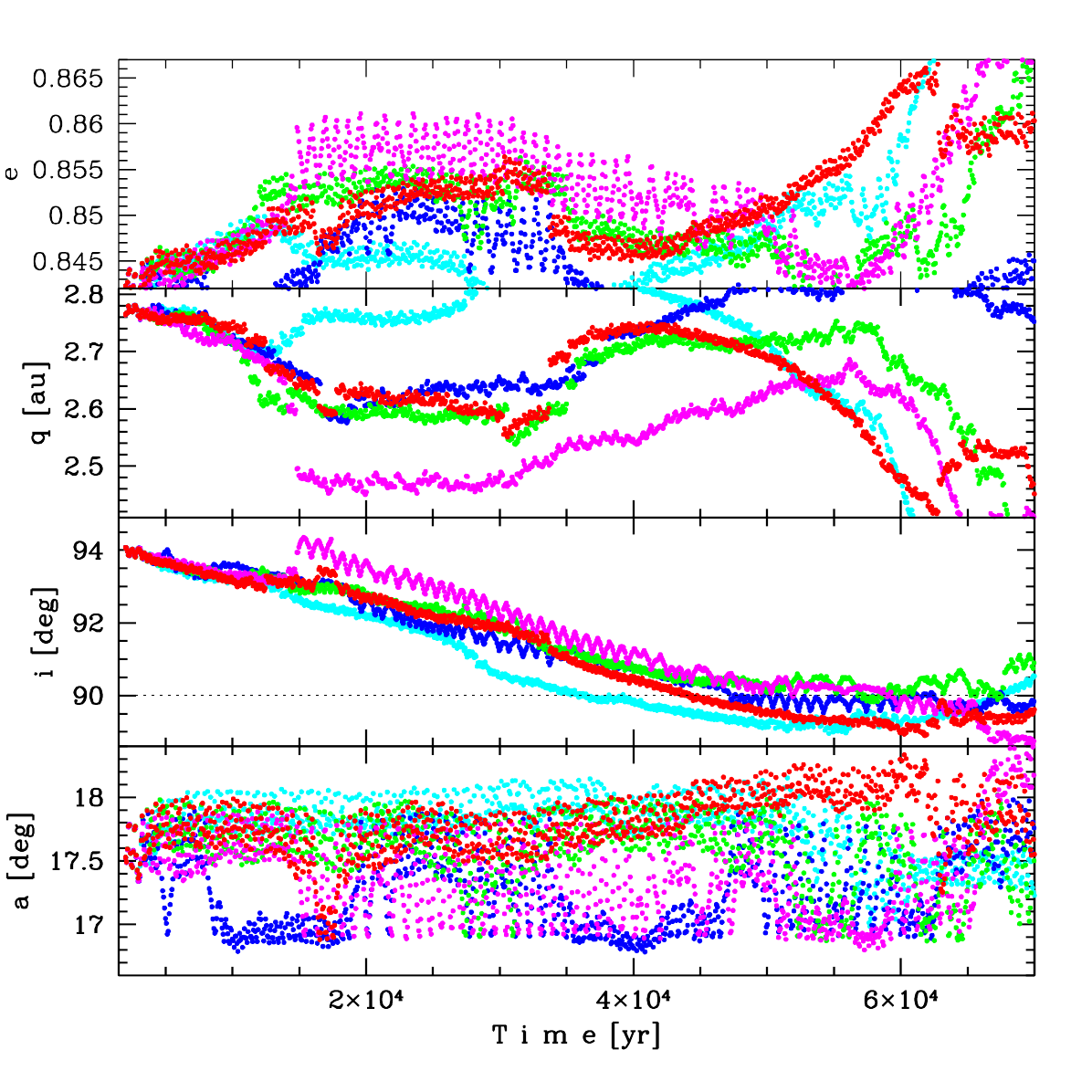}
	\caption{ Differences in the future GR evolution between C/2014~Q3 (left panel) and C/2002~K4 (right panel) where the evolution of nominal orbits is given by red ink and next four VC are also shown. We limited to the period 70\,kyr (horizontal axes) and to range of orbital elements for nominal orbit (vertical axes) to show some more details of evolution for $e$, $q$, $i$, and $a$; the moment of orbital flipping is pointed by horizontal dotted thin line indicating $i=90$\degr.}\label{fig:2014q3_2002k4-nominal} 
\end{figure*}

As previous two numbered comets, 161P/ Hartley-IRAS is also classified as NEO in JPL database, with Earth MOID of 0.4484\,au and Jupiter MOID of 0.1149\,au. Currently, this comet has the shortest orbital period of 21.4\,yr of any comet we have studied. We obtained the first close approach to Earth a bit closer than 0.1\,au, in October 3374 and this event is visible in the right panel of Fig.~\ref{fig:161p-b5-q-i-planets}. Table~\ref{tab:161P-planet-approaches} lists the most interesting (deepest) approaches to planets over the next 1,500\,yr. 

We found that 161P evolution is spectacular because 95\% of 1001\,VCs are sungrazers just after 13\,kyr from now with $q<0.05$\,au and about 80\% have $q<0.005$\,au; see right plot of Fig.~\ref{fig:12p122p161p_qmin_time}. The decrease of $q$ towards sungrazing states happens due to the Kozai mechanism. In Sect.~\ref{sub:evol-below-limit}, we discuss these results in more detail; for additional plots see also Sect.~\ref{sec:161P-10VCs-evo}.

The evolution of the swarm of 1001\,VCs during this 13\,kyr period in 2D plane of orbital parameters $q$ and $i$ is shown in the left panel of Fig.~\ref{fig:161p-b5-q-i-planets} where cyan dots represent the whole swarm after about 3\,kyr from now (the cross mark nominal orbit position), orange cloud of dots describe the situation after about 6.5\,kyr from now, and green dots in the background shows the distibution after about 13\,kyr from now. The deepest approaches to all planets in these 13\,kyr period for the first 100 of 1001\,VCs are visible in the right panel of this figure.  Individual VC makes from 300 to 700 revolutions around the Sun during this period. 

\cite{bai-cha-hah:1992} also considered the evolution of this comet in the context of achieving a sungrazing state. They obtained for a nominal starting orbit of 161P a decrease to $q \simeq 0.01$ with a simultaneous reduction in $i$ (their Fig. 2), where a first minimum of $q$ was reached after about 20\,kyr, and next after 80\,kyr during the future evolution of 100\,kyr with an inclination flip occurring between these two events and lasts until the end of integrations. Based on the data arc from 1983 to 2005, we obtained that the first loop of $q$ libration around $\omega$ takes place earlier than obtained by \cite{bai-cha-hah:1992} and that the $q$-decrease is deeper, typically below 0.005\,au, than they predicted. For more on the probability of evolution towards sungrazers and inclination flipping see Sects.~\ref{sec:evol-to-small-q} and \ref{sec:evol-flipping}.

\subsection{C/2014 Q3 (Borisov)}\label{subsec:evo-2014q3}

It is the comet with the longest period among those studied here. Individual VCs make typically 400--1200 revolutions around the Sun during 100\,kyr evolution. 
This comet is classified as HTC in JPL Database, with Earth MOID of 0.8662\,au and Jupiter MOID of 1.354\,au.

We found that close approaches to giant planets below 0.2\,au are rare in the swarm of 1001\,VCs. Upper right panel of Fig.~\ref{fig:three_HTCs-evo} shows that the most frequent among all planets are close encounters with Saturn. Similarly, encounters below 0.7\,au are sporadic and the most probable are encounters with Saturn. 

As can be expected from above, the evolution of the swarm is slow with moderate dispersion at the end of evolution, see Table~\ref{tab:2002k4-stat_at-the-end} and upper left panel of Fig.~\ref{fig:three_HTCs-evo}, where only one VC is outside the plot ($i$=139\degr). In a typical VC evolution, after an initial small decrease in $q$, it slowly increases, and the median of the distribution after 100\,kyr is $\sim$1.9\,au, whereas $i$ statistically tends to increase during the first 50\,kyr and next to decrease slowly. Upper left panel in Fig.~\ref{fig:three_HTCs-evo} also shows that for many VCs from the swarm there is a change from direct to retrograde type of motion (starting $i$=89.9\degr) at the beginning of evolution. However, the median of $i$ at the end of evolution is 86.1\degr. After 100\,kyr of evolution, the medians of $a$, $Q$, and $P$ for a cloud of 1001\,VCs are 33.0\,au, 64.1\,au, and 189\,yr, respectively. Middle upper panel of Fig.~\ref{fig:three_HTCs-evo} shows that $q$ rarely drops below 1.0\,au  and never below 0.5\,au during 100\,kyr of future evolution of 1001\,VCs.

\subsection{C/2015 GX (PanSTARRS)}\label{subsec:evo-2015gx}

This comet has initial $q$, $e$, and $i$ between those of C/2014~Q3 and C/2002~K4, and $a$, $Q$, and $P$ smaller of both (see Table.~\ref{tab:objects}). About 60\% of 1001~VCs during the future evolution have $q$ larger than its initial value, as seen in the middle row of Fig.~\ref{fig:three_HTCs-evo} (left panel). Of the remaining $\sim 40$\% of VCs, only 2.3\% reaches the state $q_{\rm min}<0.1$ including 8~VCs (0.8\%) below 0.005\,au, which is an intermediate situation between C/2014~Q3 and C/2002~K4 (see previous and next sections).

The evolution of this comet in the initial period (2-15\,kyr) is mainly ruled by Jupiter, then we have more and more close approaches to Saturn and much less often to Uranus and Neptune. Towards the end of the evolution, we have well-defined core of distributions in all orbital elements (see green marginal histograms in the middle row of Fig.~\ref{fig:three_HTCs-evo} (central panel) and Table~\ref{tab:2002k4-stat_at-the-end}), with the median and nominal VC differing in the orbital element values, in particular in $i$ by about 2\%, while in $a$, $Q$, and $P$ at the level of 20--30\%. Ignoring the effects of comet activity and aging, according to our results C/2015~GX will evolve to a larger values of $q$, $a$, $Q$, and $P$ than the respective orbital elements of the present orbit whereas eccentricity a slightly decreased.  

Two VCs in the swarm have hyperbolic orbits at the end of evolution. In the first case, $q$ increases during the VC evolution. After the close approach to Jupiter around 56.9\,kyr (0.018\,au) the eccentricity increases from 0.85 to 1.06, while $q$ increases from 2.34\,au to 3.24\,au, and $i$ decreases from 90.5\degr ~to 80.4\degr. In the second case, a hyperbolic orbit is achieved after passing through a stage of decreasing $q$ to a value of 0.055\,au (after 82.5\,kyr) with a simultaneous increase of $i$ to about 160 deg. Less than 2\,kyr later (when the comet perihelion slightly increased to 0.07\,au), a close approach to Jupiter on 0.01\,au causes the orbit to change from already near parabolic ($e=0.9945$) to hyperbolic ($e=1.0045$). 

\subsection{C/2002 K4 (NEAT)}\label{subsec:evo-2002k4}

\begin{table}
	\caption{Statistics of orbital elements for 1001\,VCs at the end of dynamical evolution (after 100\,kyr) for HTCs with osculating $q>1.6$\,au. Statistics are based on all VCs with $e<1$.} \label{tab:2002k4-stat_at-the-end} 
	\centering
	\setlength{\tabcolsep}{2.0pt} 
	\begin{tabular}{ccccccc}
		\hline 
		Orbital	  &  minimal & 10\%    & 50\%    & 90\%    & maximal  & nominal \\
		element	  &  value   &         & median  &         & value    & orbit   \\
		$[1]$     & $[2]$    & $[3]$   & $[4]$   &$[5]$    & $[6]$    & $[7]$   \\
			\hline\hline
		\\
		\multicolumn{7}{c}{C/2014 Q3 (Borisov)    }	\\ \\
		$q$ [au]  & 0.849    &  1.58   & 1.87    &   2.06  &      3.37    & 1.68  \\
		$e$~~~~~~~~~& 0.8709   &  0.9278 & 0.9451  & 0.9668  &    0.8879  & 0.9333\\
		$i$ [deg] & 78.4     &   84.3  &  86.1   &   88.3  &    139.4     & 85.4  \\
		$a$ [au]  & 14.7     &   24.0  &  33.0   &   57.2  &    140.5     & 25.2  \\
		$Q$ [au]  & 27.8     &   46.3  &  64.1   &  113.   &    279.~~    & 48.7  \\
		$P$ [yr]  & 56.1     &  117.~~~~&  189.~~~~&  433.   &   1665.~~~~  & 127.~~~ \\
		
		\\
		\multicolumn{7}{c}{C/2015 GX (PanSTARRS) }	\\ \\
		$q$ [au]  & 0.0040   &  1.55   & 2.32    &   2.99  &      4.73    & 2.52  \\
		$e$~~~~~~~~~& 0.5896   &  0.8246 & 0.8693  & 0.9180&    0.9998  & 0.8843\\
		$i$ [deg] & 43.5     &   88.1  &  94.2   &  100.2  &    156.4     & 92.3  \\
		$a$ [au]  &  5.97    &   13.0  &  17.1   &   25.0  &    291.~~    & 21.8  \\
		$Q$ [au]  & 10.1     &   23.8  &  32.0   &   47.3  &    581.~~    & 41.1  \\
		$P$ [yr]  & 14.6     &   46.8  &  70.9   &  125.   &   4977.~~~~  & 102.~~ \\
		\\
		\multicolumn{7}{c}{C/2002 K4 (NEAT)   }	\\ \\
		$q$ [au]  & 0.00233  &  0.684  & 1.87    &   2.68  &      3.89    & 1.13  \\
		$e$~~~~~~~~~& 0.5459   &  0.8422 & 0.9007  & 0.9709  &      0.99989 & 0.9301\\
		$i$ [deg] & 29.7     &   84.7  &  89.1   &   94.5  &    161.9     & 88.9  \\
		$a$ [au]  & 4.22     &   13.1  &  17.5   &   27.0  &    734.5     & 16.2  \\
		$Q$ [au]  & 7.69     &   24.5  &  33.0   &   52.6  &   1465. ~~~  & 31.2  \\
		$P$ [yr]  & 8.66     &   47.5  &  73.2   &  140.~~~~&  19904. ~~~~~~~& 65.0 \\
		\hline
	\end{tabular}
\end{table}

Today, this comet has the largest $q$ of any comet we have studied here. In the first 50\,kyr of evolution, we deal with perturbations from three giant planets: Neptune, Uranus, and Saturn in time order. The deepest close-ups are shown in the bottom row of Fig.~\ref{fig:three_HTCs-evo} (right plot) for the first 100 of 1001\,VCs. The first changes in the sense of motion in this swarm occur after about 20\,kyr (from retrograde to prograde, see left plot in the same row). One can see there that the typical behaviour is to decrease $q$ and $i$ slowly relative to the starting orbit. During 20\,kyr -- 50\,kyr the swarm disperses, a little more towards smaller 'q' and 'i', but the nominal orbit slightly increases in $q$. In this period, approaches to Jupiter began to dominate in number over others, and in some cases approaches to Earth-like planets started to occur. Towards the end of the evolution, some of the VCs with extreme $i$ are outside the left plot in the bottom row of Fig.~\ref{fig:three_HTCs-evo} (compare with Table~\ref{tab:2002k4-stat_at-the-end}). At the end of evolution, the nominal orbit is marked with a greenish cross in this figure and the ranges of orbital elements is shown in Table~\ref{tab:2002k4-stat_at-the-end}. 

After about 80\,kyr, a few VCs come close to the Sun (see middle plot in the bottom row of Fig.~\ref{fig:three_HTCs-evo}). During this evolution, about 6\% of comets approach the Sun below 0.1\,au, of which 2.2\%  are sungrazers with $q<0.005$\,au. We also noticed three cases with hyperbolic orbits. One, occurred in a sungrazing state with $q$ around 0.001\,au, as a result of slightly deeper than the previous approaches to Mercury (below 0.1\,au), after around 99\,kyr. However, the decrease of $q$ below 0.005\,au had already occurred more than 10 orbital revolution earlier. Therefore,  a comet would have disintegrated before than achieved $e>1$. The other two cases resulted from approaches to Jupiter at 0.003\,au, and 0.331\,au, which occurred after 82\,kyr and 95\,kyr, respectively. These two give a probability of 0.2\% of leaving the Solar System during 100\,kyr of evolution for C/2002~K4.

Ultimately, we conclude that this comet will remain an HTC in the future 100\,kyr, with 0.68\,au $<q<$ 2.7\,au, where the median of this distribution falls at the value of 1.9\,au (the $q$ for nominal orbit is 1.1\,au) for $q$ (Table~\ref {tab:2002k4-stat_at-the-end}), and its orbit will be prograde with the probability of about 0.65. 

The differences in the dynamical evolution of C/2002~K4 and C/2013~Q4 are presented comparatively in Fig.~\ref{fig:2014q3_2002k4-nominal}. 
Currently, the comet C/2013~Q4 is in the range of weak high-order mean motion resonances (MMRs) with Saturn and Neptune, which were detected and presented in more detail in Sect.~\ref{sec:evol-MEGNO}, describing short-term chaotic behaviour. By examining the long-term future evolution of the VCs of this comet (Fig.~\ref{fig:2014q3_2002k4-nominal}, left), it is possible to detect more strongly interacting lower order MMRs in the timespan of 100\,kyr. These include 1:4 S, 1:5 S and 1:6 S resonances with Saturn, of which the 1:5 S resonance lying around $a$=28.02 au is the most likely, while the others may occur on longer time scales. 

The comet C/2002~K4 exhibits weak high-order mean motion resonances with Saturn, Uranus and Neptune, which were also described in more detail in Sect.~\ref{sec:evol-MEGNO}. With evolution on longer time scales up to 100 kyr, the orbital elements of this comet in turn evolve into areas where further MMRs appear. These are more strongly interacting lower-order resonances. C/2002~K4 seems to experience a variety of long-standing libration MMRs with planets. We found that in the range of semimajor axis between 17\,au and 18\,au can be: the 1:6 resonance with Jupiter ($a$ about 17.17\,au), 2:5 resonance with Saturn ($a$ around 17.65\,au), and, potentially series of resonances with Uranus as 7:6, 8:7, and 9:8 ($a$ around 17.32, 17.55, and 17.74\,au, respectively). At least some of them can occur in these five VCs presented in the Fig.~\ref{fig:2014q3_2002k4-nominal} (right panel). The most significant appears to be the U 8:7 resonance, which does not impact the comet yet, but is already close to the nominal orbit (see Sect.~\ref{sec:evol-MEGNO}), so the comet could easily migrate into its region of effect.

\begin{figure}
	\centering
	\includegraphics[width=8.8cm]{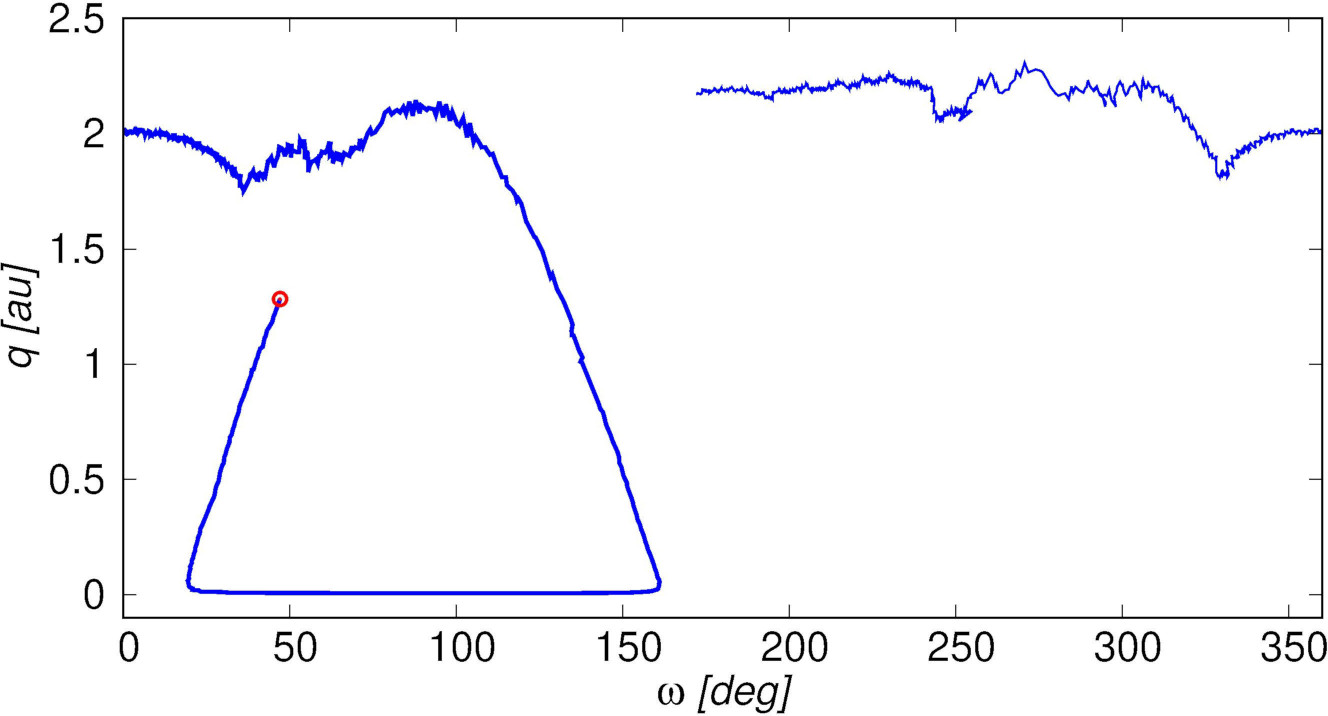}
	\includegraphics[width=8.8cm]{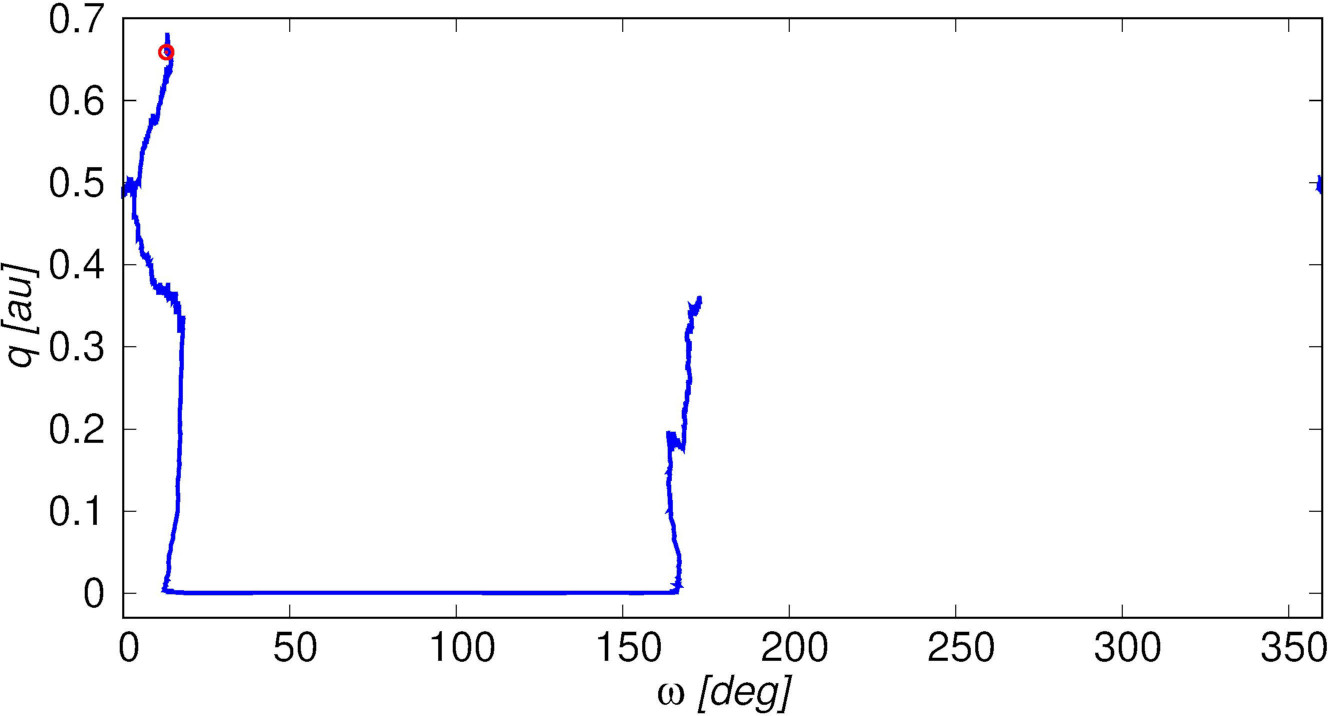}
	\caption{Orbital evolution of the comets 161P/Hartley-IRAS (top) and 122P/de Vico (bottom) showed on the ($\omega,q$) plane over a timespan of 100\,kyr. The starting point of each dynamical evolution is indicated by a red circle. See the text for further details.} \label{fig:161p_Kozai} 
\end{figure}

\begin{figure}
	\centering
	\includegraphics[width=8.8cm]{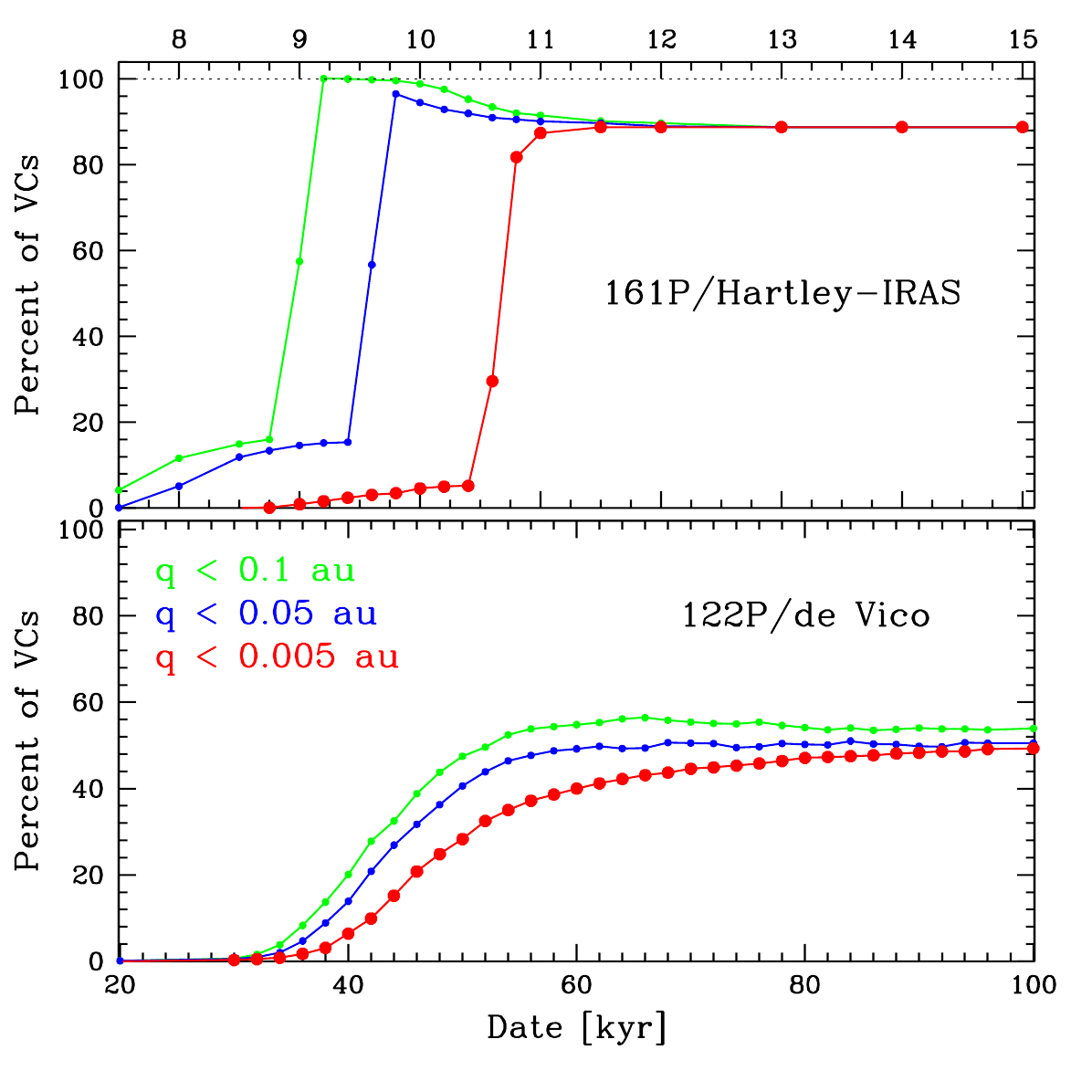}
	\caption{Percent of VCs achieving $q$ below 0.10\,au (green symbols) 0.05\,au (blue) and 0.005\,au (red) during the evolution of 161P/Hartley-IRAS (upper panel) and 122P/de Vico (lower panel). We stop the evolution when VC achieves $q<0.005$\,au, that is, each orbit when $q$ decreases below $q_{\rm lim}$ is frozen after this moment.} \label{fig:161p_sungrazers} 
\end{figure}

\section{Evolution to small perihelion distances -- sungrazing state}\label{sec:evol-to-small-q}

As we have shown earlier, drops to very small $q$ during the future evolution to 100\,kyr are common phenomena in the evolution of 122P/de Vico and 161P/Hartley-IRAS. Therefore we describe these two cases in more detail in Sect.~\ref{sub:evol-below-limit}. However, it is a rare phenomenon in the four remaining cases. Drops below 0.005\,au occur in 2.2\% and 0.4\% for C/2002~K4 and 12P, respectively, and we had no case of $q$ decreasing below 0.5\,au for C/2014~Q3. For C/2015~GX, we noted only 0.8\% of VCs exhibited decreasing below 0.005\,au, which is an intermediate number of sungrazers between C/2014~Q3 and C/2002~K4.  

We confirm that the decrease in $q$ to sungrazing states in the future dynamical evolution of comets 122P and 161P (see Sect.~\ref{sub:evol-below-limit}) proceeds via the Kozai mechanism (see next section). This has also been previously identified by others in dynamics of certain HTC  \citep[see for example][]{carussi-etal:1987, bai-eme:1996}, and particularly in dynamics of these two comets \citep{bai-cha-hah:1992}. 

\subsection{Kozai mechanism}\label{sub:evol-Kozai}

The dynamical behaviour known as Kozai (or Kozai-Lidov) resonance subsequently induces large, long-term variations in the eccentricity of the object. As is well known, when a Kozai resonance occurs for objects with a high value of orbit inclination to the ecliptic ($i>32.9$\degr), the value of $\omega$ librates around 90\degr ~or 270\degr. In a similar cyclic manner, the value of $q$ changes \citep{DermottandMurray1999, Naoz2016}. It is convenient to illustrate this on the $(\omega, q)$ plane. The occurrence of a Kozai resonance will result in curves surrounding the points (90\degr, $q_0$) or (270\degr, $q_1$) on this plane, where $q_0$ and $q_1$ are certain values of the perihelion distance. 

As we illustrated in the previous section, Kozai resonance can play an important role in the long-term dynamical evolution of HTCs. These comets experience complex gravitational interactions with planets, leading to significant changes in their orbits. The Kozai resonance, which is a type of secular perturbation, can cause oscillations in the orbit of a comet, particularly in its inclination and its distance from the Sun. This resonance can even switch their orbits from prograde to retrograde (see Sect.~\ref{sec:evolution-classical} and \ref{sec:evol-flipping}). In HTCs, such resonances work alongside other dynamical phenomena, like close planetary encounters, shaping their paths over long period of time \citep{bai-eme:1996}.

Top panel of Fig.~\ref{fig:161p_Kozai} shows the dynamical evolution of comet 161P/Hartley-IRAS on the $(\omega, q)$ plane, where the time-series data used is 100\,kyr. The beginning of the dynamical evolution of the object is marked with a red circle. The figure suggests that the object is in Kozai resonance, as indicated by the characteristic loop round $\omega = 90$\degr, although it is incomplete and, after about $\sim 16$\,kyr, the object exits this resonance. To identify which of the major planets mainly causes the 161P to enter this resonance, several test integrations were performed. Specifically, we repeated the simulations, with the modification that the gravitational influence of each giant planet was sequentially turned off. As a result of these tests, we found that the effect persists in all cases except one. Namely, it was observed that the object is not in the Kozai resonance only when Uranus is excluded from the calculations. It is also worth noting that the other planets, especially Jupiter, affect the timescale of the Kozai resonance. Both, our numerical simulations and theoretical estimates \citep{Kiselevaetal1998} indicate that when only one planet, in this case Uranus, acts as the source of the Kozai resonance, the time-scale of the resonance is of the order of 100\,kyr. In the case where the comet is affected by all planets, the time-scale is of order 10 kyr. This shows the substantial impact that other major planets have on the dynamics of this comet while it is in Kozai resonance.

For 161P, we additionally checked the orbital loops in the $(\omega, q)$ plane for a random selection of dozens of VCs. In all cases, we obtained graphs suggesting that these clones remain in the Kozai resonance, although in some cases, we observed librations around $\omega=270$\degr.

Among the comets analysed here, we found that comet 122P may also be in Kozai resonance (see bottom panel of Fig.~\ref{fig:161p_Kozai}). For this comet, a similar analysis was performed as was done with the comet 161P, i.e., a giant planet was successively removed from our solar system model and then it was checked whether the Kozai resonance still occurred. We found that this resonance only occurs when Jupiter is removed from our solar system model. When any of the other major planets were removed, this resonance did not appear. This may suggest that Jupiter is not the primary cause of the resonance observed in the dynamics of this comet. Finally, we checked its orbit on the $(\omega, q)$ plane for a random selection of dozens of VCs. In all cases we obtained similar behaviour as showed in the bottom panel of Fig.~\ref{fig:161p_Kozai}. 

\subsection{Probabilities of comet survival}\label{sub:evol-below-limit}

In this study, we adopt the minimal value of  $q_{\rm lim} = 0.005$\,au $=1.08$\,R$_\odot$ as the limit of the perihelion distance below which the comet  will not survive its approach to the Sun. We call this {\it $q$-limit for survival} and we were able to calculate the probability of survival at each moment of future dynamical evolution for the sample of 1001\,VCs.
It is worth mentioning that \cite{fer-gal-you:2016} assumed in their dynamical studies that a comet ‘collides’ with the Sun when it reaches a $q$ below 0.0173\,au. Therefore, one can consider our results of the VC percentage being destroyed due to the approach with the Sun below $q_{\rm lim}$ as a lower limit of the probability of comet destruction.

Using the examples of the two comets mentioned above (122P and 161P), in which the decrease in the perihelion distance below this limit is a common phenomenon during their future evolution, we show what conclusions regarding their future follow from this assumption. Fig.~\ref{fig:12p122p161p_qmin_time} shows the distribution of $q_{\rm min}$ achieved by the VC~swarms of 122P/de Vico (middle plot) and 161P/Hartley-IRAS (right plot) during their future evolutions in the GR~model of motion. 

We obtained $q_{\rm min}$ values in the range between 0.0001\,au -- 0.0049\,au (median) -- 0.5217\,au for 122P. According to our assumptions, it turn out that the probability of cometary survival is equal to 50\% at the end of dynamical calculations (after 100\,kyr), see lower panel in Fig.~\ref{fig:161p_sungrazers}. First event of  $q_{\rm min}$ below $q_{\rm lim}$ take place around 18\,kyr from now. The most drops below $q_{\rm lim}$ occur between 30\,kyr and 60\,kyr in the swarm of VCs (Figs~\ref{fig:12p122p161p_qmin_time} and \ref{fig:161p_sungrazers}).

For 161P we have even more interesting picture of future evolution. Values of $q_{\rm min}$ are here between 0.00149\,au -- 0.00418\,au (median) -- 0.0272\,au. and first such event take place after about 6\,kyr from now. However, all clones reach its $q_{\rm min}$ just before 9.5\,kyr from now. It means that we can conclude that before 6\,kyr of evolution from now the probability of comet survival is equal 1. Then the probability gradually decreases to 0.112 for the next 3\,kyr, where all 112~VCs that survive achieve the minimal distance to the Sun in the range of $0.005$\,au~$< q_{\rm min} < 0.0272$\,au becoming then a sungrazing object with eccentricity very close to 1 during their future evolution of about 13\,kyr. After this stage, their perihelion distance grows and eccentricity decreases differently.  

A complementary image is obtained in Fig.~\ref{fig:161p_sungrazers} which shows the increase of percentage of comets with $q$ below 0.10\,au, 0.05\,au, and 0.005\,au during the period 5--13\,kyr of 161P evolution. We obtained that 88.8\% of VCs will experience sungrazer state with $q<q_{\rm lim}$ in the first 13\,kyr of evolution, so they will be below the limit assumed for cometary disintegration. So, we can conclude that the probability of comet survival is at most about 12\% after the first 13\,kyr of 161P evolution. The green curve peaked a few kyr earlier and indicates that 98.9\% achieved $q < 0.1$\,au after some of them passed the state of minimal $q>q_{\rm lim}$ (10.1\% of all VCs) and  $q<q_{\rm lim}$ (4.6\% of all VCs) (see Fig.~\ref{fig:12p122p161p_qmin_time}), however, most of them (84.2\% of all VCs) will only reach the state $q<q_{\rm lim}$ in the next few kyr from this peak (compare with blue and red curves).  

\section{Probabilities of flipping orbits}\label{sec:evol-flipping}

\begin{figure*}
	\centering
	\includegraphics[width=5.9cm]{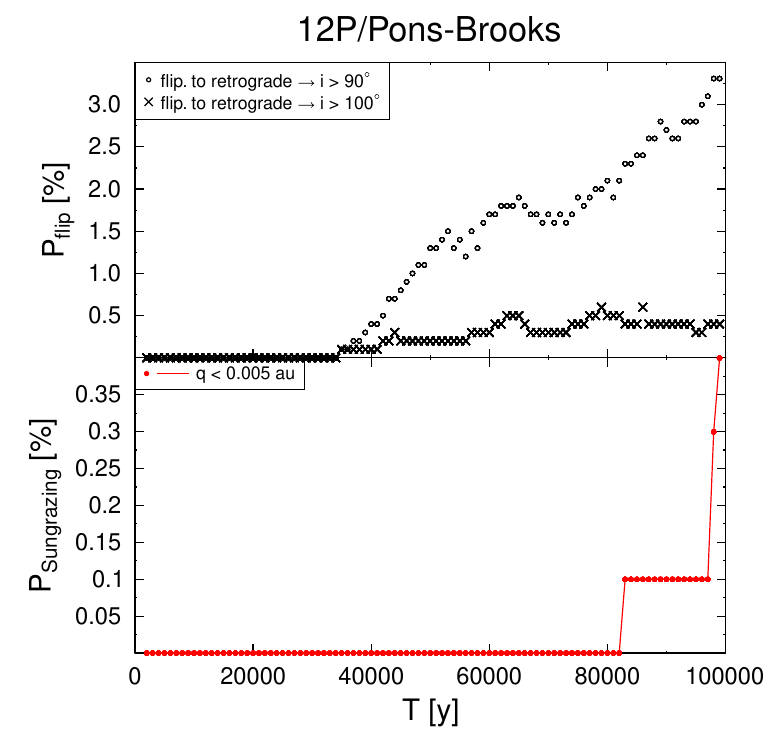}
	\includegraphics[width=5.9cm]{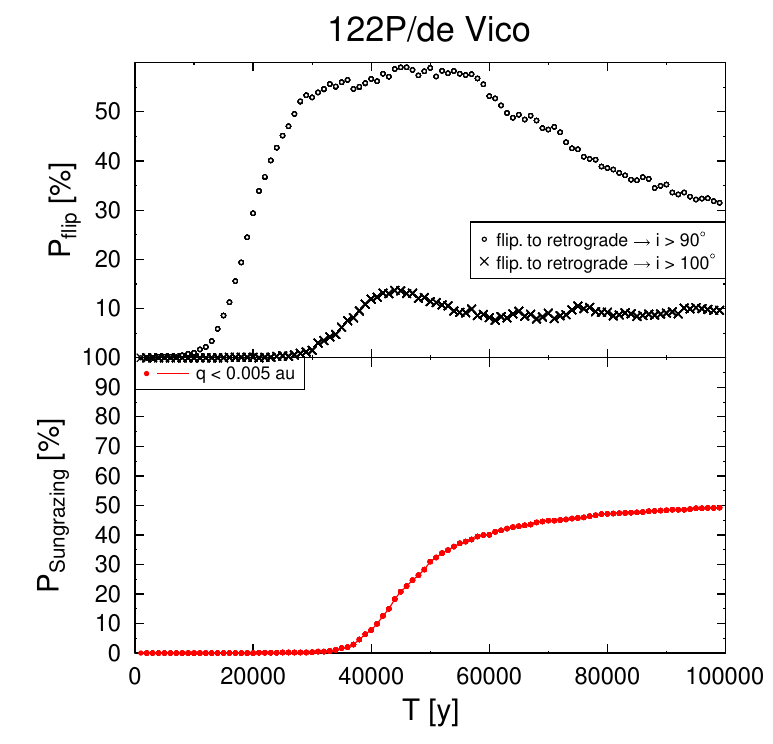}
	\includegraphics[width=5.9cm]{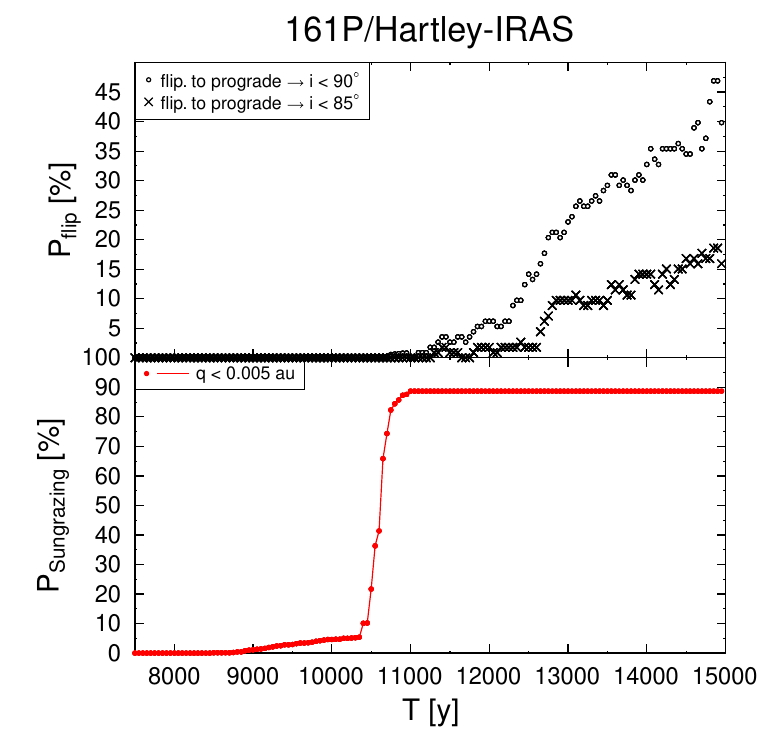}
	\includegraphics[width=5.9cm]{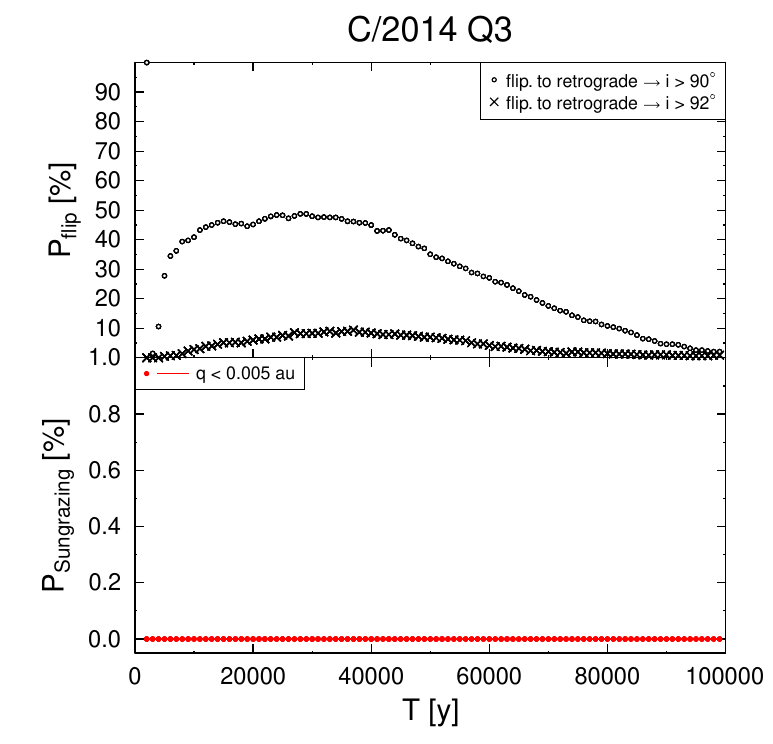}
	\includegraphics[width=5.9cm]{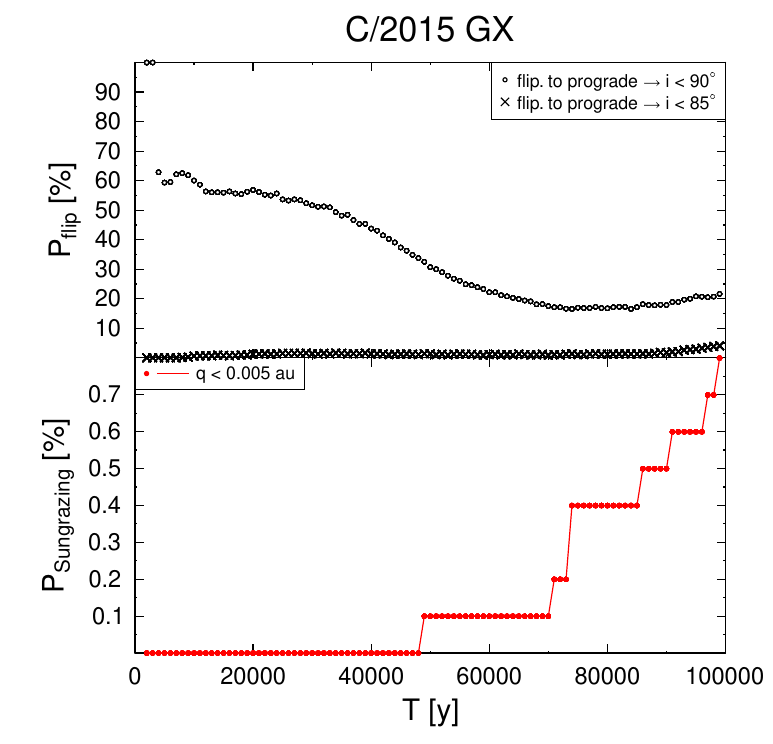}
	\includegraphics[width=5.9cm]{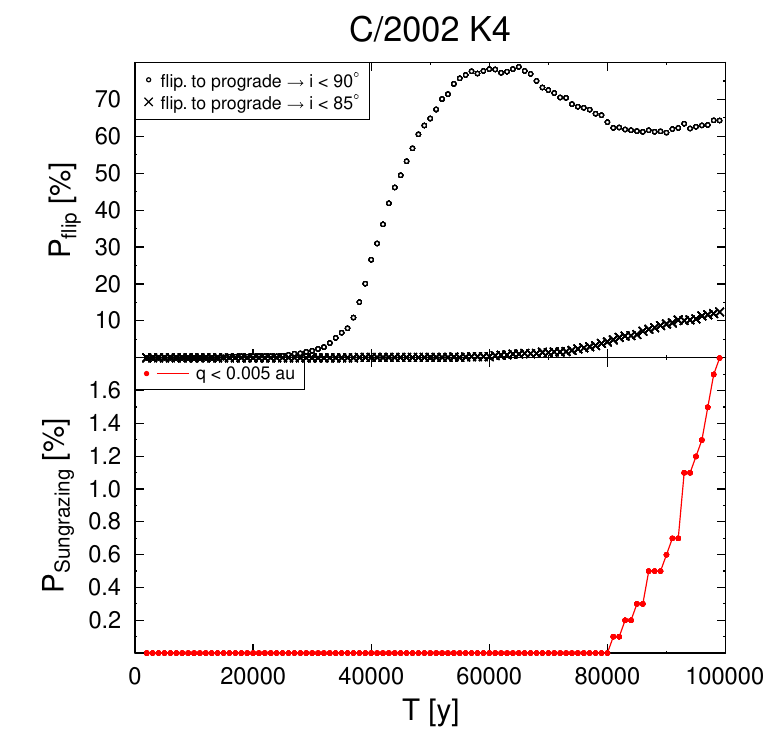}
	\caption{Orbit flipping probabilities of investigated HTCs. Upper row: 12P , 122P, and 161P. Lower row: C/2013~Q3, C/2015 GX, and C/2002 K4.}	\label{fig:flipping}
\end{figure*}

As was qualitatively shown earlier, a change from prograde to retrograde motion and vice versa occurs in the evolution of each of the six comets studied. Here, we assess the probability of such an event for each. Some comets studied are formally classified as 'prograde' (12P/Pons-Brooks, 122P, C/2014 Q3 (Borisov)) and some as 'retrograde': 161P, C/2002 K4 (NEAT), C/2015 GX (PanSTARRS). Thus, when we use the term 'flipping' here, we are dealing with reorientating the orbit from or to retrograde. We have used conditional probability formalism to estimate the probability of such events over time because, in addition to flipping, there is an exclusion factor for our comets when they pass too close to the Sun. We then assume their elimination from the system by sungrazing. Admittedly, we have found a few cases of comets being ejected into hyperbolic orbit (ejection), but this is an insignificant number in statistical terms. Therefore, we can use a similar approach to estimate the flipping probability as in the work of \cite{Kankiewicz2020}, where this was done for retrograde asteroids incidentally ejected from the system. The only difference is that the main elimination factors are not ejections, but sungrazing scenarios. The formalism and formulae used in the calculations are explained below.

\vspace{0.1cm}

Let us follow a case where the event corresponds to an orbit transition from prograde to retrograde, hence to a state with $i > 90^{\circ}$. We assume that $P(R)$ is the probability of a clone remaining (surviving) in the system. By the way $P(R)=1-P_{Sungrazing}$, where $P_{Sungrazing}$ is the probability of elimination from the system, i.e. $P(q < 0.005)$. Next, for each moment of time, we count all matching events to estimate $P((i > 90^{\circ}) \cap {R})$. Finally, we need to compute the conditional probabilities. We always give them in relation to time. Therefore, the formula for the conditional probability of flipping is given as follows:

\begin{equation}
P_{flip}=P((i > 90^{\circ})|{R})=\frac{P((i > 90^{\circ}) \cap {R})}{P(R)}=\frac{P((i > 90^{\circ}) \cap {R})}{1-P(q < 0.005)}.
\end{equation}

\noindent In the case where flipping occurs in the reverse direction, i.e. from a retrograde orbit to a prograde orbit, we use the analogical formula, but we calculate the probabilities with reference to the event $i < 90$\degr.

If we want to use a more restrictive criterion to be sure that explicit flipping has occurred, we can use other angles as arguments, i.e., depending on the need, these could be the limits $i > 100^{\circ}$ or $i < 85^{\circ}$ for the reverse case. An important point is that the widely used retrograde orbit definition is related to the ecliptic, not the invariable plane of the Solar System -- we have an angular difference of about 1.6 degrees between these planes. Hence, orbits close to 90 degrees in this range would be treated with caution regarding the judgement of flipping and categorised as near-polar.

Fig.~\ref{fig:flipping} shows the flipping probabilities estimated for our six comets. For each comet, two plots were prepared, with the top one always showing how the probability of flipping (conditional) evolves. The bottom plots show the probability of sungrazing $P(q < 0.005)$, i.e. the elimination of the VC from the system.


12P/Pons-Brooks is currently a comet in a prograde orbit, but there is a probability at around 2\% after 60\,kyr and exceeding 3\% after 100\,kyr that its inclination will exceed 90\degr ~(Fig.~\ref{fig:flipping}, upper left plot). If we change the threshold to $100^{\circ}$ , corresponding to a clearly retrograde orbit, the probability of such an event is at most around 0.5\% after 60\,kyr. A factor that minimally affects these scenarios is the termination of the life of this comet by sungrazing, with a probability at most 0.4\% after 100\,kyr, as described in Sect.~\ref{subsec:evo-12p}. Therefore, the most plausible possibility seems to be that the comet will remain in a prograde orbit, with a low chance of flipping and an even smaller chance of sungrazing.

Comet 122P/de Vico is more interesting than the previous one in terms of its sensitivity to flipping. It is in a prograde orbit but is only a few degrees away from reorientation at $90\degr$. After just about 30 kyr, the probability of this event will already exceed 50\%; however, on a time scale of 40 kyr, more than 10\% of the VCs evidently have a retrograde orbit, with $ i> 100\degr$. The sungrazing scenario, described in Sect.~\ref{subsec:evo-122p}, is gradually starting to come into play at these time scales. At the final considered time scale of 100 kyr, the comet will survive with a half probability. Finally, considering the full-time scale, we can expect a conditional flipping probability of more than 30\% (Fig.~\ref{fig:flipping}, middle plot in the upper row).

Comet 161P Hartley/IRAS, according to the scenario discussed in Sect.~\ref{sub:evol-below-limit} with a high probability could become a sungrazer as early as 13\,kyr and this probability strongly affects the prediction of possible flipping. 161P is a retrograde comet, and as flipping here, we consider the probability of its passage into a prograde orbit. We have assumed an additional threshold for a prograde orbit as $i<85\degr$, since the cases of reaching even smaller inclinations were marginal. The results of the corresponding estimates can be found in Fig.~\ref{fig:flipping}, upper right plot.
It can be seen that a non-zero probability of flipping appears only after 9\,kyr from now when the fraction of surviving VCs is ten times smaller. Assuming the comet nevertheless survives the passages close to the Sun, it could flip its orbit to a prograde with a conditional probability of more than 45\% and $\sim 20$\%, for the threshold of $i<90\degr$ and $i<85\degr$, respectively.


Comet C/2014 Q3 (Borisov) is actually in a near-polar orbit with an extreme inclination of 89.9 degrees, which is formally a prograde orbit. Practically as early as the start of the simulation, many VCs jump into orbits exceeding 90 degrees, so the $P_{flip}$ values are relatively large at the start (Fig.~\ref{fig:flipping}, lower left panel). A lot may be explained by imposing a threshold of 92 degrees on the inclination -- the cases of going into such retrograde orbits are already marginal, as can be seen, and the jumps to 95 degrees are so marginal that they would be impossible to show. Furthermore, all VCs survived the simulation with no sungrazing events. Hence, the remaining probabilities in the figure are identical. C/2014 Q3 will be in a near-polar orbit in the future, with minor fluctuations in inclination near $90\degr$, most likely to remain in a prograde orbit by definition.


C/2015 GX (PanSTARRS) is a comet with a retrograde and near-polar orbit. Analysing the dispersion of the studied VC inclinations, it can be said that they fluctuate slightly below $90\degr$ at the beginning of the integration. According to the criterion, this can be described as flipping, but the dispersion statistics (Tab.~\ref{tab:2002k4-stat_at-the-end}) say that the motion is mostly retrograde, with the probability of jumping below $90\degr$ to a prograde orbit decreasing over time to near the 20\% value. Only 5\% of VCs have a chance of achieving orbits with an inclination below $85\degr$ (Fig.~\ref{fig:flipping}, middle plot in the lower row). The possibility of sungrazing state, as described in Sect.~\ref{subsec:evo-2015gx} is at 0.8\% at the end of the simulation and only marginally affects the flipping statistics. In general, the orbit will remain near-polar for a long time.

C/2002 K4 (NEAT) is a comet in a retrograde orbit, so here we analyse its probability of reorientation to a prograde orbit (Fig.~\ref{fig:flipping}, lower right plot). For the first 30\,kyr, the probability of flipping is marginal, but after 60\,kyr it reaches almost 80\%, eventually dropping to 65\% after 100\,kyr. Therefore, the chance of flipping is high, and the inclination could fall below $85\degr$ with a final probability of 12\% under the unlikely condition that the comet does not end its life as a sungrazer. The probability of such an event is low and appears to be non-zero after 80\,kyr. More detailed statistics on the evolution of this comet are given in Sect.~\ref{subsec:evo-2002k4}. Flipping from a retrograde to a prograde orbit is clear and is highly probable here.

Summarising the statistical analysis of flipping, it was possible to detect a noticeable number of such events for the VCs studied. In several cases, the probability reaches high values, even though sungrazing is conditionally included. As mentioned in the previous sections, Kozai resonance may be an important evolutionary factor in the whole process (Sect.~\ref{sub:evol-Kozai}). Comet 12P flips its orbit with low probability, but in the case of 122P (Sect.~\ref{subsec:evo-122p}) and 161P (detailed scenario in Sect.~\ref{sub:evol-below-limit} and Sect.~\ref{sub:evol-Kozai}) the values are even significant. In light of the preceding discussion, flipping is particularly apparent for 122P. C/2014 Q3 and C/2015 GX will remain near-polar comets. C/2004~K4, on the other hand, is a comet that will evidently undergo flipping in the future, and the triggering factor may be the influence of the large Solar System planets described in detail in Sect.~\ref{subsec:evo-2002k4}.

\section{Lyapunov time estimates --  methodology and results}\label{sec:evol-LT}

\begin{figure*}
	\centering
	\includegraphics[width=5.9cm]{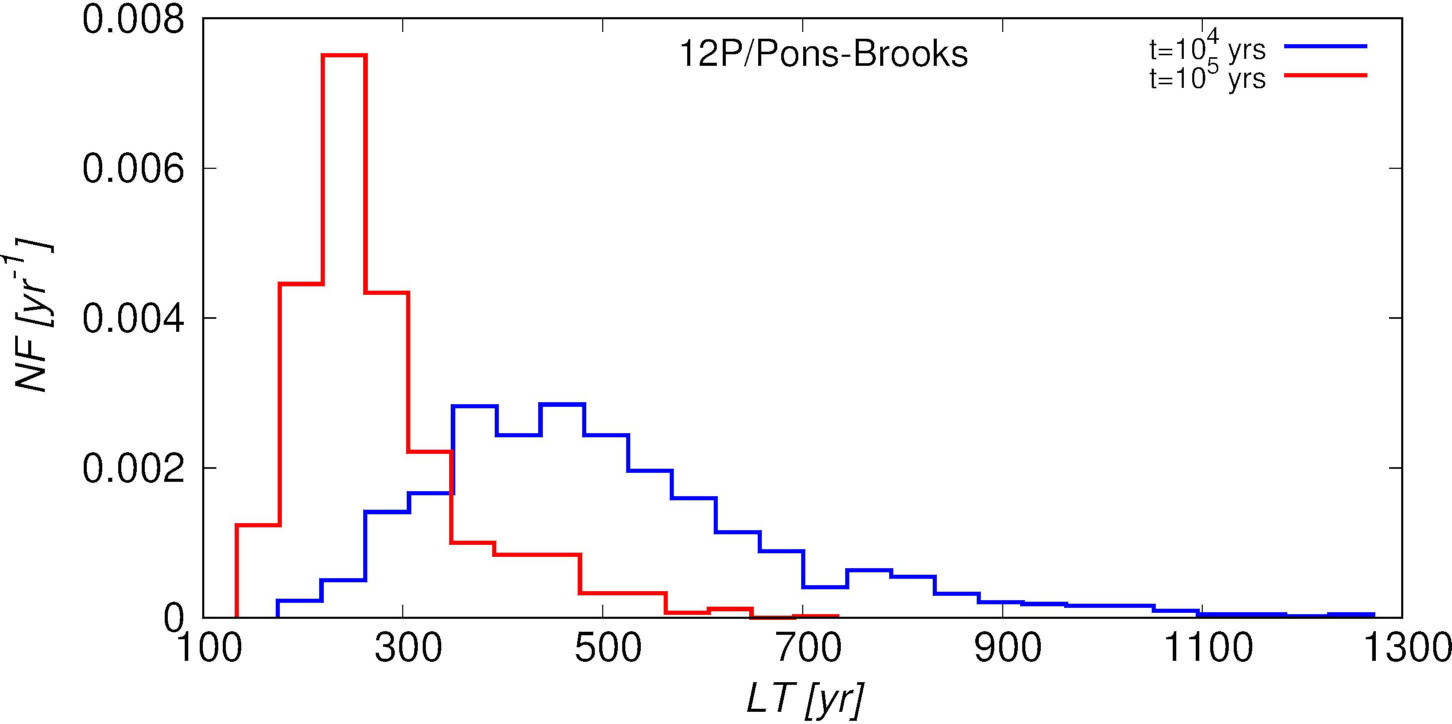}
	\includegraphics[width=5.9cm]{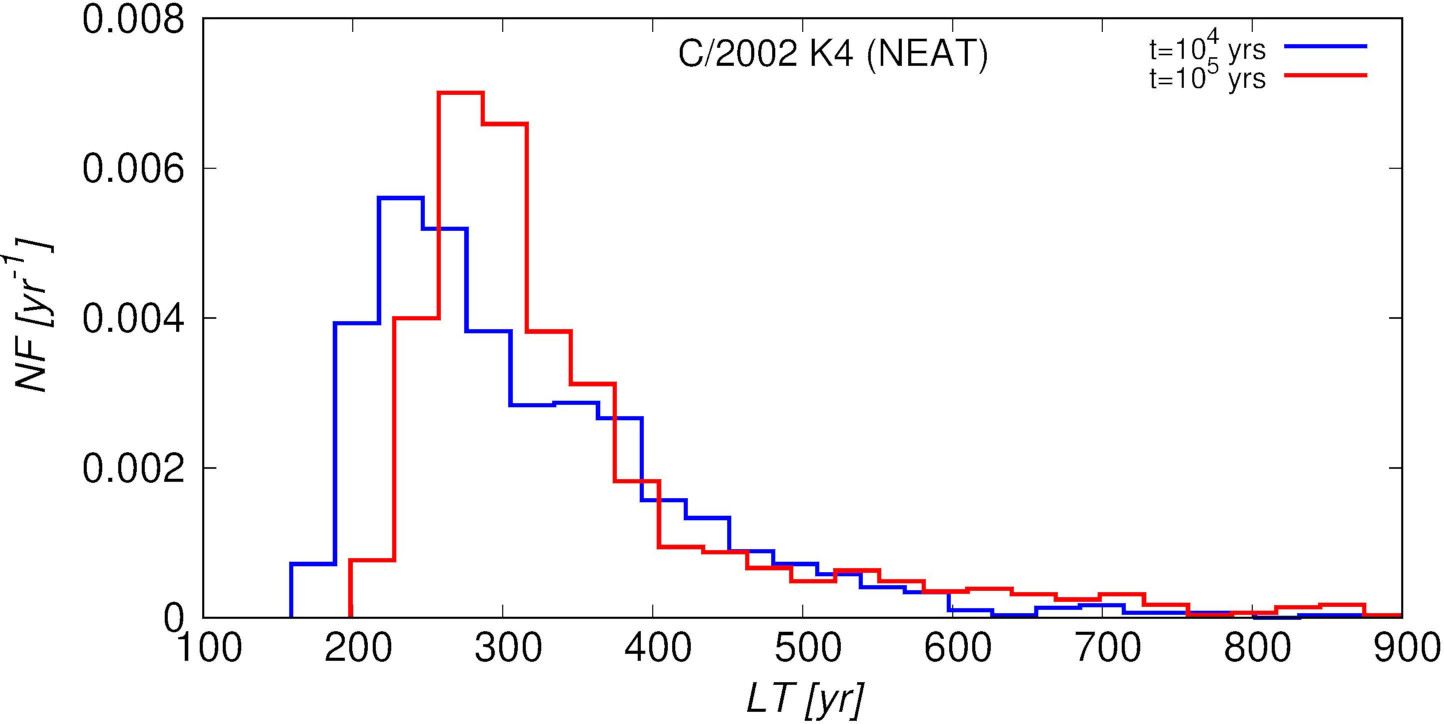}
	\includegraphics[width=5.9cm]{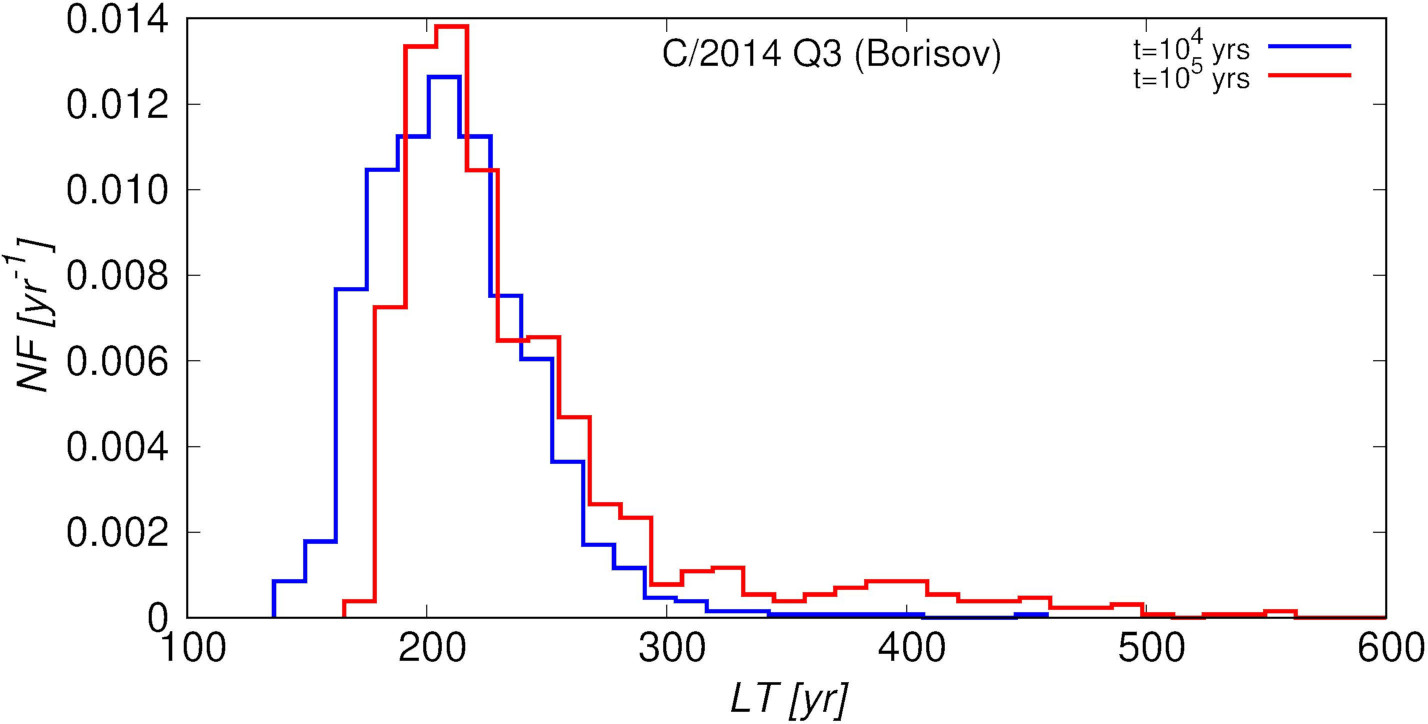}
	\includegraphics[width=5.9cm]{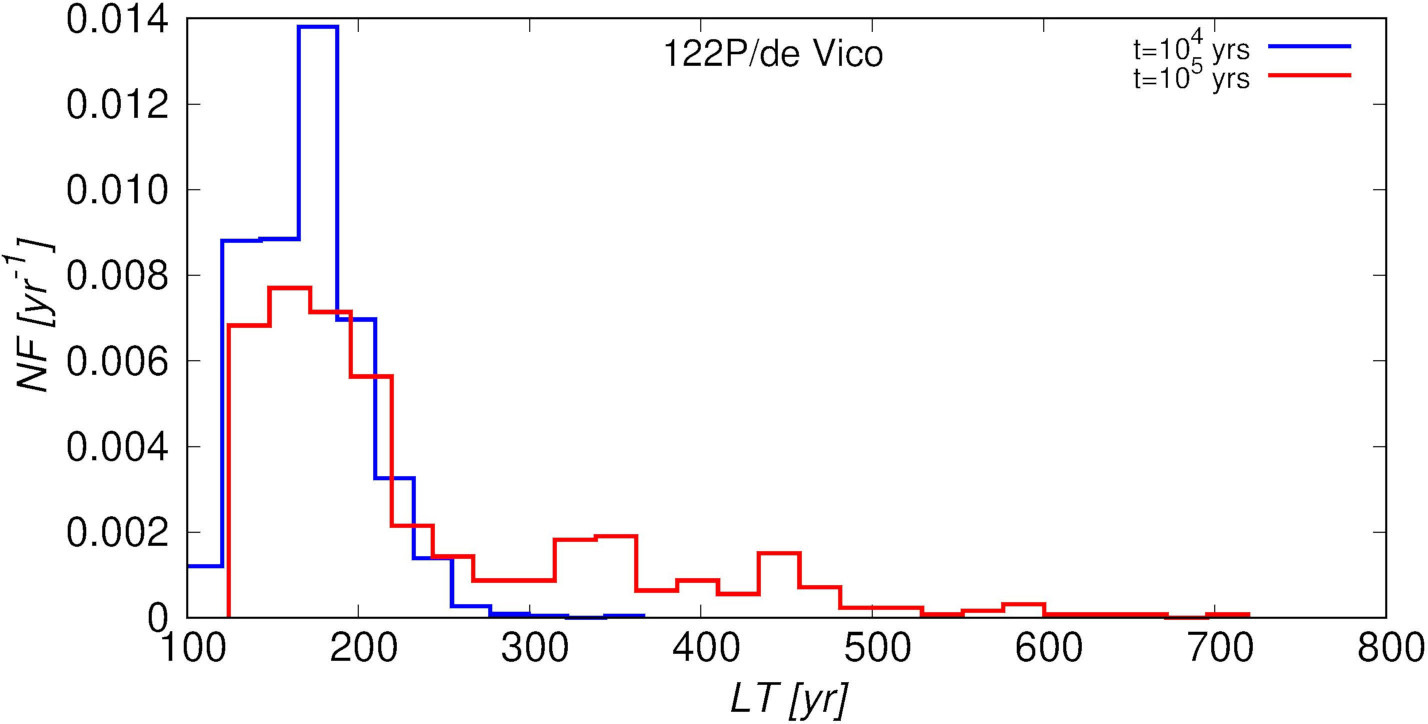}	
    \includegraphics[width=5.9cm]{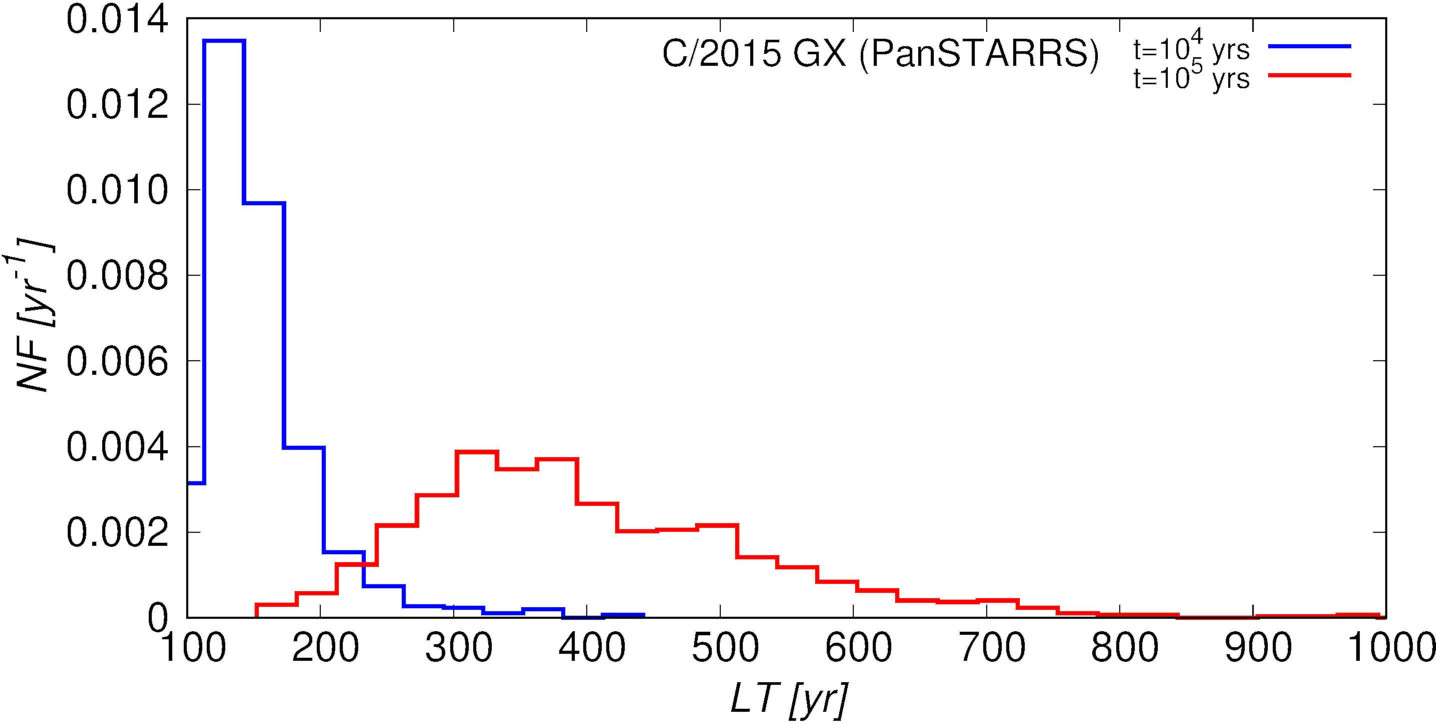}
	\includegraphics[width=5.9cm]{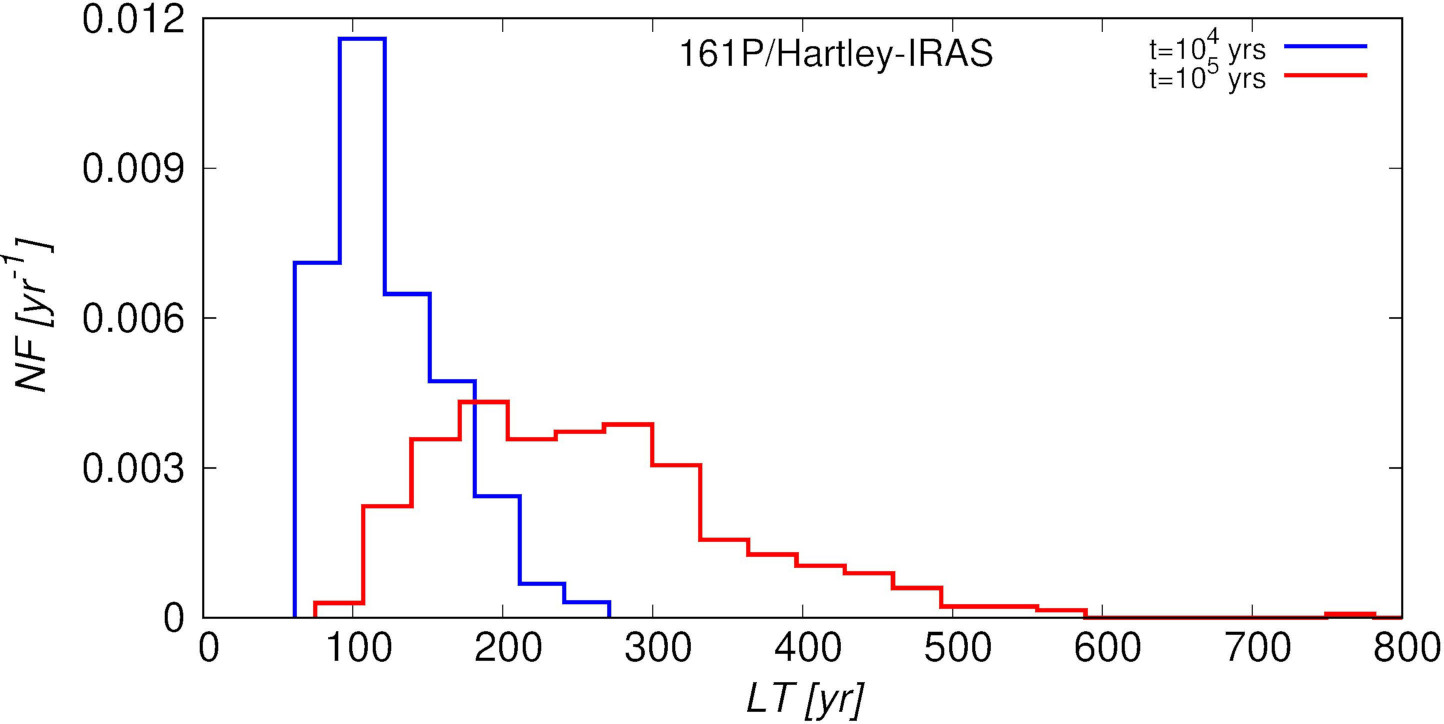}
	\caption{Statistical distribution of the Lyapunov time calculated for the swarm of VCs after 10\,kyr (blue lines) and 100\,kyr (red lines) for GR orbits representing HT~objects analysed in this study. The histograms were made for VCs excluding sungrazing cases.}	\label{fig:Lapunov}
\end{figure*}

\begin{table*}
	\caption{\label{tab:HTCs-LT} Parameters describing the statistical distribution of
		Lyapunov time calculated for 1001\,VCs and MEGNO mean values after 10\,kyr (columns[2]--[9]) and 100\,kyr (columns[10]--[16]) for GR~orbits of HT~objects anlysed in this study. The parameters are: the mean Lyapunov time ($\langle LT\rangle$ ), the standard deviation ($\sigma$), the 1st, 2nd (Median), and 3rd quartiles ($Q_1$, $Med$, $Q_3$) and the skewness ($\gamma_Q$).	The values of all dimensional parameters shown in the table are in Julian years. The MEGNO values are taken from the analysis of $a$--$i$ maps (Sect.~\ref{sec:evol-MEGNO}). See text for details.}
	\centering
	\setlength{\tabcolsep}{4.0pt} 
	\begin{tabular}{cccccccccccccccc}
	\hline \hline 
	Name     &    \multicolumn{7}{c}{after 10 kyr} &  \multicolumn{8}{c}{after 100 kyr}  \\
	&$\langle LT\rangle$ & $\sigma$&  $Q_1$ & $Med$ & $Q_3$& $IQR$  & $\gamma_Q$ & MEGNO   &$\langle LT\rangle$ & $\sigma$&  $Q_1$ & $Med$ & $Q_3$& $IQR$ & $\gamma_Q$  \\
	&[yr] & [yr]& [yr] & [yr] & [yr] & [yr]    &     & mean value   &[yr] & [yr]& [yr] & [yr] & [yr]  & [yr] &  \\
	$[1]$    & $[2]$  & $[3]$  & $[4]$    &$[5]$ & $[6]$  & $[7]$  & $[8]$    & $[9]$  &$[10]$ &  $[11]$ &$[12]$ &$[13]$ &$[14]$ &$[15]$ &$[16]$   \\
	\hline
	\hline \\
     \multicolumn{16}{c}{ $LT$ statistics excluding sungrazing cases} \\
	12P/Pons-Brooks         & 527 & 177 & 401 & 495 & 613 & 212  & 0.11    &~~8.4$\pm$3.0   &297 &86 &243 &276 &326 &84  & 0.21 \\
	C/2002 K4 (NEAT)        & 332 & 108 & 254 & 302 & 387 & 133  & 0.28    &13.1$\pm$4.6    &373 &148 &287 &325 &393 & 106 &0.29 \\
	C/2014 Q3 (Borisov)     & 217 &~~34 & 193 & 213 & 235 &~~42  & 0.05    &17.3$\pm$4.3    &250 &71 &209 &227 &262 &53 & 0.32\\
	122P/de Vico            & 183 &~~32 & 163 & 181 & 201 &~~38  & 0.05    &25.4$\pm$5.5    &247 &108 &171 & 204& 298&128  & 0.47 \\
	C/2015 GX (PanSTARRS)   & 168 &~~43 & 141 & 159 & 182 &~~41  & 0.12    &26.8$\pm$7~~~   &425 &151 &328 &398 &498 &170 & 0.18 \\
	\\
    161P/Hartley-IRAS       & 139 &~~40 & 110 & 130 & 165 &~~55  & 0.27    &~~5.8$\pm$1.6   &302 &260 &198 &267 &328 & 130 &-0.06 \\
    \\
     \multicolumn{16}{c}{ $LT$ statistics including sungrazing cases} \\
    161P/Hartley-IRAS       & 171 &~~89 & 116 & 141 & 192 &~~76  & 0.35    &~~5.8$\pm$1.6   &283 &194 &188 & 256& 326 & 137 &-0.004 \\
	122P/de Vico            & 183 &~~32 & 163 & 181 & 201 &~~38  & 0.05    &25.4$\pm$5.5    &230 & 87 &175 & 203& 254 & 72 & 0.22 \\
	\hline
\end{tabular}
\end{table*}

We begin the discussion of the stability of the analysed comets with Lyapunov time estimates. Although the dynamics of HT~comets have been discussed in numerous studies, the orbital stability -- particularly the estimation of the Lyapunov time for this type of comet -- has not been extensively explored. Some insights into the estimation of the Lyapunov time for HTCs can be found in \citet{Shevchenko2007}, where the author proposes, that for HTC or LPC, the lower bound of the Lyapunov time is generally greater than approximately half the orbital period of the comet in question. For comet Halley, a comprehensive analysis of the Lyapunov exponent was performed, taking into account the stability of its orbital clones. Various estimates of this time range from several decades to over 500 years \citep{Shevchenko2007, Munoz:2015, Boekholt:2016, Perez:2019, Kaplan:2022}. The Lyapunov time estimates we obtained for a sample of 6~HTCs analysed are of the same magnitude order as those for Halley's comet (see below).

The Lyapunov time ($LT$) is one of the key parameters used to identify the chaotic behaviour. $LT$ represents the characteristic time over which dynamics remain predictable in a chaotic regime of dynamical evolution. A short $LT$ indicates rapid divergence between nearby trajectories, highlighting the chaotic nature of the comet trajectory, while a longer $LT$ suggests a more stable, predictable orbit. This parameter is defined as the inverse of the maximal Lyapunov exponent (MLE), which can be estimated as follows \citep[see e.g.,][chap.~9]{DermottandMurray1999}.
\begin{equation}
\centering
\lambda = \lim_{t \to \infty} \frac{1}{t-t_0} \ln \frac{d(t)}{d(t_0)}
\end{equation}
\noindent where $d(t)$ and $d(t_0)$ represent the distances between two close trajectories at any given time ($t$) and at the initial time ($t_0$), respectively. To make the definition of the MLE parameter complete, one would need to take the limit as $d(t_0) \to 0$ in the above formula, which may be difficult to achieve from a numerical perspective.

Another problem is estimating an appropriately large time of integration in order to achieve convergence of LT values (see e.g. \citealt{Perez:2019} and references therein). Compared to the approach used for comet 1P/Halley by \citeauthor{Perez:2019}, our study shares several similarities: we used variational equations and perturbing planet data from JPL Ephemeris. Concerning the differences, instead of a sphere constraining the initial conditions near the nominal position, we used a swarm of 1001 VCs that have a dispersion corresponding to a 6-dimensional error ellipsoid, allowing us to examine a large number of VCs matched to the original observations. With a sufficiently large sample, we obtained a good statistical representation, but there are potentially elimination factors, such as sungrazing or less frequent VC ejections from the system. We made an effort to account for these factors in the results for individual objects. 

In numerical calculations the convergence of LT depends on several factors, for example: the used model of the Solar System, the starting data and also the optimisation of the algorithm itself that estimates the rate of expansion of close trajectories. During their long evolutionary period, comets change their dynamic regime. Therefore, for comparison purposes, our estimations are always limited to a given fixed time interval. In most of our cases, an interval of 10 ky seems sufficient to obtain convergence and a good approximation of the LT value. Extending this time interval, e.g. to 100 kyr, is not necessarily beneficial, as the object may migrate into areas of different dynamical regimes. 

Because of the numerical difficulties mentioned above, several methods have been developed to estimate the LT. One method is the renormalisation technique, also known as the two-particle method \citep{Benettin:1976}, while another involves the use of variational equations \citep{ReinandTamayo2016}. 
The renormalisation method can often result in an incorrect estimate of $LT$ \citep{Tancredietal2001}, so we opted to use the latter method to estimate $\lambda$ and thus $LT$. This estimation was performed, as in the case of MEGNO parameter (see Sect. \ref{sec:evol-MEGNO}), using the publicly available REBOUND software \citep{Rein2012}, the IAS15 integrator \citep{Rein2015}, and a finite time interval of 10\,kyr (see above). All calculations were performed in the GR regime.

In literature, the estimates of Lyapunov time or exponent for the bodies of interest are commonly based on using one (nominal) orbit. However, each individual orbit is subject to uncertainty in determining its orbital parameters, which may imply uncertainty in the $LT$ determination \citep[see e.g.][]{Murisonetal1994}. 

In order to determine the Lyapunov time and its uncertainty, a swarm of 1001\,VCs, constructed as described in Sect.~\ref{sec:evolution-classical}, was used to trace their orbital evolution over 10\,kyr and to get estimate of $LT$ form each individual VC. The statistics and histograms for orbital evolution over 100\,kyr are also presented for comparison purposes. Using sets of $LT$ values, we computed various statistical parameters, including the mean value of $LT$ ($\langle LT\rangle$), its standard deviation, $\sigma$, the 1st, 2nd (Median), and 3rd quartiles ($Q_1$, $Med$, $Q_3$), as well as the skewness. The last parameter is the Bowley skewness ($\gamma_Q$) calculated according to the formula: $((Q_3 - Med) - (Med - Q_1))/(Q_3 - Q_1)$. It ranges from $-1$ to $+1$ and a value of $\gamma_Q=0$ indicates a symmetrical histogram; $\gamma_Q>0$ signifies a right skewed distribution, and $\gamma_Q<0$ denotes a left skewed histogram. Another useful parameter that characterises a probability distribution is its 'width', which indicates how the values of a given parameter (in our case, $LT$) cluster around either its mean or median. In the case of clustering around the mean, the width is represented by $\sigma$, whereas for clustering around the median, it can be calculated, for example, the interquartile range defined as $IQR=Q_3-Q_1$.

Statistical analysis of $LT$ was performed only for those VCs for which $q>q_{\rm lim}$. We also divide the comets into two groups while discussing the results. The first group consists of objects whose majority of VCs did not evolve to $q<q_{\rm lim} = 0.005$\,au (5 comets) during the period of 10\,kyr (see Sect. \ref{sec:evolution-classical}). The results of these comets are displayed and described in order of decreasing $\langle LT\rangle$. Note that this group includes 122P in which swarm has 50\% of VCs experiencing a sungrazing state of $q<q_{\rm lim}$ at the end of the period of 100\,kyr. (Sect.~\ref{sec:evol-to-small-q}). Since we show for comparison also statistics after the period of 100\,kyr, this comet is a special case in this group with an increasing chance to be destroyed starting from 30\,kyr and reaching as much as 50\% at the end of 100\,kyr. We consider separately comet 161P, as it will most likely disintegrate within a period of 13\,kyr. In this case, approximately only 10\% of the swarm of 1001\,VCs survives at the end of this period and the $LT$ statistics are poorer than for the other five HTCs.

The results of our calculations of $LT$ after 10\,kyr and 100\,kyr described above are presented in Table \ref{tab:HTCs-LT} and Fig.~\ref{fig:Lapunov}, where vertical axes show normalised frequency counts ($NF$). This parameter can help to compare two different histograms and is defined as a the number of occurrences of a given values in a bin, divided by the total number of elements in the dataset and a length of this bin. The table clearly shows that the orbits of all examined comets exhibit chaotic behaviour, with $\langle LT\rangle$ values varying from approximately 100 to several hundred years. The right-skewed distribution, clearly visible in each histogram in Fig.~\ref{fig:Lapunov}, is a common feature. Below, we describe our results in more detail.

\subsection{Halley-type comets: avoiding the sungrazing state for the first 10\,kyr}\label{sub:LT_5comets}

Discussing the results after 10\,kyr, we rank the analysed comets fulfilled the subject criterion in terms of stability from the most stable to the least stable orbits: 12P/Pons-Brooks, C/2002 K4 (NEAT), C/2014 Q3 (Borisov), 122P/de Vico, and C/2015 GX (PanSTARRS). The same HTC order we apply in the next section where MEGNO analysis is shown (Sect.~\ref{sec:evol-MEGNO}).

The orbit of 12P/Pons-Brooks, with $\langle LT\rangle$ value of 527\,yr and $Mod$ of 495\,yr, is the most stable among those we analysed. The distribution of $LT$ is characterised by a notable scatter of values for individual VCs ($\sigma=177$\,yr, $IQR=212$\,yr). Due to the long though weak tail until 1300\,yr, a small right skewed asymmetry of this distribution can be observed ($\gamma_Q=0.11$).

C/2002 K4 (NEAT) has a second largest value of $\langle LT\rangle$ at 332\,yr and $Med$ at 302\,yr. For this comet, we also have a notable dispersion in $LT$ values ($\sigma=108$\,yr, $IQR=133$\,yr). Consequently, the mean and median $LT$ values are notable higher than the average $LT$ are reached by many VCs in a swarm, with a few approaching nearly 100\,yr. Therefore, this object has the most right skewed distribution with $\gamma_Q=0.28$ among the analysed HTCs.

Value of $\langle LT\rangle=$217\,yr was obtained for C/2014~Q3 (Borisov). Similar value we got for median ($Med=233$\,yr). The distribution of $LT$ values is characterised by a relatively low dispersion ($\sigma=34$\,yr, $IQR=42$\,yr). Similarly, as for 122P, we also obtained an almost symmetrical distribution of $LT$ ($\gamma_Q=0.05$). However, we observe a fairly long, but weak right tail extending up to $LT$=500\,yr.

For 122P/de Vico we have a notable smaller value of the $\langle LT\rangle$ about 183\,yr as well as the median ($Med=163$\,yr). The histogram width is also smaller ($\sigma=32$\,yr, $IQR=38$\,yr) and its skewness is small $\gamma_Q=0.05$. However, the right tail of the $LT$ distribution, which is almost invisible, is relatively long, extending even to 370\,yr. 

For C/2015 GX (PanSTARRS) the $\langle LT\rangle$ is estimated to be 168 years. A similar value for the median, which is 159\,yr, was obtained. The values of 43\,yr and 41\,yr was obtained for $\sigma$ and $IQR$, respectively. The histogram of $LT$ values shows a right skewed asymmetry ($\gamma_Q=012$) and fairly long but thin right tail that reaches up to 500\,yr, as in the C/2014~Q3 case.

Generally, on the longer time scale of 100\,kyr, the results seem different for these five comets. However, in the case of two of them, C/2002~K4 and C/2014~Q3, we have similar estimates of the mean and median $LT$ as for 10\,kyr. This may indicate a similar dynamical regime on the longer and shorter time scales. For the three remaining comets, we have different values for the mean and median $LT$ compared to those obtained after 10\,kyr, which in turn suggests that each of these three comets has moved into a different dynamical regime compared to the one at the beginning of its evolution. The most notable observation is the distinct increase in uncertainty of $LT$ estimates (mean and median). Only comet C/2014~Q3 still stands out with similar estimates in both time scales and relatively small uncertainties. This result can be understood as being consistent with a picture of the smallest number of close approaches to giant planets over the next 100\,kyr of future evolution of this comet (upper right panel in Fig.~\ref{fig:three_HTCs-evo}) compared to similar pictures for the remaining four comets.

It is worth noting that the width of the LT histogram can depend on many factors. One of them is the significant dependence of the LT on the initial conditions, in particular, the semi-major axis of a given VC. This issue is discussed in \citet{Murisonetal1994}, where the dependence of LT determination on the semi-major axis, as well as on the eccentricity and inclination of the orbit was analyzed.

\begin{figure*}
\centering
\includegraphics[width=8.4cm]{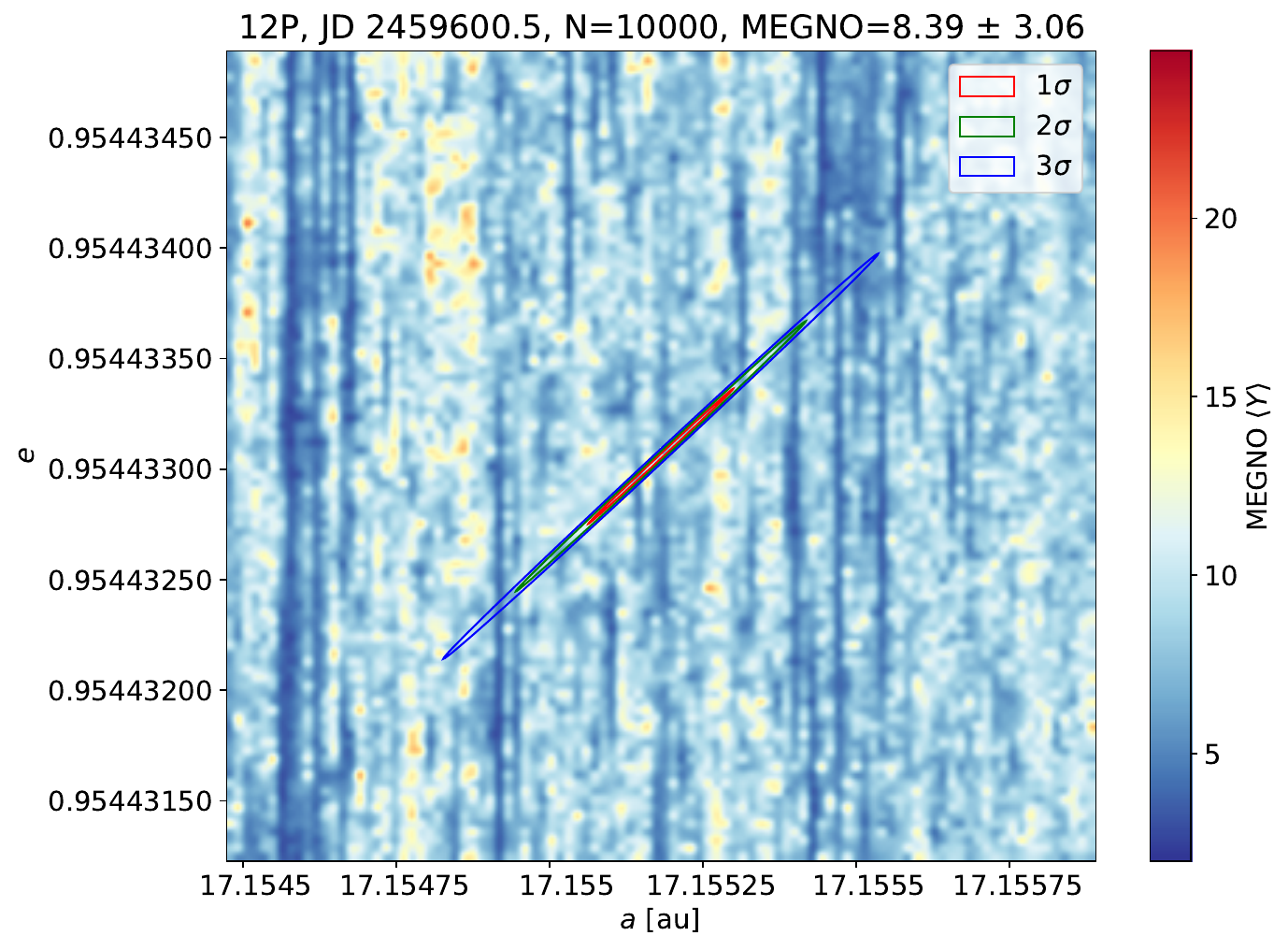}
\includegraphics[width=8.4cm]{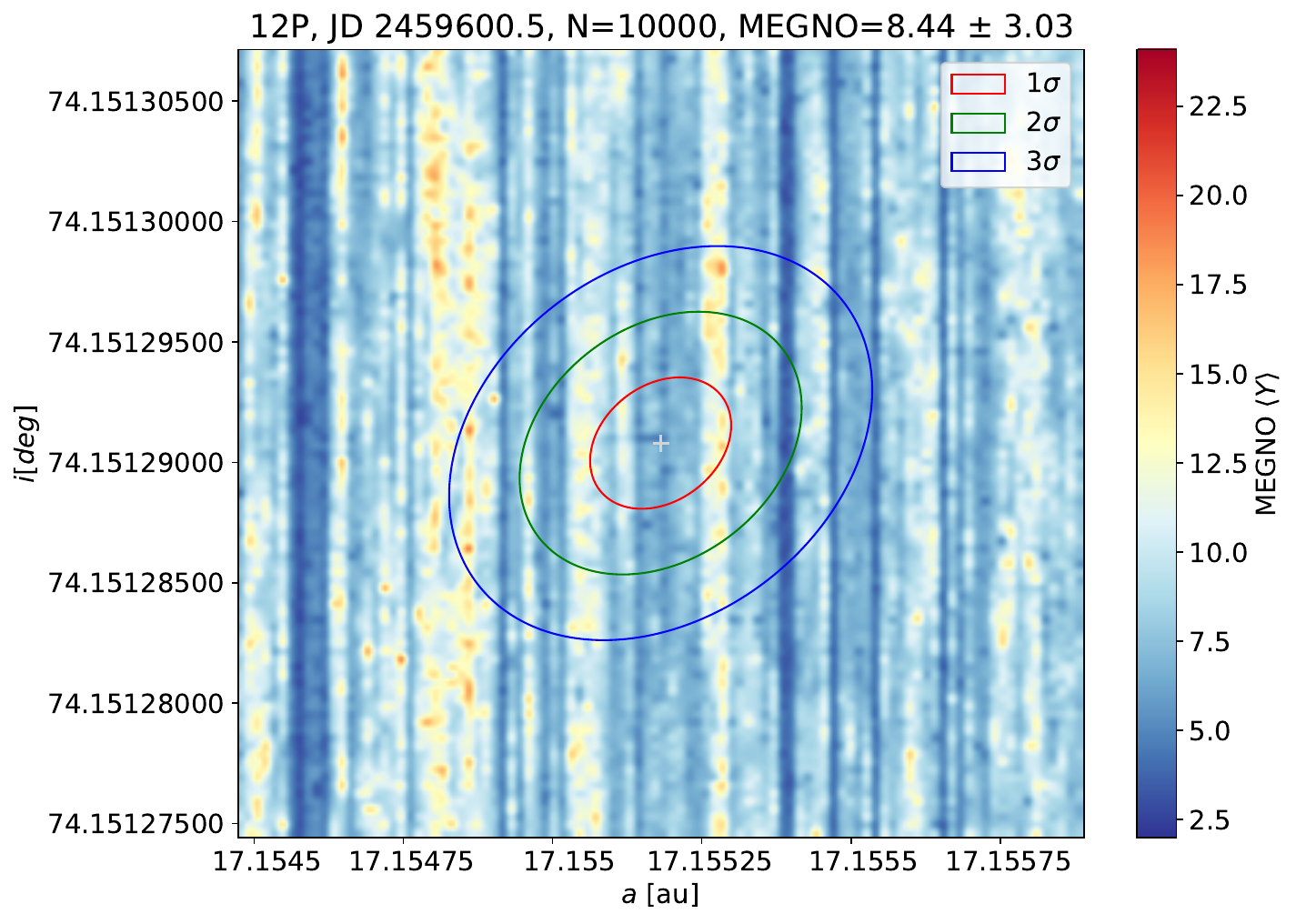}
\caption{MEGNO maps of comet 12P based on data from our orbital solution (gravitational variant, 'GR, 2020-2024'). Left panel: $a-e$ plane. Right panel: $a-i$. In this example, the elements $a-e$ were highly correlated, thus the covariance ellipse on the left panel is very narrow.
} \label{fig:MEGNO12a}
\end{figure*}
\begin{figure*}
\centering
\includegraphics[width=8.4cm]{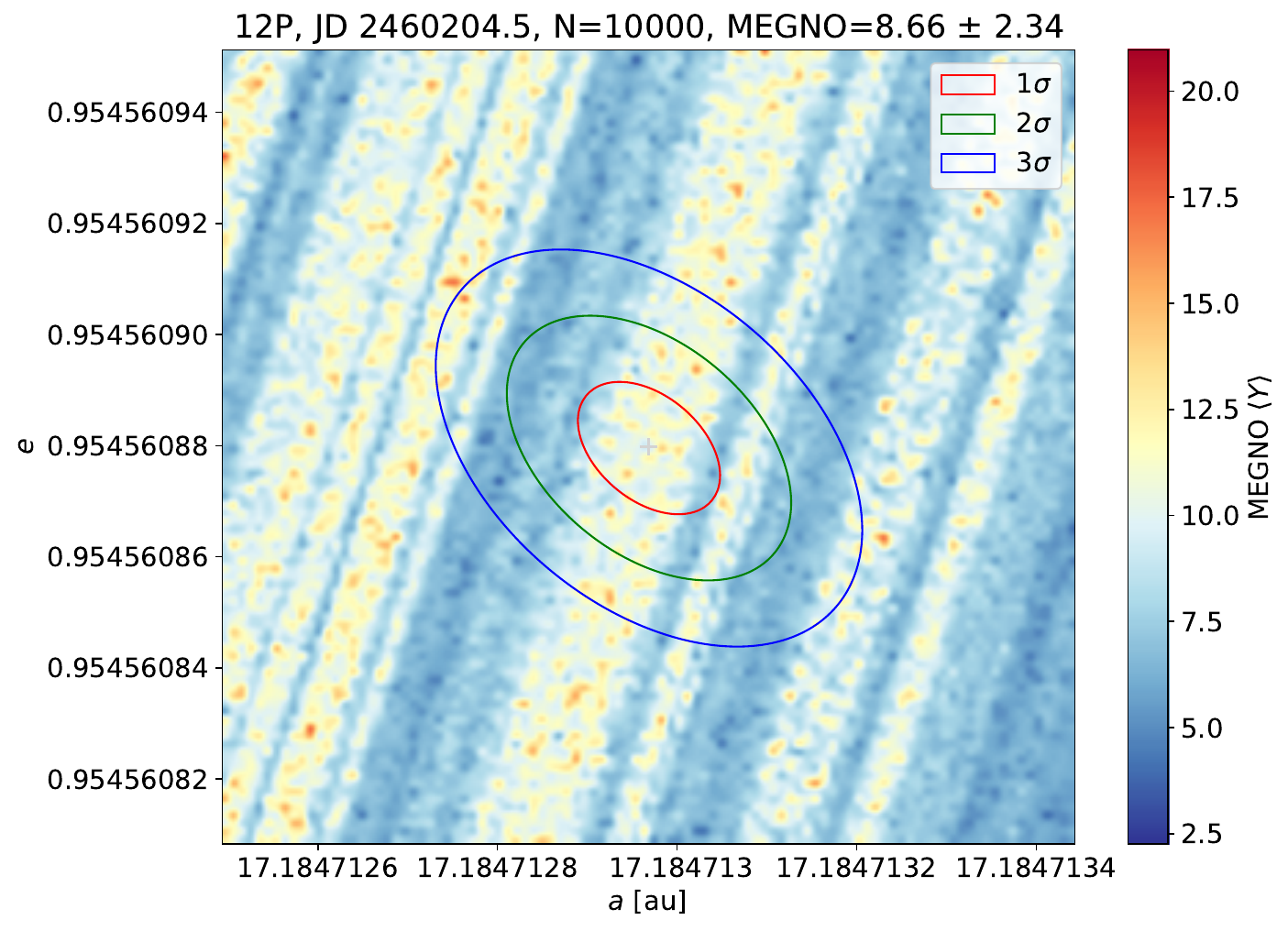}
\includegraphics[width=8.4cm]{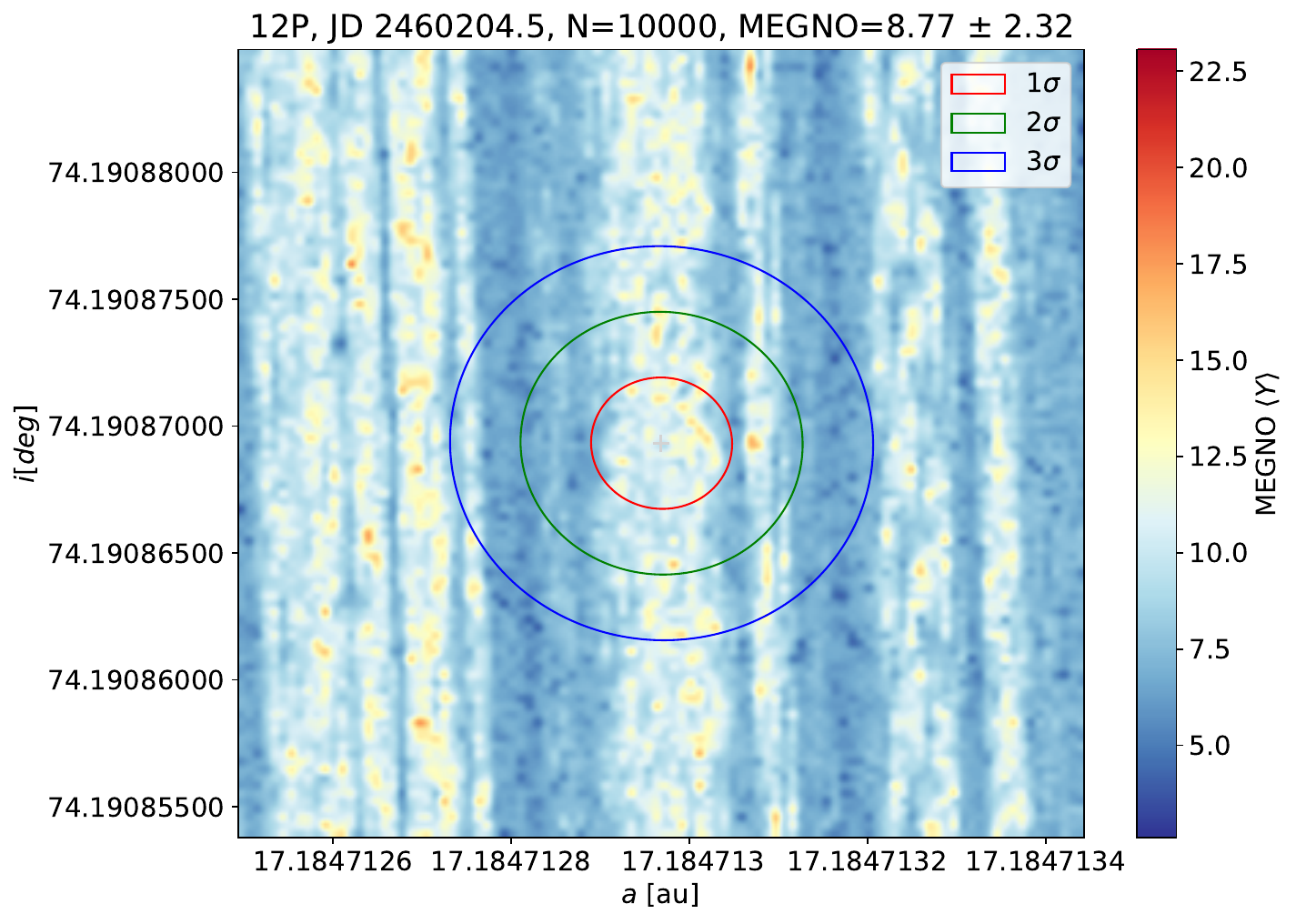}
\caption{MEGNO maps of comet 12P based on data from our orbital solution (non-gravitational variant, 'NG, 2020-2024'). Left panel: $a-e$ plane. Right panel: $a-i$.} \label{fig:MEGNO12}
\end{figure*}

\begin{figure*}
\centering
\includegraphics[width=8.4cm]{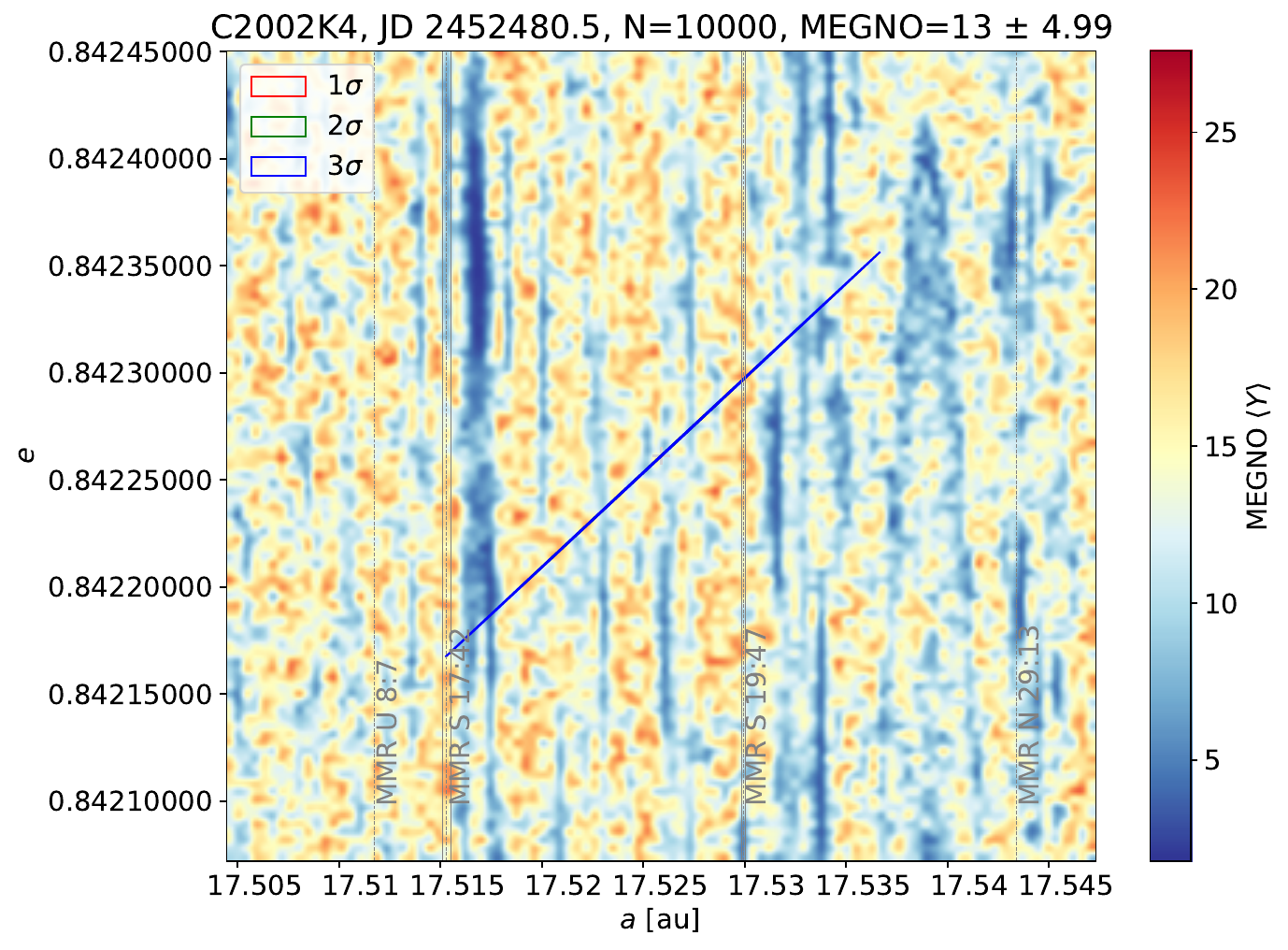}
\includegraphics[width=8.4cm]{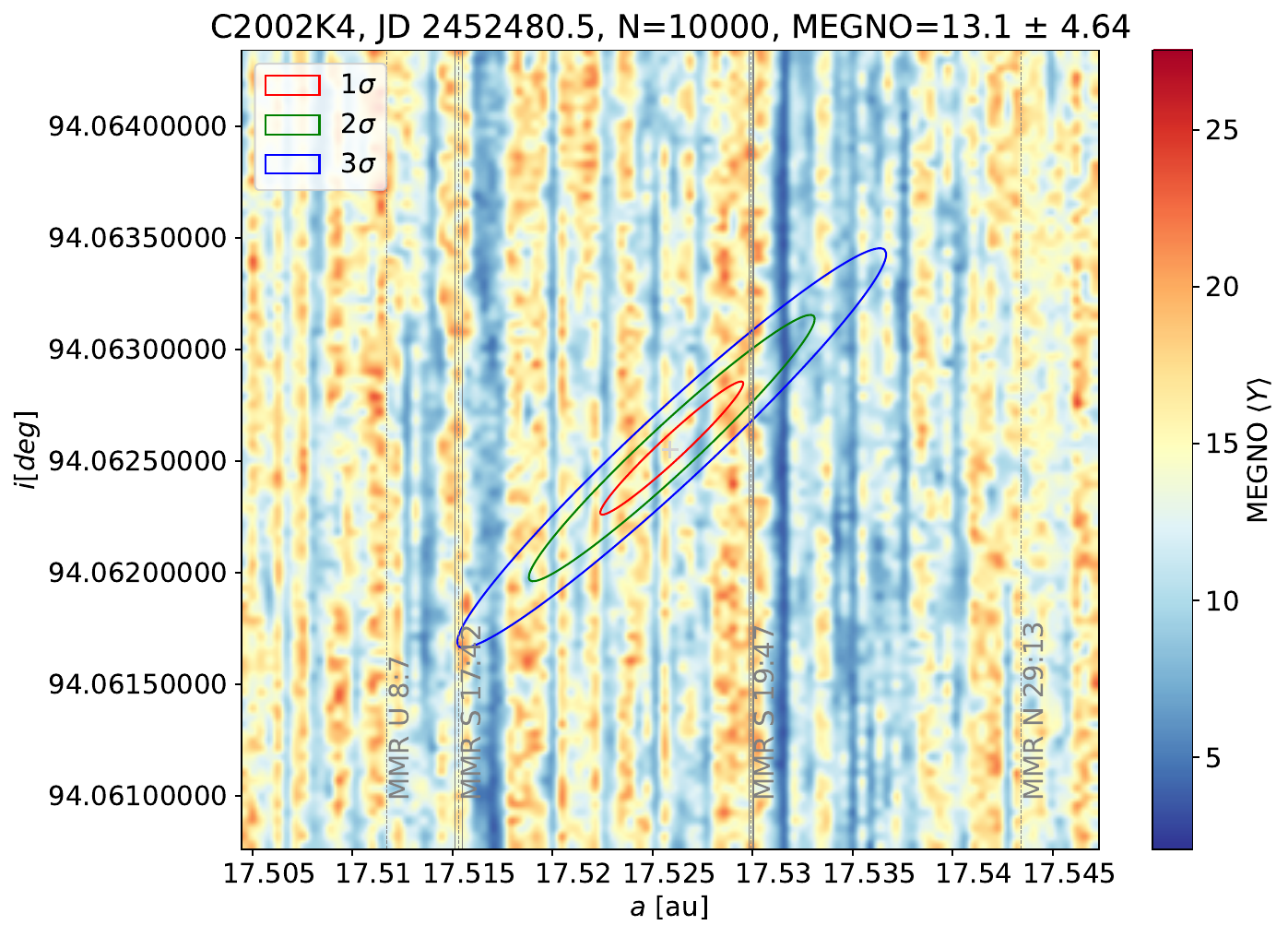}
\caption{MEGNO maps of comet C/2002 K4 based on data from our orbital solution (variant GR). Left panel: $a-e$ plane. Right panel: $a-i$. The strongest MMR resonances are indicated.} \label{fig:MEGNOc2002k4}
\end{figure*}
\begin{figure*}
\centering
\includegraphics[width=8.4cm]{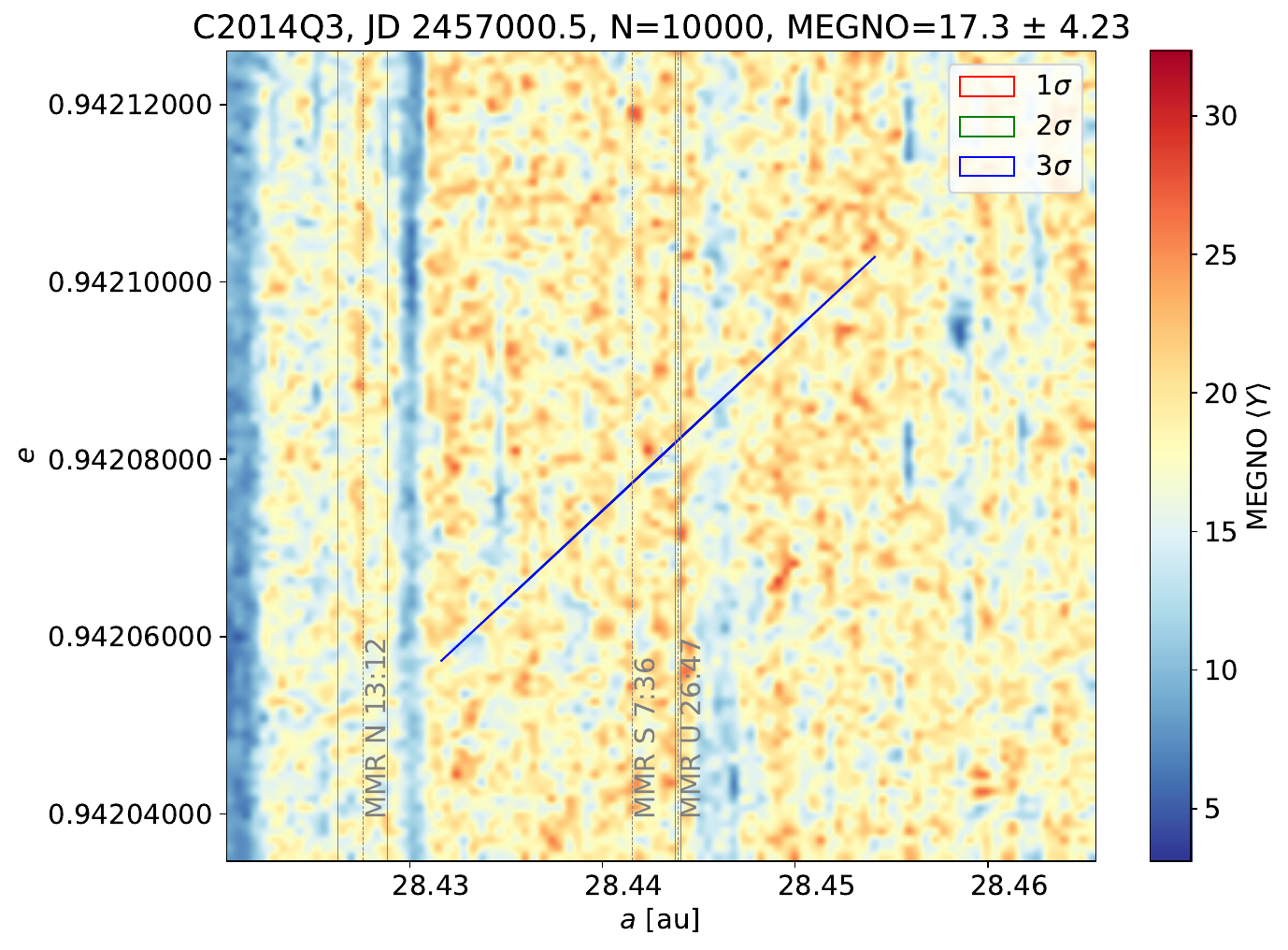}
\includegraphics[width=8.4cm]{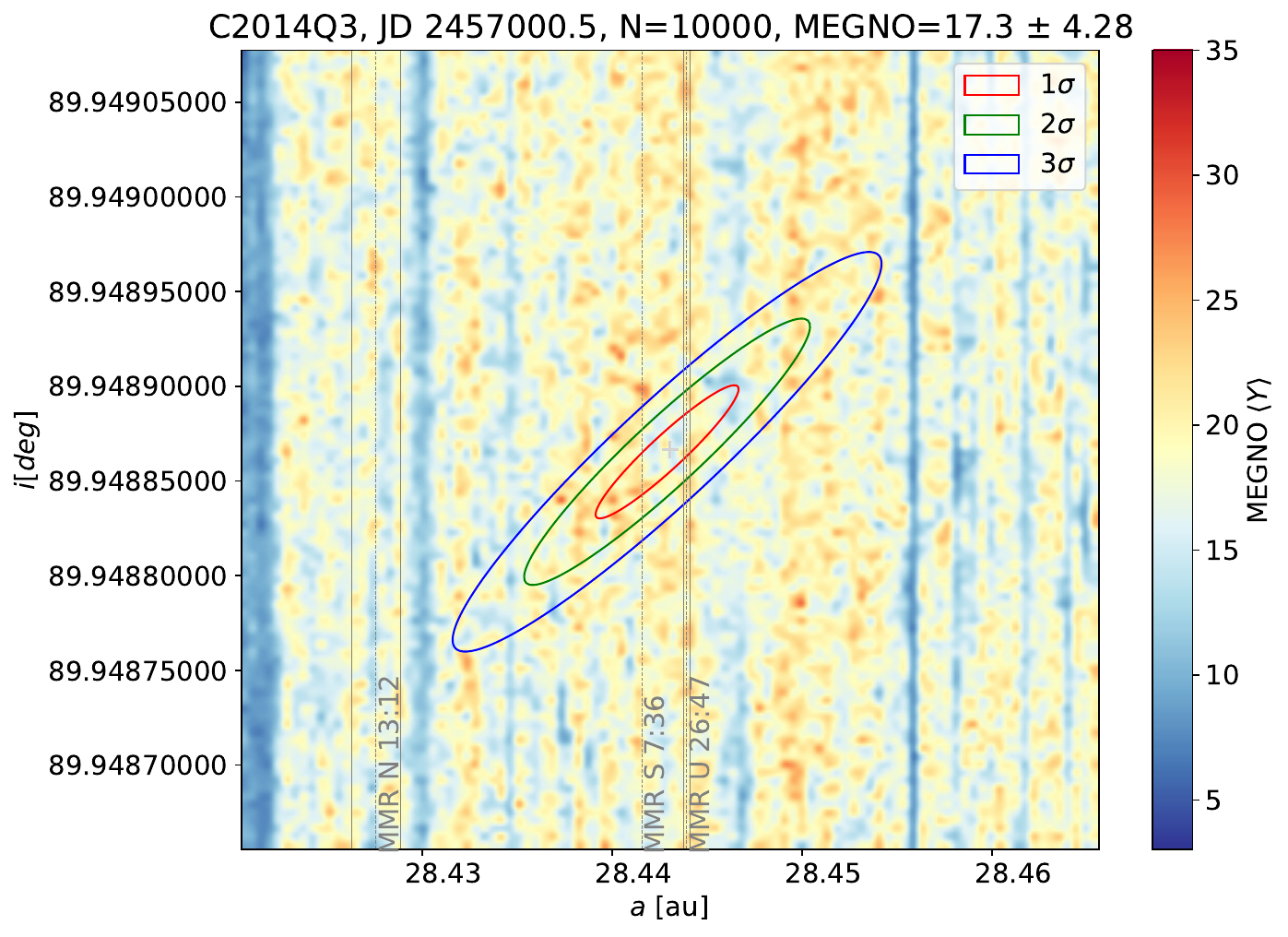}
\caption{MEGNO maps of comet C/2014 Q3 based on data from our orbital solution (variant GR). Left panel: $a-e$ plane. Right panel: $a-i$. The strongest MMRs in the vicinity of nominal solution are indicated.} \label{fig:MEGNOc2014q3}
\end{figure*}
\begin{figure*}
\centering
\includegraphics[width=8.4cm]{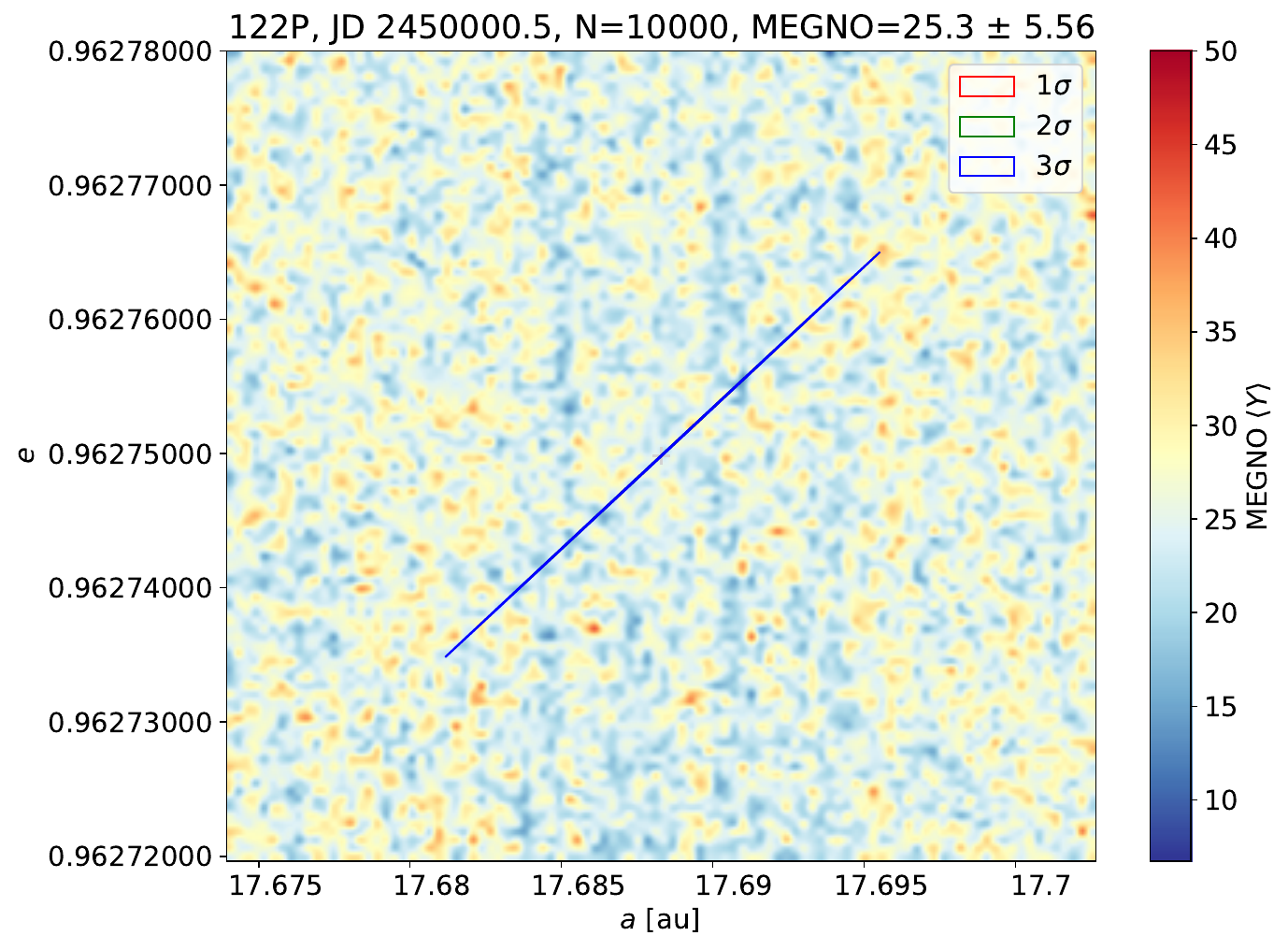}
\includegraphics[width=8.4cm]{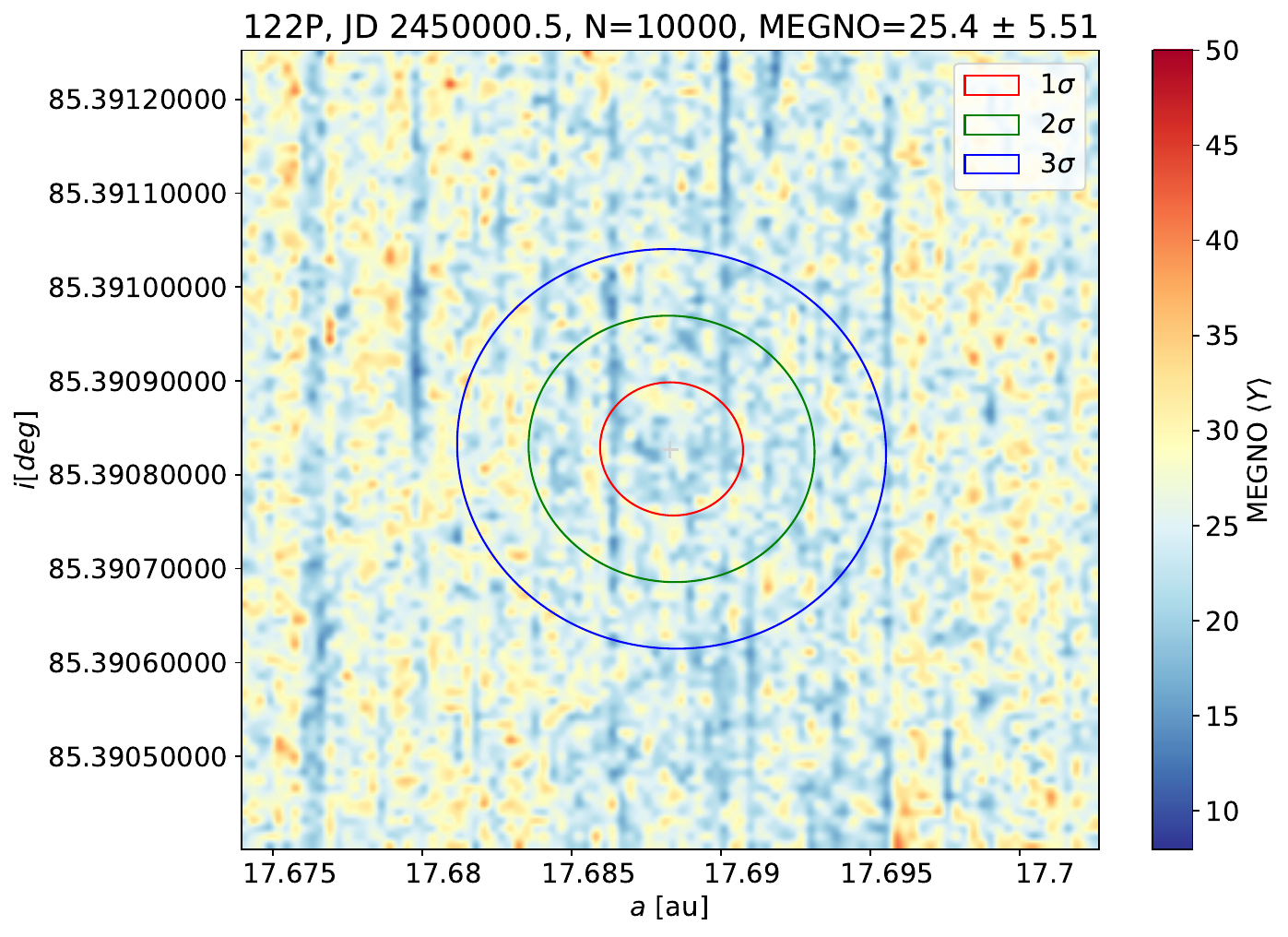}
\caption{MEGNO maps of comet 122P based on data from our orbital solution (variant GR). Left panel: $a-e$ plane. Right panel: $a-i$.} \label{fig:MEGNO122}
\end{figure*}
\begin{figure*}
\centering
\includegraphics[width=8.4cm]{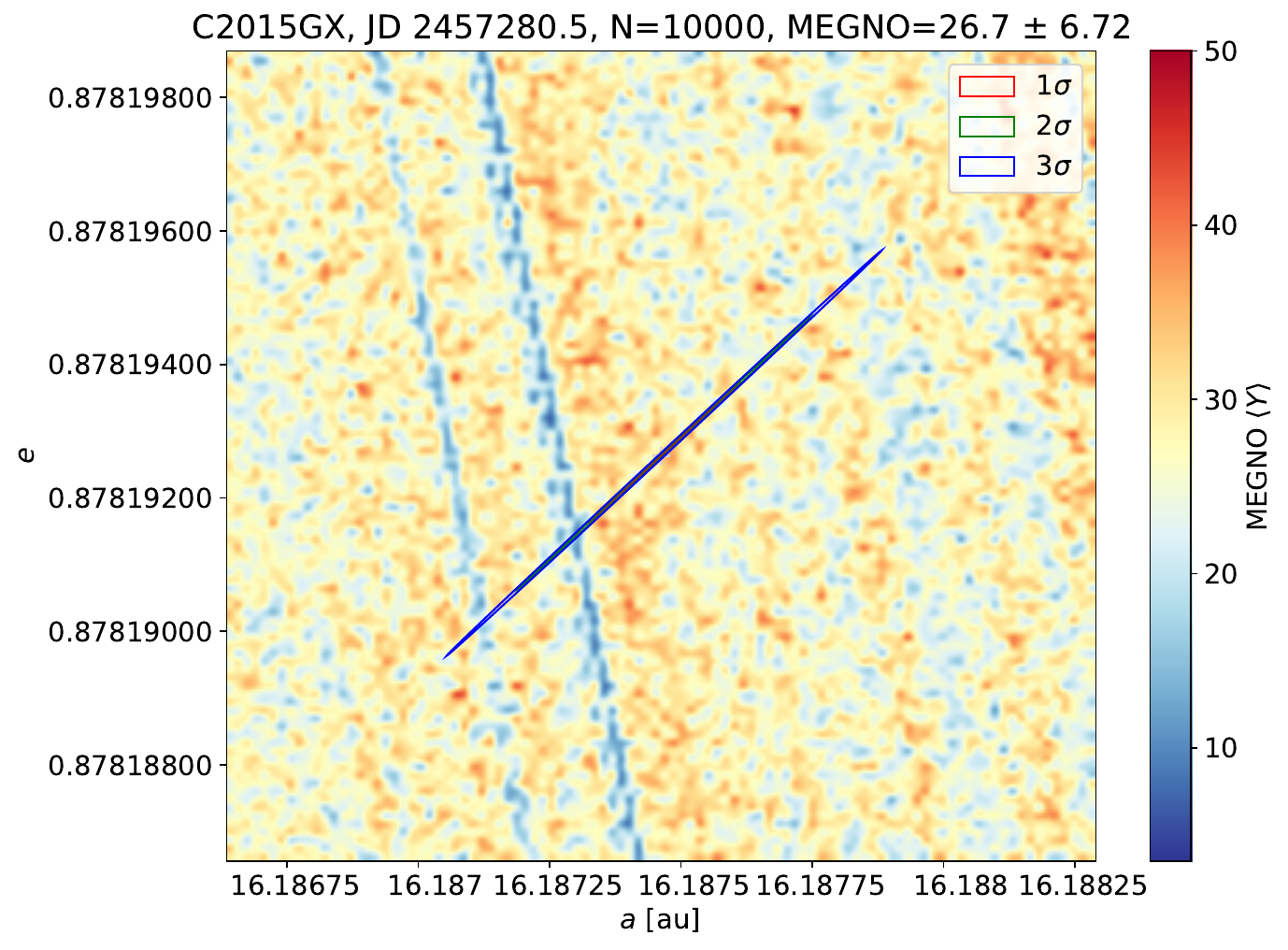}
\includegraphics[width=8.4cm]{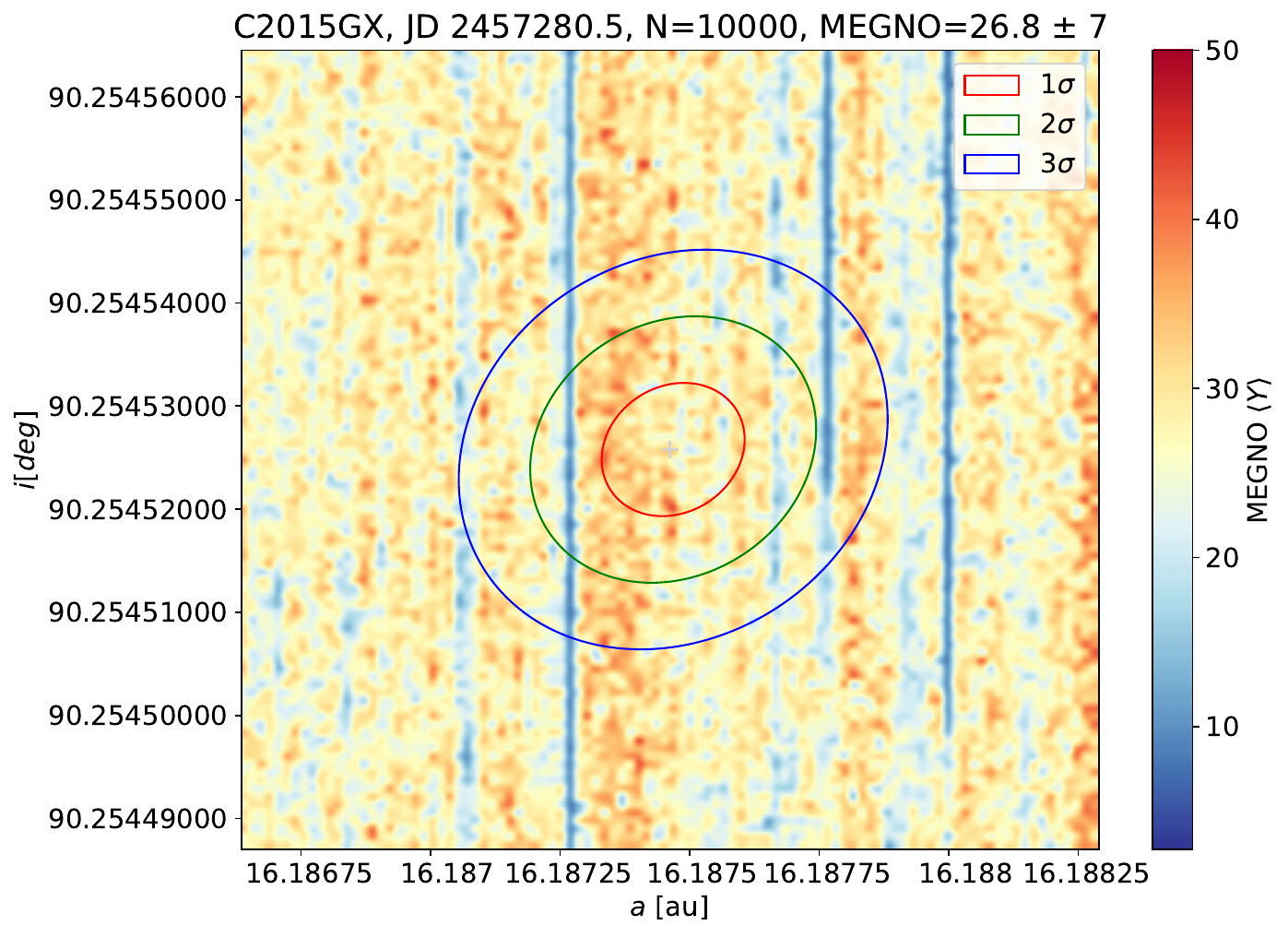}
\caption{MEGNO maps of comet C/2015 GX based on data from our orbital solution (variant GR). Left panel: $a-e$ plane. Right panel: $a-i$.} \label{fig:MEGNOc2015gx}
\end{figure*}
\begin{figure*}
\centering
\includegraphics[width=8.4cm]{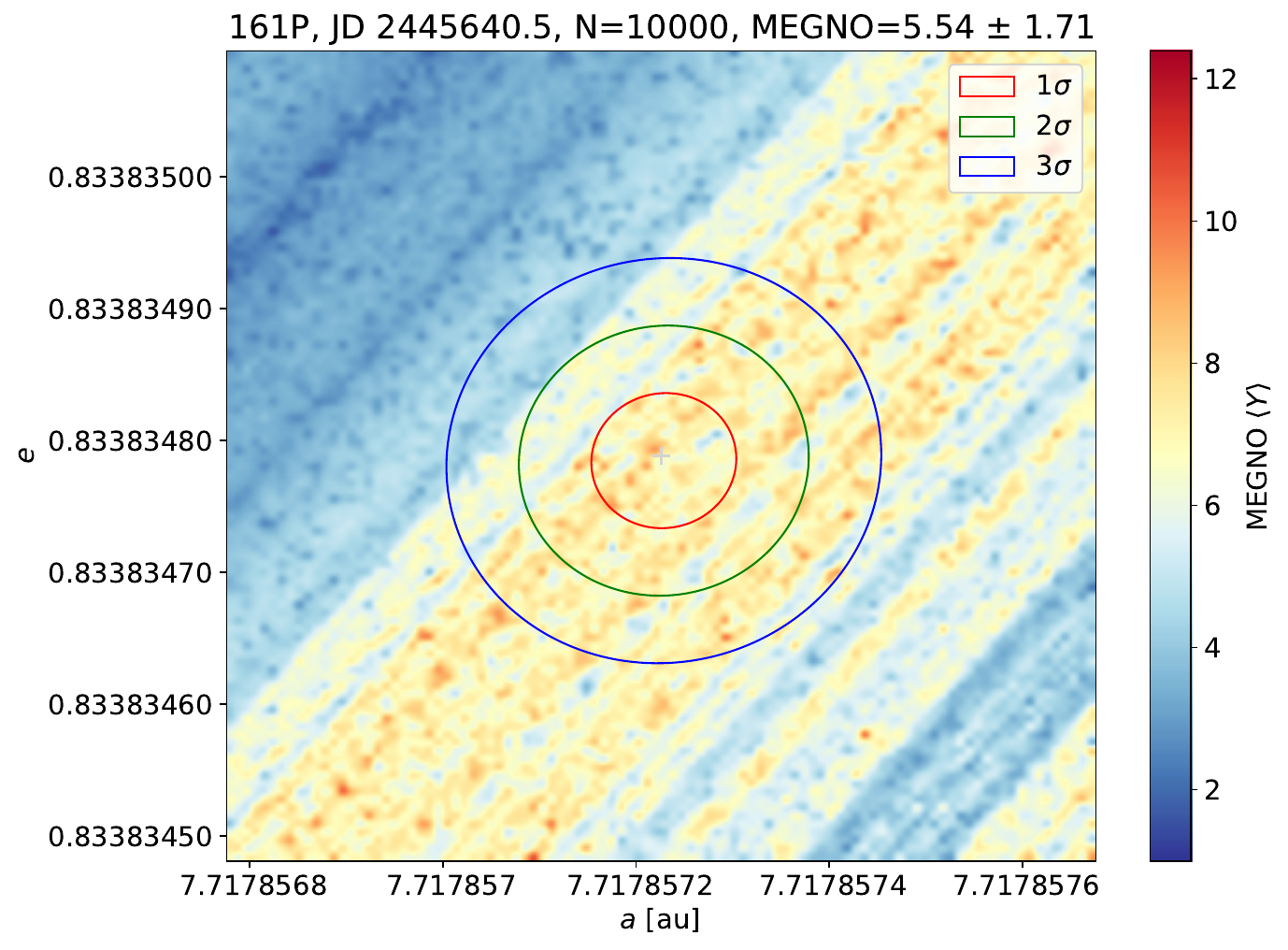}
\includegraphics[width=8.4cm]{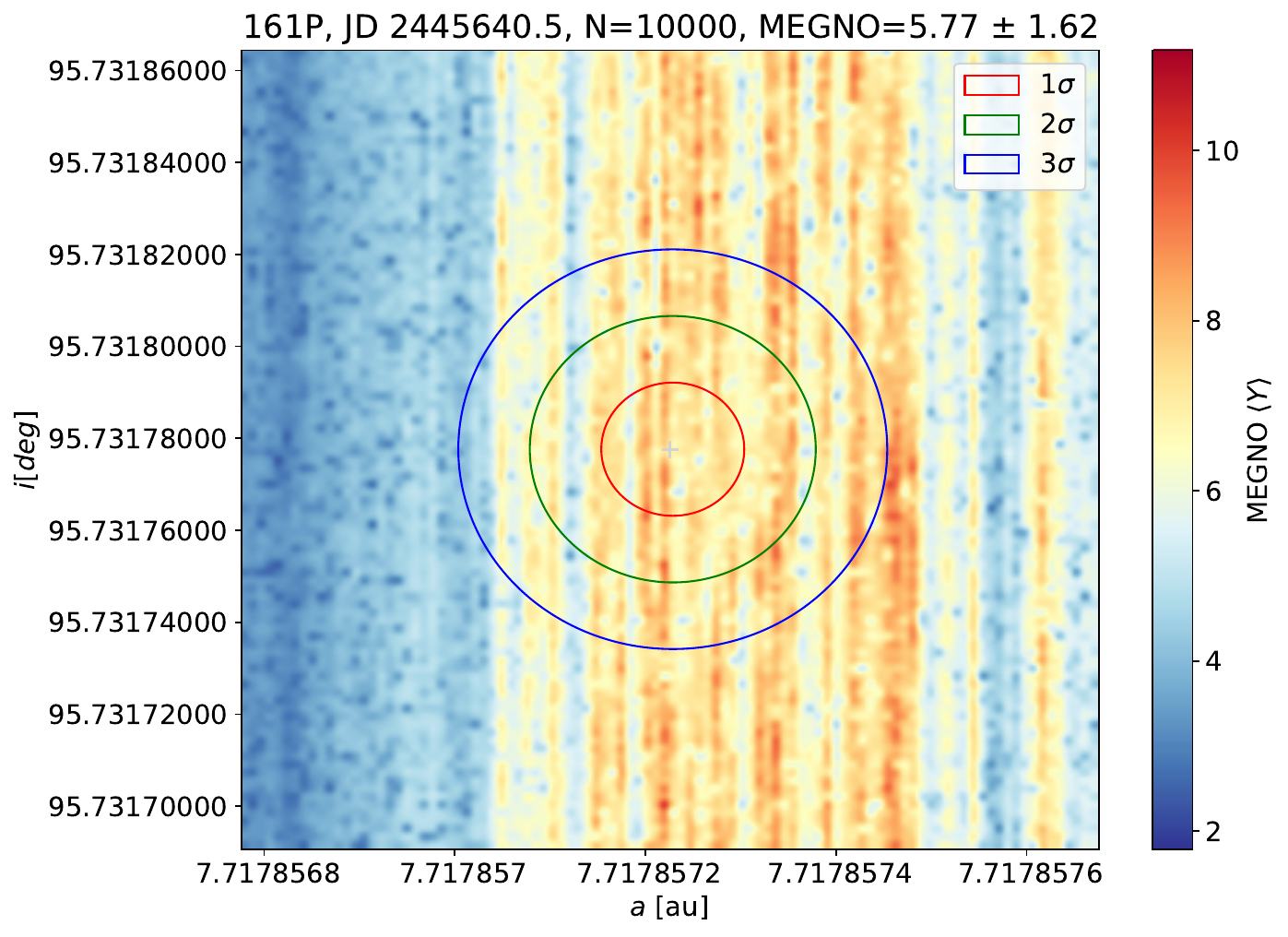}
\caption{MEGNO maps of comet 161P based on data from our orbital solution (variant GR). Left panel: $a-e$ plane. Right panel: $a-i$.} \label{fig:MEGNO161}
\end{figure*}

\subsection{Unique case of 161P/Hartley-IRAS}\label{sub:LT_161P}
Statistics after 10\,kyr for 161P/Hartley-IRAS are based on a 10-times poorer VC sample relative to the remaining HTCs, because here in the short period between 8.5\,kyr and 9\,kyr from now about 90\% of VCs were removed from the analysed swarm due to experiencing episodes of $q$ below the assumed $q_{\rm lim}$ (Sect.~\ref{sec:evol-to-small-q}). In at least part, this reflects the incorrect $LT$ distribution for this comet. We obtain low $\langle LT\rangle$ value for 161P/Hartley-IRAS: $\langle LT\rangle=$139\,yr, $Med=130$\,yr, $\sigma = 40$\,yr, and $IQR = 55$\,yr. The distribution is characterised by a distinct, thick tail on the right side, resulting in relatively high skewness ($\gamma_Q = 0.27$), which is the second highest among the comets studied here.

An analysis of 161P was previously performed by \citet{KankiewiczWlodarczyk2021}. They analysed dynamics of 100\,VCs of this comet in two regimes: GR and NG, and concluded that the mean Lyapunov time is over 600\,yr in both cases. This estimated value of $LT$ is greater than what we obtained. However, despite using slightly different data and methodology, the results are of the same order; see also Sect.~\ref{sub:MEGNO_161P}, where we discuss the differences in $LT$ estimates for this object.

\section{MEGNO indicators: methodology and results}\label{sec:evol-MEGNO}

In the previous section, parameters such as Lyapunov time, which are the inverse of the Lyapunov exponents \citep{Benettin:1976}, are used. However, there are more numerical chaos indicators based on a similar concept. Additionally, for the specific purposes of this work, the Mean Exponential Growth factor of Nearby Orbits (MEGNO) indicator was also chosen \citep{Cincotta:2000} due to its particular advantages in planetary systems applications \citep{Gozdziewski:2001}. In particular, MEGNO is considered to be a fast converging indicator, which is relevant for our sample, which potentially changes its dynamical regime during long numerical integration. Here, the MEGNO indicator was estimated on a finite time interval of 10 kyr for all comets. As the main tool, REBOUND software \citep{Rein2012} and the IAS15 integrator \citep{Rein2015} were used. In this short-term chaos estimation, the MEGNO mapping algorithm treated the VCs as massless test particles, and the sungrazing criterion ($q<0.005$\,au) was not applied. As shown by previous results (Sect.~\ref{sec:evol-to-small-q}), the risk of such an event occurring within 10 kyr appeared to be negligible (except for the 161P case).

Obtained MEGNO maps in ranges corresponding to orbit determination errors show the complex sensitivity of trajectories to changes in initial conditions near the nominal solution. Typically, they are presented here in ($a,e$) and ($a,i$) space. The nominal osculating orbits of the comet sample determined here, as well as the corresponding orbits from the JPL database shown in Appendix~\ref{sec:MEGNO-JPL}, were used comparatively in this manner to estimate the chaos. Finally, the resulting plots facilitate judgements about how strong the chaos is for the orbits in the range of observational errors. If there are areas of elements near the nominal solution where there are some chaotic zones, for example, related to MMR or approaches to planets, this can have a direct impact on the limited predictability of the trajectories. This approach also allows, in general, the detection of less and more stable orbital solutions.

The relation of the presence of chaotic areas on the map to the presence of MMR is not straightforward. Indeed, resonances act in a complex way and can play both a destabilising and a stabilising role. Therefore, they may not always be associated with larger values of the indicators marked on the charts (thus, changes may take on different tones on the colour maps as 'stripes', as detailed below).
Not all MMRs can be easily detected in our examples, so only those of the lowest order and greatest relative strength are shown. The software of \cite{Gallardo2021} (Resonance Atlas) was used for this purpose.

The MEGNO plots are shown as colour maps. Inside each plot, an approximate range of errors in the elements is shown in the form of a confidence ellipse (for our orbit determinations) or a grey cross (for orbits from the JPL database) -- the solutions in the centre cover the majority of error limits according to the $\pm 3 \sigma$ criterion. In brief, the diagrams are scaled by observational errors and the most significant results are located in the centre. MEGNO values around $\sim 2$ indicate quasi-periodic motion but, in most cases, are quite larger. Overall, the warmer the colour, the greater the chaos. Mean motion resonances (MMRs) are visible on the plots as numerous stripes of different colours. The warmer colours of the bars may indicate a local, destabilising effect of the resonance involved. Above the figures, the epoch of osculation for the elements, the total count of data points (N) and the MEGNO mean value with its standard deviation are given.

In this section, we use the order of discussing comets as in the previous section. It means that HTCs are presented in order of increasing MEGNO mean value, except 161P, which is a special case.

\subsection{HTCs not experiencing the sungrazing state until 10 kyr}\label{sub:MEGNO_5comets}


12P/Pons-Brooks is a comet for which we were able to determine orbits well in both the NG and GR regimes. Therefore, we carried out MEGNO calculations in different variants of the initial data; however, both in the GR regime. For comparison purposes, we used the initial orbit from the JPL database, which only offers an orbit in the NG variant for this comet. Consequently, for comet 12P we had two variants of starting orbits determined (Fig.~\ref{fig:MEGNO12a}, Fig.~\ref{fig:MEGNO12}), and one from JPL database (upper panel of Fig.~\ref{fig:MEGNO12JPL}), with new observational data arriving very fast nowadays due to the perihelion passage in April 2024 (see Table~\ref{tab:12P-starting-orbits} for details). However, despite these quick changes due to progress in observations and different data coverage, all orbit variants lead to a similar conclusion in terms of chaotic behaviour. This is a very stable comet, with the MEGNO obtained consistently from 8.39 to 8.77 with standard deviations from 2.32 to 3.06. This comet is in a highly inclined orbit, but slightly short of the value that prone to flipping polar orbits ($\sim 90\degr$) have. Due to the geometry of the problem (zones of planetary interaction, different variants and epochs of the studied elements), the more and less chaotic areas are arranged at different angles in the plots and in these cases, there are no very strong resonances near the nominal solution. Admittedly, a 1:6 J MMR (Jovian) resonance is located in the vicinity of 17.17\,au, but its relative strength of influence does not affect the current chaotic behaviour of the comet, nor is it visible here. This is because the area of this otherwise strong resonance is located outside the range of the current propagation of our VCs. However, it cannot be ruled out that in long-term evolution the comet may evolve to the area affected by this resonance. It is worth mentioning here that \cite{carussi-etal:1987}, starting from observational data available earlier, obtained that 12P is in a 1:6\,J resonance, but the latest observations and the orbits determined from them revise this conclusion. Nevertheless, the presence of this resonance on a longer time scale is visible, as shown in \ref{subsec:evo-12p}.
\vspace{0.1cm}

For the next comets we prepared MEGNO maps always in two variants: for the VC cloud based on the orbit determined by us, and, in addition, for the orbit available at JPL on July 2024 (plots presented in Appendix~\ref{sec:MEGNO-JPL}).


C/2002 K4 (NEAT), with MEGNO values from 13.0 to 13.5 is the second comet (excluding the 161P) in terms of stability (Fig.~\ref{fig:MEGNOc2002k4}). Lots of weak, overlapping resonances do not significantly affect its chaotic behaviour. However, only a few of the strongest with Saturn, Uranus and Neptune could be identified: S\,17:42, U\,8:7, \mbox{N 29:13}. These are formally retrograde resonances (RMMR).


C/2014 Q3 (Borisov) shows remarkably even results for the chaotic indicators (MEGNO: 17.3, see Fig. \ref{fig:MEGNOc2014q3} for more details). Remarkable are the numerous small, sometimes overlapping resonances. Near lying resonances have been identified: with Uranus (high order U 26:47), with Neptune (N 13:12) and potentially even with Saturn (S 7:36). This comet, in a previous simulation using 1001\,VCs, recorded numerous close approaches primarily with Saturn, and in smaller numbers with Uranus, Neptune and Jupiter. It should be noted that close approaches are a very different factor affecting chaos than MMR, as much depends on their strength and regularity. Both factors acting together obviously shape the picture of chaotic orbital behaviour.


For comet 122P/de Vico, MEGNO ratios were estimated for two orbit variants: the one determined in this work and the one available in the JPL database (Figs. \ref{fig:MEGNO122} and the middle panel of \ref{fig:MEGNO12JPL}). The orbit osculation epochs, which are also the starting epochs, differ slightly from each other, but in both cases, similar MEGNO values of 23.6$\pm$5.2 to 25.4$\pm$5.51 were obtained. Therefore, both orbit variants lead to similar quantitative conclusions about the chaoticity.
Again, there are no nearby significant resonances, apart from a weak MMR of 2:5 S, lying in the vicinity of 17.65\,au and thus slightly below the studied semimajor axis ranges. A potential reason for the relatively chaotic behaviour could be the regular approaches to Jupiter.


The orbit of the C/2015 GX (PanSTARRS) appears to be the most sensitive to changes in initial conditions, and the MEGNO indices are always in the range between 26.7 and 28.2, regardless of the orbit variant studied (details: Figs. \ref{fig:MEGNOc2015gx} and \ref{fig:MEGNOc2014q3JPL}). In summary, this comet is the most chaotic within the analysed HTCs.

The methodology described above appears to give coherent results for the comets studied. Most importantly, despite the different sets of orbital elements (with different errors and covariance parameters) and osculating epochs for the initial conditions, it was possible to reflect the chaos and MEGNO indicators in a consistent way. Combined with the determination of LT values, described in the previous section, it was possible to estimate the relative chaotic nature of the comets studied: from the most sensitive orbit of C/2015 GX, through 122P, C/2014 Q3, C/2002 K4 to the very stable, different variants of the 12P orbit. In the context of mean-motion resonances, a few are found in the direct vicinity of the nominal orbits (and, therefore, within the error range of the elements) and are visible in the chaos maps. In particular, this is evident for high-order resonances from the giant planets (Saturn, Uranus, Neptune) for comets C/2014 Q3 and C/2002 K4.

The exception in the correlation between MEGNO mean values and LT mean values is the sungrazing-sensitive comet 161P, initially in a very stable motion regime (see below).


\subsection{161P/Hartley-IRAS}\label{sub:MEGNO_161P}

Comet 161P/Hartley-IRAS appears to be the object with the smallest MEGNO values among the studied examples (Figs.~\ref{fig:MEGNO161} and the lowest panel in Fig.~\ref{fig:MEGNO12JPL}). The resulting average MEGNO values of 5.54$\pm$1.71 ($a$-$e$ plane) and 5.77$\pm$1.62 ($a$-$i$ plane) for our orbital solution confirm the analogical values of 5.66$\pm$1.36 and 6.03$\pm$1.21 calculated for the standard orbit from the JPL SBDB database. That means a consistent solution in terms of the value of the MEGNO indicator. For this particular comet, the values for MEGNO over a finite interval of time came out surprisingly low, which may be an indication that for a time on the order of the next thousands of years it temporarily resides in a stable dynamical regime, and then experiences more close approaches to planets (most to Jupiter) and transitions to other, potentially less stable areas, including sungrazing scenarios. Hence the different behaviour of the MEGNO and LT numerical estimators (described in the previous section), since convergence is achieved differently for them, even though they study the same kind of expansive close trajectories. In Sect.~\ref{sec:evol-LT}, we show that the evolution of our VC swarm of 161P in the GR regime also gives scattered statistics of $LT$ values that additionally are different after 10\,kyr than after 100 kyr. Anyway, the resulting chaos estimates report on the predictability of the orbit, but do not completely conclude on the dynamic future of this comet. Crucial in conclusion may be the fact that this comet will be in a sungrazing state with $q<0.005$\,au already on the scale of 13\,kyr with an extremely high probability (i.e. it will disintegrate; see Sect.~\ref{sec:evol-to-small-q}), as we have found when studying the evolution of its VCs (Sect.~\ref{subsec:evo-161p}).

\section {Summary and concluding remarks}\label{sec:conclusions}

The previous sections contain three complementary aspects of the future dynamical evolution of the six HTCs. Before we dealt with evolution, we focussed on a comprehensive investigation of the strategy for selecting the starting orbits for this research. Therefore, these studies also include the orbit determination from positional data and the discussion related to this issue. The main conclusions from these two-stage studies are presented below.

\begin{enumerate}[wide, labelwidth=!, labelindent=0pt, itemsep=1pt]
	
	\item We have shown that the NG parameters in the Marsden model (eq.~\ref{eq:g_r}) for 12P/Pons-Brooks based on 1882-1953, 1953-2024, and 2000-2024 are different, although small (Sect.~\ref{sec:orbit-deterination}). However, the small uncertainties of NG~parameters for different data-arcs suggest variability in these parameters between successive appearances. This indicates that it is best to rely on data from this comet most recent appearance (2020-2024) when studying its future dynamics. For the remaining comets, the NG~osculating orbits are more uncertain.
    Therefore, we carried out the dynamical evolution studies until 100\,ky forwards in the GR~regime; only for 12P we discuss the early results concerning the NG approach -- both in determining the nominal (starting) osculation orbit from positional data and in the dynamical evolution forwards in time (see last point).

	\item The evolution of HTC based only on the nominal orbit differs from the statistics based on the median distributions of a swarm of orbital elements. This may lead to qualitatively different conclusions about the direction of the dynamical evolution of individual objects (see Tables~\ref{tab:12P-stat_at-the-end}, \ref{tab:122P-stat_at-the-end}, and \ref{tab:2002k4-stat_at-the-end}). For example, for C/2014~Q3 after 100\,kyr, the median values of $q$, $a$, and $P$ are 1.87\,au, 33.0\,au, and 189\,yr, respectively, indicating that typically the swarm of orbits representing this comet motion evolves towards increasing of all three orbital elements. However, the nominal orbit reaches values of 1.68\,au, 25.2\,au, and 127\,yr showing a different trend (respective current values are: 1.65\,au, 28.4\,au, and 152\,yr). In turn, for C/2002~K4 after 100\,kyr we have medians $q$, $a$, and $P$ of 1.87\,au, 17.5\,au, and 73.2\,yr, respectively, indicating an evolution towards a clear decrease of $q$ (now we have $q=2.76$\,au) with almost unchanged $a$ and $P$ (now: 17.53\,au, 73.10\,yr), but the nominal orbit shows an even greater decrease to $q=1.13$\,au and some a decrease of $a$ and $P$ to values of 16.2\,au and 65\,yr.

	\item  In the evolution of HTCs with $q<1.3$\,au (comets: 12P/Pons-Brooks, 122P/de Vico, 161P/Hartley-IRAS, Figs.~\ref{fig:12p_q_i_ng}, \ref{fig:122p-b5-q-i-planets}, and \ref{fig:161p-b5-q-i-planets}) numerous close approaches to terrestrial planets occur frequently, whereas in the case of the three remaining comets with $q>1.6$\,au (C/2002~K4, C/2014~Q3, and C/2015~GX, right column in Fig.~\ref{fig:three_HTCs-evo}), close approaches to terrestrial planets below 0.1\,au are rare (only for an individual VC from a cloud of 1001\,VCs).

	\item  Our research confirms that HTCs can be an effective source of comets with perihelion close to the Sun ($q$<0.2\,au) and consequently -- sungrazing comets. We obtained high probabilities of reaching a sungrazing state causing at the most likely comet destruction for comets 161P/Hartley-IRAS and 122P/de Vico (Sect.~\ref{sub:evol-below-limit}), small non-zero probabilities for 12P/Pons-Brooks,  C/2002~K4 (NEAT), and C/2015~GX (PanSTARRS), and zero for C/2014~Q3 (Borisov). We found that 161P with the probability of 0.9 becomes sungrazer with $q$ below 0.005\,au after about 10\,kyr from now, so it will probably disintegrate. We calculated the same fate for 122P with the probability of 0.5 after 100\,kyr future evolution. Both comets with high decay probabilities in the next 100\,kyr currently have $q<1.3$\,au. This, combined with the fact that half of the sample studied consists of three comets with $q>1.6$\,au, and for all of them we found very low sungrazer state probabilities (from zero to 2.2\%), may suggest a much higher probability of destruction of HTCs with currently small $q$. 
	
	\item Comets 161P and 122P reach the sungrazing state mainly because they are in the Kozai resonance.  The occurrence of this resonance for these objects was previously noted by \citet{bai-cha-hah:1992}. However, it seems that these objects are captured in this state temporarily, and if they aren't destructed earlier due to low values of $q$, they may leave it. Our analysis shows that, in the case of both comets, Jupiter is not the primary factor responsible for placing these objects in the Kozai resonance. In the case of the remaining four comets, we have not been able to find clear evidence of the Kozai resonance in their dynamic evolutions.
	
	 \item During the dynamical evolution in the GR regime, we did not observe any significant changes (that is causing exit from the HTCs population) in the Tisserand parameter value, $T_J$, even in the case of the evolution to the sungrazing state. For example, for 161P, $T_J$ was always close to 0.6 during the 15\,kyr of future evolution while the probability of destruction due to the sungrazing state rose to 90\%. However, the $T_J$ scatters at the end of the evolution within the VC swarms for each HTC were clear, for example, for 12P and 122P at the end of {100\,kyr} of the future evolution we obtain $T_J$ between 0 and 1, while for C/2002~K4 and C/2014~Q3 -- in the narrower range between 0 and 0.5 (compare with starting $T_J$ values shown in Table~\ref{tab:objects}). 

\item The flipping orbit phenomenon occurs frequently but with different intensities for each of the six studied comets. The flipping scenario was described in Sect.~\ref{sec:evol-flipping} using conditional probability formalism. Since the factor that potentially eliminates VCs from the Solar System is the occurrence of sungrazing, it was primarily considered in this analysis. The most statistically significant examples of flipping occur for comets 122P and 161P. The flipping in our examples takes place, of course, in both directions, from a prograde to a retrograde orbit or vice versa. The conditional probabilities of flipping are sometimes at the level of a few tens of percent, so they are plausible scenarios. Comets in near-polar orbits give many flipping detections during the integration, but these are often minor changes in inclination, as their values remain within a narrow range (C/2014 Q3 and C/2015 GX). C/2002 K4, on the other hand, has a more probable chance of transit into a prograde orbit.

	\item  Almost all VCs of the three studied comets with $q>1.6$\,au (C/2014~Q3, C/2015~GX, and C/2002~K4) experience the flipping orbit phenomenon during their future evolution until 100\,kyr, and their inclinations remain large to the ecliptic plane except for only a few VCs where the sungrazing phenomenon occurs. One can see that for 80\% of VCs (after cutting off 10\% of the wings of the distributions, see Table~\ref{tab:2002k4-stat_at-the-end}) the inclinations are within the range 84\degr ~-- 101\degr ~at the end of studied evolution for these three comets.
	
	\item The ejection of VCs into hyperbolic orbits seems to be a rare event in the future evolution of studied HTCs. This event has never occurred in the evolution of 1001\,VCs of C/2014~Q3, whereas for the remaining comets only in single cases, it was always caused by a close approach to Jupiter. The highest probability of ejection from the Solar System in the next 100\,kyr was obtained for 122P -- but still, it was only 0.5\%.

	\item  We present an analysis of the stability of the GR orbits of the studied comets using the Lyapunov time calculated individually for each VC. Based on this the $LT$ distribution is described in detail for each comet, considering mean, median, standard deviation, and skewness metrics like Bowley skewness. The results show that all examined comet orbits exhibit chaotic characteristics, with $\langle LT\rangle$ values ranging from about 100 to several hundred years. Excluding the special case of 161P and based on $\langle LT\rangle$ values after 10$^4$\,yr  the comets can be ranked in stability from the least stable (C/2015 GX) to the most stable (12P; see Table~\ref{tab:HTCs-LT}).  The $LT$ distribution of each comet, skewed to the right, varies in shape, with most showing long tails indicating rare cases of high $LT$ values for individual VCs.

	\item Determining the MEGNO parameter and preparing its maps allow us to study the vicinity of the nominal orbits in terms of quantitative chaos. Since the MEGNOs are fast converging indicators, the conclusions are about the current dynamic state considering a small interval of 10\,kyr. We, therefore, obtained a short-term forecast of chaos. The rank we obtained using the MEGNO indicator entirely agrees with the Lyapunov time ranking of chaotic behaviour for five comets (previous point and Sect.~ \ref{sec:evol-LT}--\ref{sec:evol-MEGNO}). Thus, we conclude that both indicators ($LT$ and MEGNO) give a consistent ranking of HTCs in determining their chaotic nature, excluding the special case of 161P because of its high probability of reaching the sungrazing state with $q$ below 0.005\,au, the $q$-limit assumed as a limit for comet disintegration. As a secondary result, it was possible to investigate the sensitivity of input data affected by observational errors to the dependence on weak MMRs currently occurring near the nominal orbits. At present, these resonances do not significantly change the dispersion of the chaotic indicators in the short-term orbital evolution of 10\,kyr.
	
	\item Other MMRs, including more strongly interacting lower order resonances, can appear on longer time scales, as shown for selected VCs in Sect.~\ref{sec:evolution-classical}. MMRs are occasionally confirmed later by short-period MEGNO estimates at the beginning of each simulation. On longer time scales of the order of 100\,kyr, we have frequent transitions of individual VCs into areas of strong resonances with Jupiter and Saturn, as detailed in Sect.~\ref{sec:evolution-classical}. The result can, therefore, be inferred about the short-period (present) and potential long-period (future) impact of the various resonances.

	\item We present preliminary results of NG evolution for 12P (Sect.~\ref{sub:12P-NG-evo}), because only for this comet can NG effects be reasonably determined assuming constant parameters A$_1$ and A$_2$, and A$_3=0$. However, we show that different sets of NG parameters are obtained depending on the analysed data arc (point 1.). This is also why the evolution based on the constant NG parameters on a much longer time scale should be treated as very coarse.  A certain statistical relaxation of this strong assumption about constant NG parameters is to base the NG evolution results on a swarm of 250~VCs as is done here, where for each VC we have a unique set of eight parameters ($A_1$, $A_2$, and six orbital elements) fitted to positional data.In the NG regime, we found that about 10\% of VCs go through an evolutionary path from HTCs through JFCs to ETCs, and many of them end their lives in the Sun. Obtained ETC inclinations are in a wide range from a few to about 70\degr. However, the orbital inclinations of the actual ETC population (about 50 comets) are not greater than 30\degr. On the other hand, numerical simulations have not yet achieved a satisfactory production rate of ETC orbits \citep{lev-dun:1997, har-bai:1998, Valsecchi:1999, levison-etal:2006}.
    Therefore, if some of them evolve from the HTC population, the problem of deficiency will be at least partially alleviated. However, the results obtained here should be considered today only as a guide for further research in this direction. In the future, we plan to build an improved recipe for NG effects in the dynamic evolution study of HTCs (see also Sect.~\ref{sub:12P-NG-evo}).
\end{enumerate}

\section*{Data availability}

The starting data underlying all results discussed in this article are available in Appendices~A and B. The other data underlying this article will be shared on reasonable request to the corresponding author.

\section*{Acknowledgements}

This research has made use of positional data of comets provided by the International Astronomical Union's Minor Planet Center.

During the preparation of this work, the resources of the Centre of Computing and Computer Modelling of the Faculty of Natural Sciences of Jan Kochanowski University in Kielce were used.

We thank Andrzej M. Sołtan and  Piotr A. Dybczy\'nski for discussions and anonymous referee for a helpful review.





\bibliographystyle{mnras}
\bibliography{HTCs_literatura}

{\onecolumn{

\appendix

\section{Osculating orbits of 12P/Pons-Brooks}
	\begin{table*}
{\footnotesize{		
	\caption{\label{tab:12P-starting-orbits} Orbital elements of osculating heliocentric orbits for 12P/Pons-Brooks obtained in this study; for comparison NG orbit listed in JPL in 2024 July 8 was also presented.
		The successive columns signify: $[1]$ -- orbit description where GR means gravitational orbit, NG -- non-gravitational, and next the range of data arc used for orbit determination is coded, $[2]$ -- Epoch, i.e. osculation date, 	$[3]$ -- perihelion time [TT], $[4]$ -- perihelion distance, $[5]$ -- eccentricity, $[6]$ -- argument of perihelion (in degrees), 	equinox 2000.0, $[7]$ -- longitude of the ascending node (in degrees), equinox 2000.0, $[8]$ -- inclination (in degrees); equinox 2000.0, $[9]$--$[10]$ -- radial and transverse NG~parameters, the last column gives $rms$. Last observations used: 2024~July~05 (here and for JPL's orbit retrieved from JPL Database on July 8). }
	\centering
	\setlength{\tabcolsep}{1.0pt} 
	\begin{tabular}{ccrrrrrrrrrc}
	\hline\hline
				Orbit       & Epoch         & T                   & q            &  e         & $\omega$      & $\Omega$    &  i           & A$_1$      & A$_2$     & A$_3$     & $rms$        \\
		code                & \hspace{-2mm}$[$\footnotesize{yyyymmdd}$]$&$[$\footnotesize{yyyymmdd.dddddd}$]$& $[$au$]$     &            &  $[$\degr$]$  & $[$\degr$]$ & $[$\degr$]$  & \multicolumn{3}{c}{~[$10^{-8}$\,au\,day$^{-2}$]} & [arcsec]   \\
				$[1]$       & $[2]$       & $[3]$                 & $[4]$        & $[5]$      & $[6]$         & $[7]$       & $[8]$        &  $[9]$     &   $[10]$  & $[11]$ & $[12]$      \\
				\hline\hline 
				\\
                GR/ & 18840125 & 18840126.216617 &      0.77573046 &      0.95500706 &     199.175034 &     255.774760 &      74.040554 & -- & -- & -- &        2.84 \\
		1883-1884   &          &   $\pm$0.000198 & $\pm$0.00000522 & $\pm$0.00000983 &  $\pm$0.000684 &  $\pm$0.000272 &  $\pm$0.000151 &    &    &         \\
\\		
				GR/ & 19540518 & 19540522.881006 &      0.77366848 &      0.95482888 &     199.027650 &     255.891502 &      74.176810 & -- & -- & -- &        1.40 \\
		 1953-1954  &          &   $\pm$0.000448 & $\pm$0.00000183 & $\pm$0.00000661 &  $\pm$0.000211 &  $\pm$0.000274 &  $\pm$0.000069 &    &    &    &         \\
\\ 
				GR/ & 18831001 & 18840126.217244 &      0.77574026 &      0.95501138 &     199.175816 &     255.774553 &      74.040825 & -- & -- & -- &        2.89 \\
         1983-1954  &          &   $\pm$0.000080 & $\pm$0.00000094 & $\pm$0.00000005 &  $\pm$0.000128 &  $\pm$0.000160 &  $\pm$0.000086 &    &    &    &         \\
                NG/ & 18831001 & 18840126.217291 &      0.77573864 &      0.95499943 &     199.175826 &     255.774837 &      74.040537 &$-$0.0036& $+$0.0641 &$-$0.1956 &  2.84 \\
         1983-1954  &          &   $\pm$0.000127 & $\pm$0.00000277 & $\pm$0.00000589 &  $\pm$0.000398 &  $\pm$0.000193 &  $\pm$0.000139 & 0.0931  & 0.0316    & 0.0818   &       \\ 
\\			
               GR/  & 20230918 & 20240421.131765 &      0.78087083 &      0.95456024 &     198.987736 &     255.855116 &      74.190858 & --      & --        & -- &        0.62 \\
         1953-2024  &          &   $\pm$0.000009 & $\pm$0.00000010 & $\pm$0.00000001 &  $\pm$0.000006 &  $\pm$0.000010 &  $\pm$0.000003 &         &           &    &             \\
               NG/  & 20230918 & 20240421.131164 &      0.78085812 &      0.95456097 &     198.988247 &     255.855193 &      74.190876 & 0.4613  & $+$0.1050 & -- &        0.55 \\
         1953-2024  &          &   $\pm$0.000012 & $\pm$0.00000021 & $\pm$0.00000001 &  $\pm$0.000009 &  $\pm$0.000009 &  $\pm$0.000003 & 0.0067  & 0.0028    &    &             \\
        JPL/ NG/    & 20230918 & 20240421.131032 &      0.78085738 &      0.95456101 &     198.988224 &     255.855211 &      74.190878 & 0.5561  & $+$0.1196 & -- &        0.60 \\
        1953-2024   &          &   $\pm$0.000026 & $\pm$0.00000044 & $\pm$0.00000003 &  $\pm$0.000019 &  $\pm$0.000017 &  $\pm$0.000005 & 0.0207  & 0.0122    &    &             \\
\\			
	\bf{			GR/} & 20220121 & 20240421.295636 &      0.78170914 &      0.95443306 &     198.988926 &     255.793591 &      74.151291 & --      & --         & -- &        0.54 \\
	\bf{	 2020-2024}  &          &   $\pm$0.000014 & $\pm$0.00000024 & $\pm$0.00000030 &  $\pm$0.000019 &  $\pm$0.000010 &  $\pm$0.000003 &         &            &    &             \\
	\bf{				NG/} & 20220121 & 20240421.295103 &      0.78170667 &      0.95443581 &     198.988882 &     255.793626 &      74.151310 & 0.7635  & $-$0.1827  & -- &        0.51 \\
	\bf{         2020-2024}  &          &   $\pm$0.000023 & $\pm$0.00000027 & $\pm$0.00000031 &  $\pm$0.000020 &  $\pm$0.000011 &  $\pm$0.000003 & 0.0172  & 0.0143     &    &             \\
				\hline
	\end{tabular}   }}
\end{table*}

\newpage
\section{Osculating orbits of five other HTCs}
\begin{table*}
	{\footnotesize{		
			\caption{\label{tab:5HTCs-starting-orbits} Orbital elements of osculating heliocentric orbits for five HTCs obtained in this study. Column description as in Table~\ref{tab:12P-starting-orbits}  }
			\centering
			\setlength{\tabcolsep}{1.0pt} 
			\begin{tabular}{ccrrrrrrrrrc}
				\hline\hline
				Orbit       & Epoch         & T                   & q            &  e         & $\omega$      & $\Omega$    &  i           & A$_1$      & A$_2$     & A$_3$     & RMS        \\
				code                & \hspace{-2mm}$[$\footnotesize{yyyymmdd}$]$&$[$\footnotesize{yyyymmdd.dddddd}$]$& $[$au$]$     &            &  $[$\degr$]$  & $[$\degr$]$ & $[$\degr$]$  & \multicolumn{3}{c}{~[$10^{-8}$\,au\,day$^{-2}$]} & [arcsec]   \\
				$[1]$       & $[2]$       & $[3]$                 & $[4]$        & $[5]$      & $[6]$         & $[7]$       & $[8]$        &  $[9]$     &   $[10]$  & $[11]$ & $[12]$      \\
				\hline\hline 
				\\
\multicolumn{12}{c}{122P/de Vico}		\\		%
       \bf{     GR/}& 19951010 & 19951006.023760 &      0.65889213 &      0.96274983 &      12.976731 &      79.618077 &      85.390827 & -- & -- & -- &        1.03 \\
\bf{    1995-1996}  &          &   $\pm$0.000047 & $\pm$0.00000032 & $\pm$0.00000498 &  $\pm$0.000111 &  $\pm$0.000113 &  $\pm$0.000073 &    &    &    &    \\
                NG/ &          & 19951006.022121 &      0.65887400 &      0.96250787 &      12.972241 &      79.613134 &      85.389296 & 1.690   & 2.202   & 2.771 &        0.98 \\
        1995-1996   &          &   $\pm$0.000432 & $\pm$0.00000578 & $\pm$0.00008072 &  $\pm$0.001253 &  $\pm$0.001532 &  $\pm$0.000297 & 0.710   & 0.654   & 0.668 &             \\
			\\
\multicolumn{12}{c}{161P/Hartley-IRAS}		\\		%
      \bf{      GR/} & 19831102 & 19840108.704473 &      1.28243938 &      0.83383479 &      47.110399 &       1.498723 &      95.731778 & -- & -- & -- &        0.69 \\
\bf{     1983-2005}  &          &   $\pm$0.000118 & $\pm$0.00000038 & $\pm$0.00000005 &  $\pm$0.000044 &  $\pm$0.000033 &  $\pm$0.000015 &    &    &    &    \\
        \\
\multicolumn{12}{c}{C/2014 Q3 (Brorsen)}		\\		%
          \bf{~GR/} & 20141209 & 20141119.069142 &      1.64741063 &      0.94208038 &      47.382040 &      63.129588 &      89.948867 & -- & -- & -- &        0.41 \\
\bf{2014-2015}      &          &   $\pm$0.000125 & $\pm$0.00000128 & $\pm$0.00000774 &  $\pm$0.000053 &  $\pm$0.000045 &  $\pm$0.000035 &    &    &         \\
                NG/ &          & 20141119.068401 &      1.64740990 &      0.94209893 &      47.381514 &      63.129592 &      89.948889 & 0.1538  & 1.248 & 0.4301 &        0.41 \\
   2014-2015        &          &   $\pm$0.001014 & $\pm$0.00000850 & $\pm$0.00005135 &  $\pm$0.000445 &  $\pm$0.000412 &  $\pm$0.000320 & 0.1870  & 0.269 & 0.4925 &            \\
\\				
\multicolumn{12}{c}{C/2015 GX (PanSTARRS)}		\\		%
          \bf{~GR/} & 20150915 & 20150826.661800 &      1.97175223 &      0.87819263 &     108.957857 &     235.515513 &      90.254526 & -- & -- & -- &        0.32 \\
\bf{2015-2016     } &          &   $\pm$0.000084 & $\pm$0.00000048 & $\pm$0.00000099 &  $\pm$0.000037 &  $\pm$0.000007 &  $\pm$0.000007 &    &    &    &    \\
\\			
\multicolumn{12}{c}{C/2002 K4 (NEAT)}		\\		%
        \bf{  ~GR/} & 20020725 & 20020712.971654 &      2.76451120 &      0.84226123 &      24.431151 &     308.098939 &      94.062551 & -- & -- & -- &        0.64 \\
\bf{2002 }   &          &   $\pm$0.001442 & $\pm$0.00001079 & $\pm$0.00003119 &  $\pm$0.000405 &  $\pm$0.000169 &  $\pm$0.000295 &    &    &    &    \\
\hline
\end{tabular}   }}
\end{table*}

\newpage
\section{Additional plots for future evolution of 161P/Hartley-IRAS}\label{sec:161P-10VCs-evo}

\begin{figure*}
	\centering
	\includegraphics[width=8.6cm]{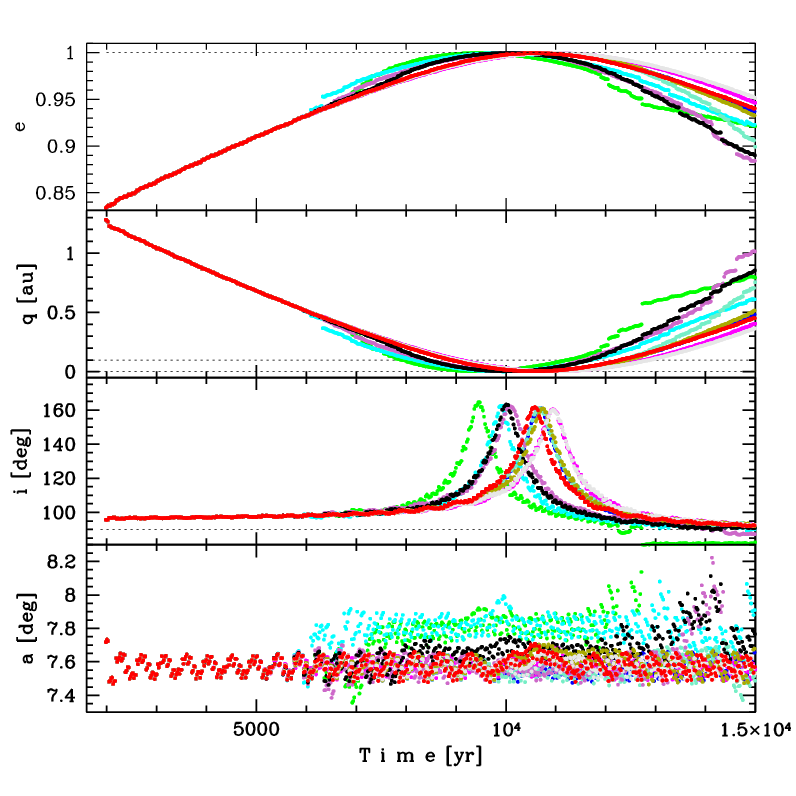}
	\includegraphics[width=8.6cm]{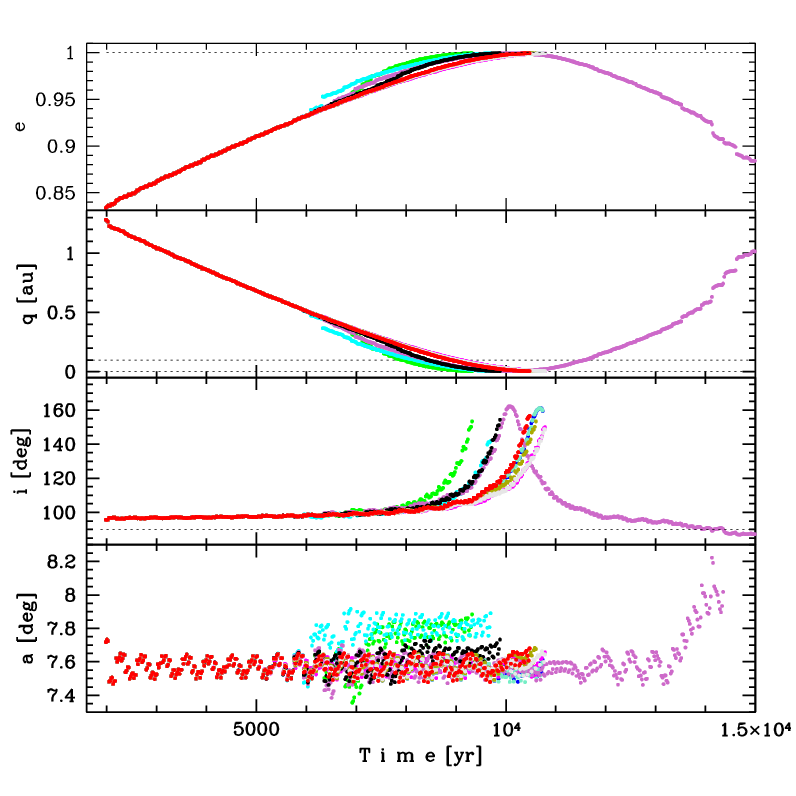}
\vspace{-0.5cm}
	\caption{Evolution of first 10 of the swarm of 1001\,VCs. Nominal orbit evolution is shown using red colour. The right panel shows the same evolution where the evolution was stopped (frozen) when $q$ decreases below 0.005\,au.}\label{fig:161P-10VCs-evo}
\end{figure*}
Fig.~\ref{fig:161P-10VCs-evo} shows the common phenomenon of reaching the sungrazing state in the future evolution of comet 161P during the 13\,kyr for the first 10 of 1001\,VCs. The characteristic signature of the growth of $e$ and $i$ with a simultaneous decrease of $q$ is visible for all VCs. In nine cases, we can see the reduction of $q$ was below 0.005\,au, which we have assumed to result in the comet destruction under the influence of a close approach to the Sun (see Sec,~\ref{sec:evol-to-small-q}). The decrease from the $q$ value of about 0.2\,au to minimal value (usually less than 0.005\,au) occurs typically after about 150 revolutions around the Sun. The evolution of two VCs (shown by green and cyan colours) differs from the others due to their deeper multiple approaches to Jupiter. Then, there would be an evolution to a state close to that before the sungrazing state -- if the comet had survived multiple close approaches grazing the Sun. With our assumptions, only one of these 10 VCs shown in Fig.~\ref{fig:161P-10VCs-evo} would have survived (see right panel of the figure).

\newpage
\section{MEGNO maps for orbits taken from JPL database}\label{sec:MEGNO-JPL}
\begin{figure*} 
	\centering
	\includegraphics[width=8.5cm]{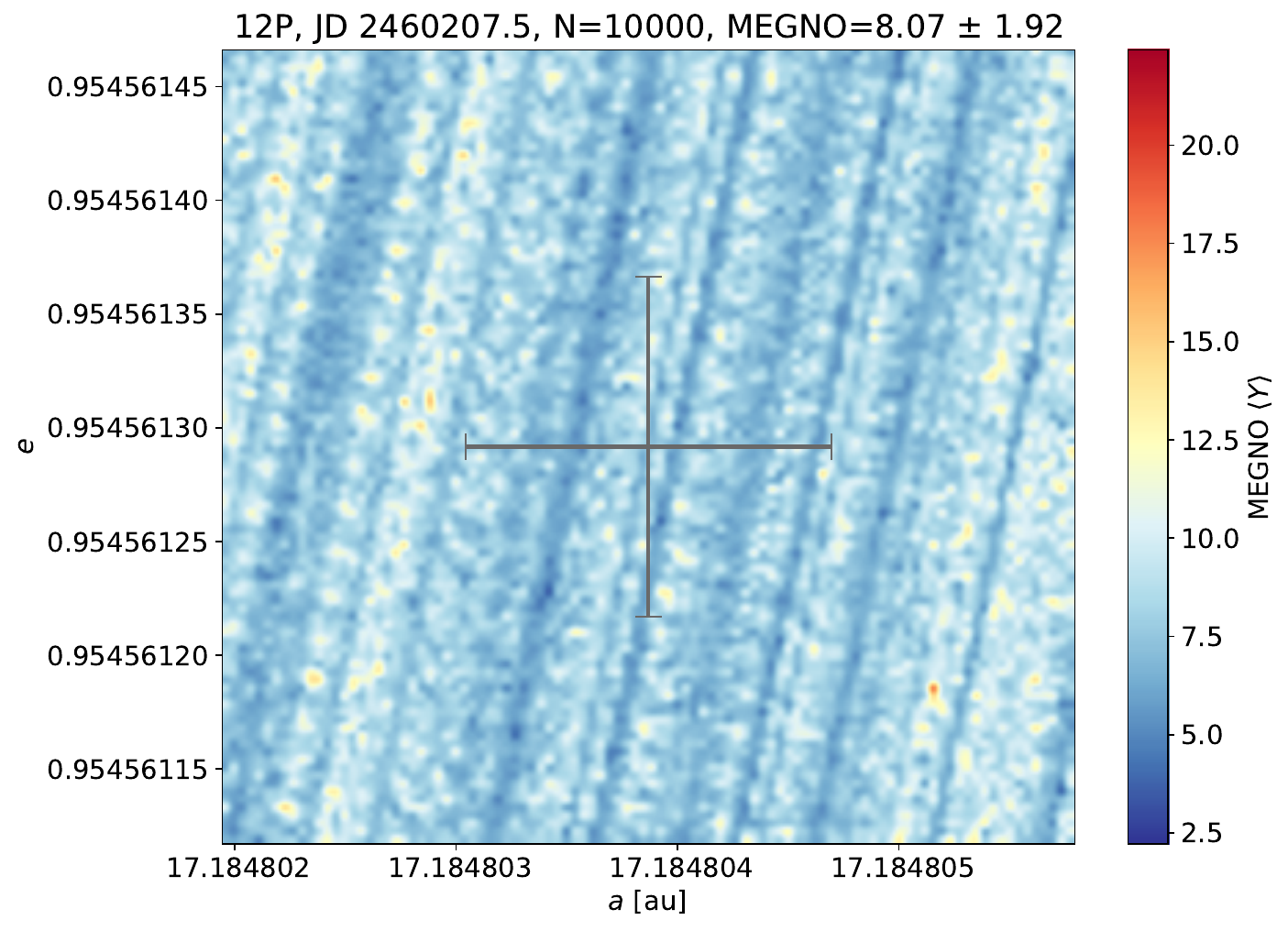}
	\includegraphics[width=8.5cm]{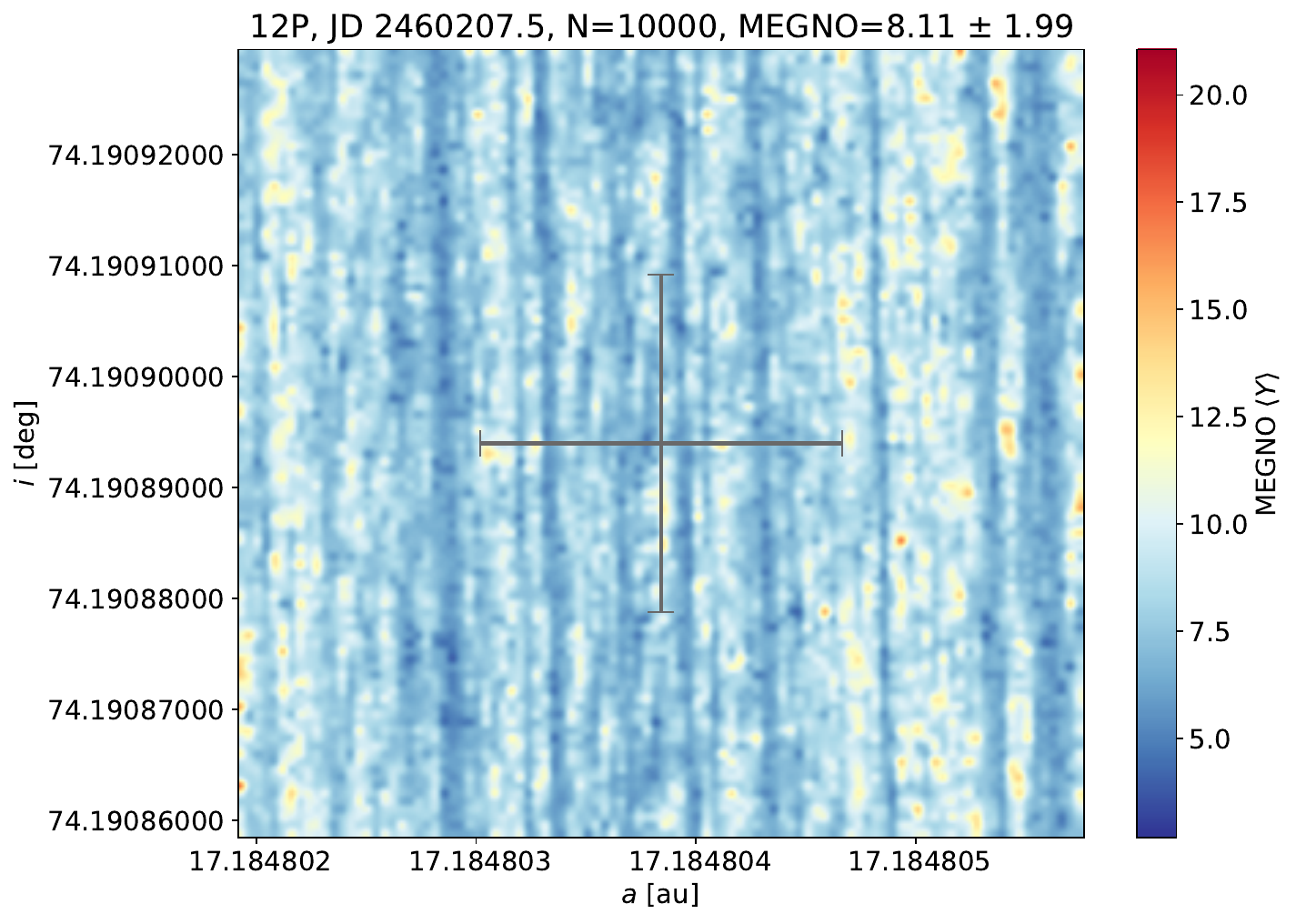}
	\includegraphics[width=8.5cm]{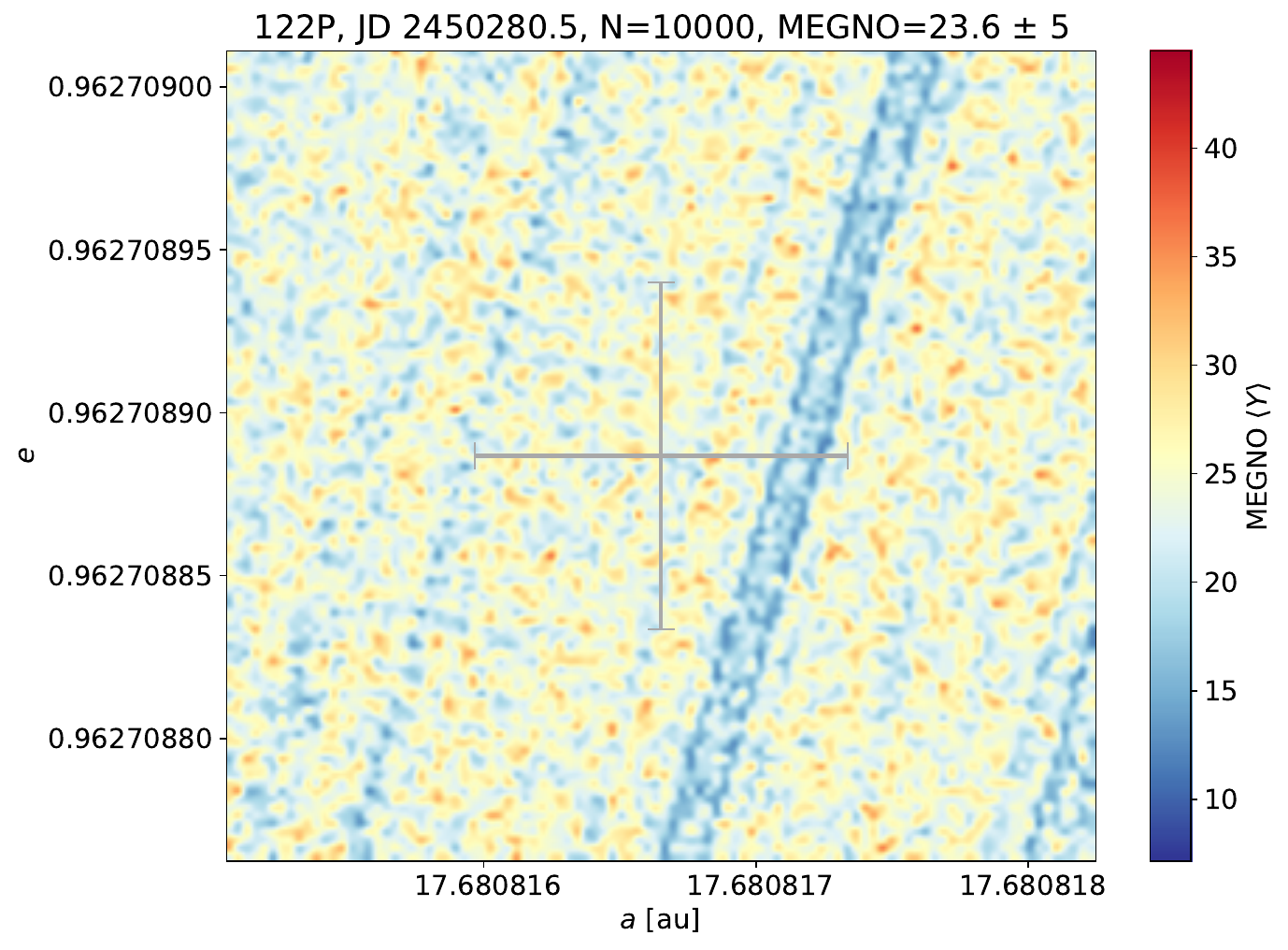}
	\includegraphics[width=8.5cm]{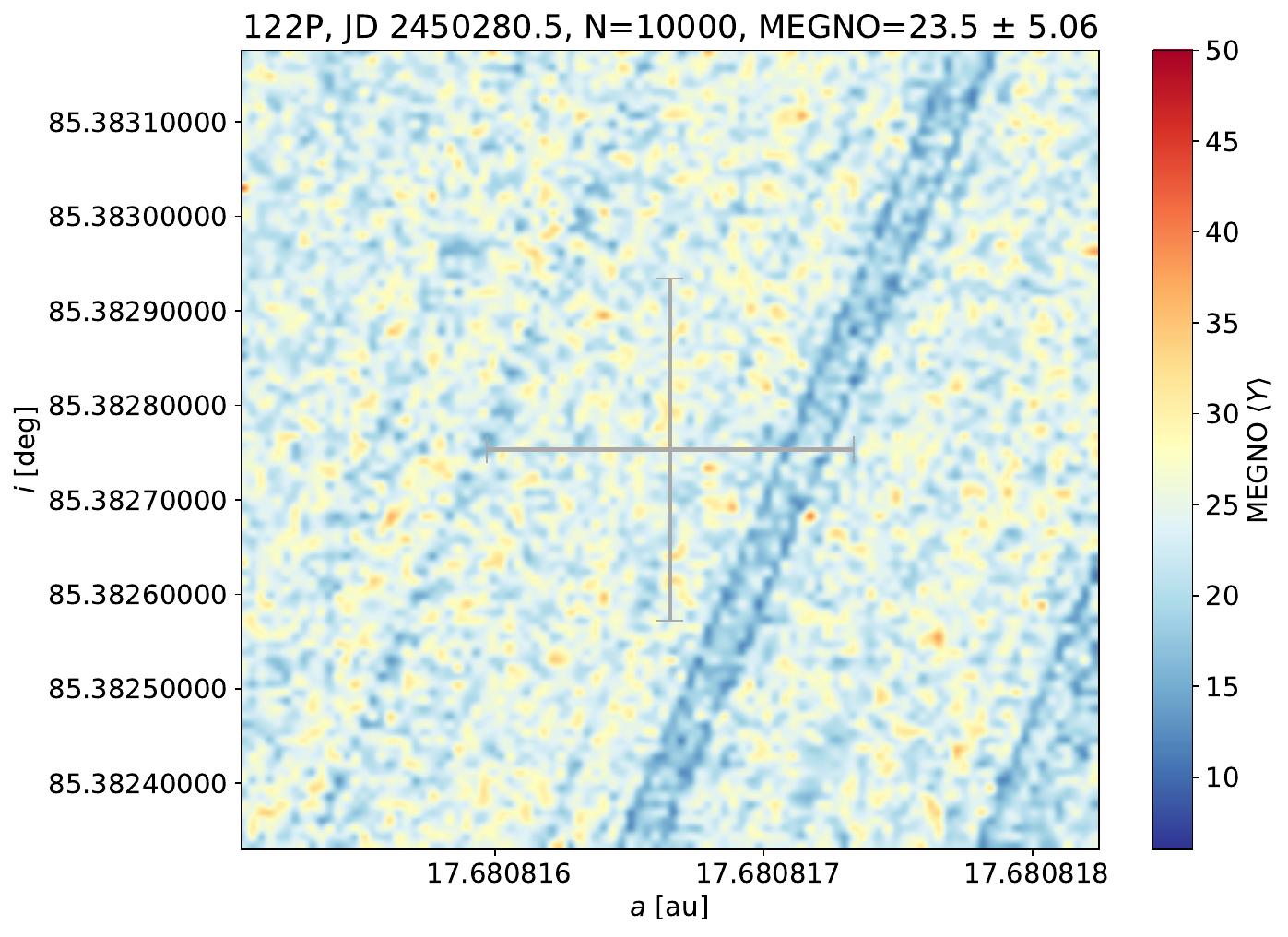}
	\includegraphics[width=8.5cm]{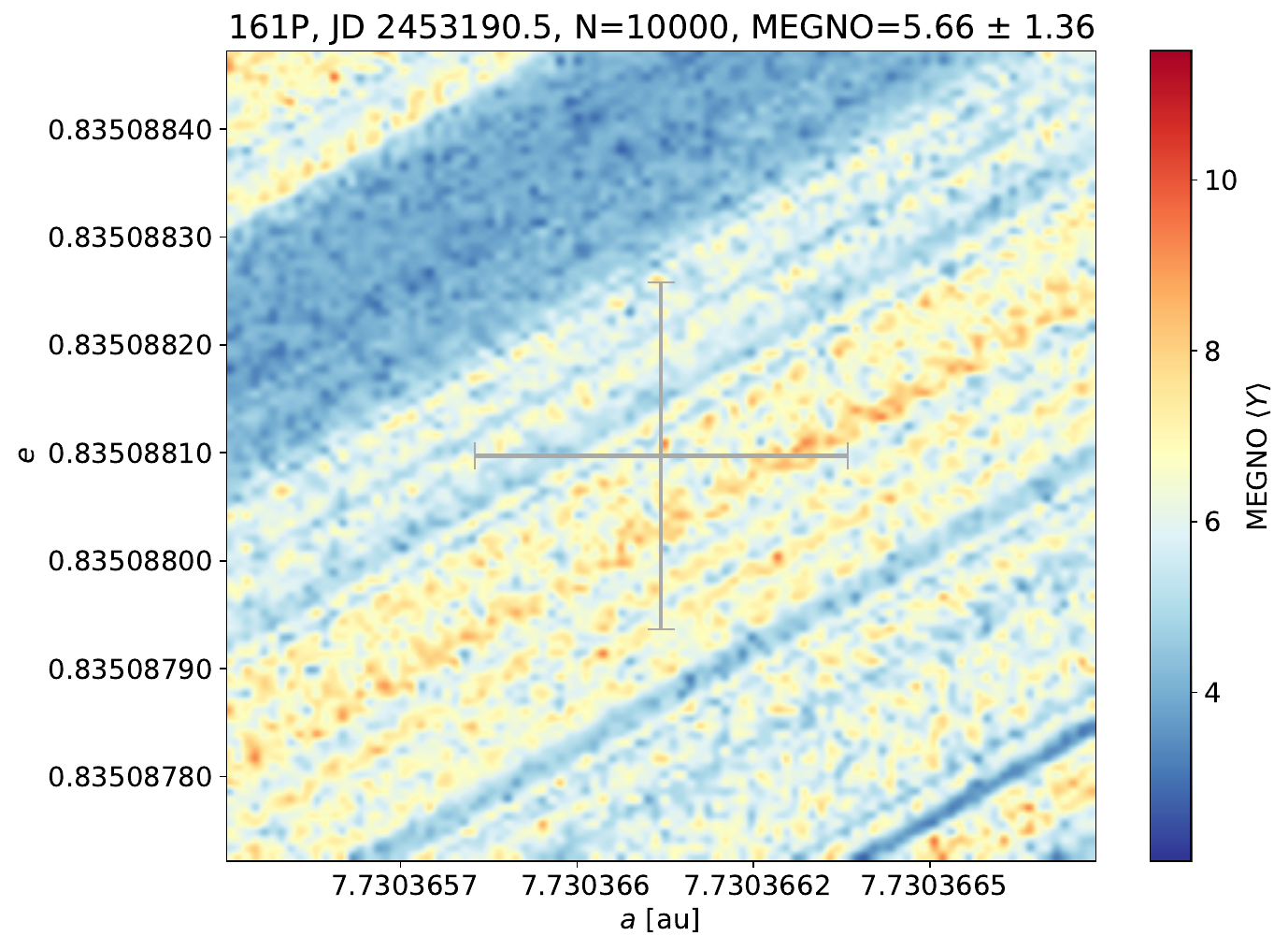}
	\includegraphics[width=8.5cm]{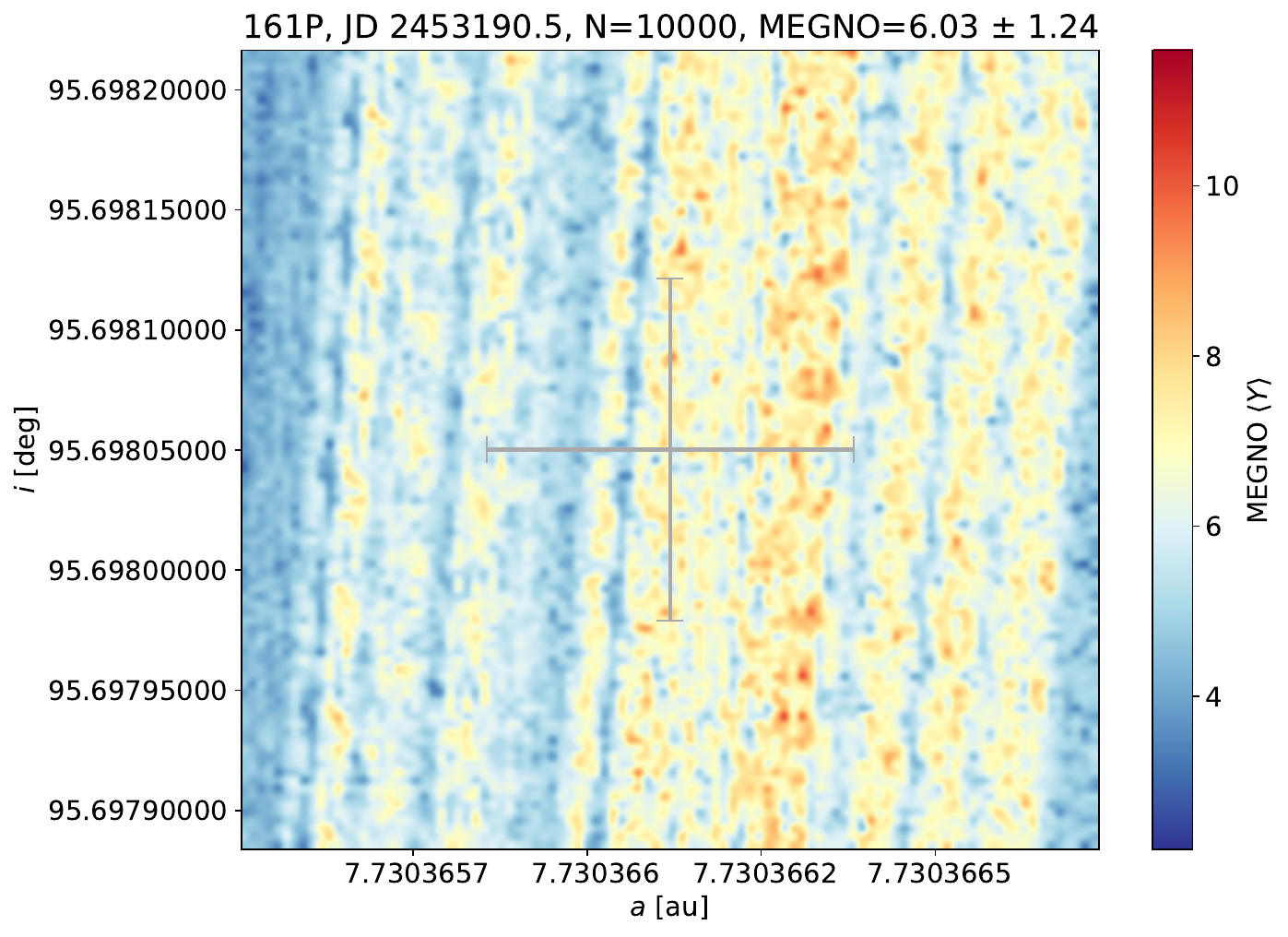}
	\caption{MEGNO maps of comets 12P (upper panels), 122P (middle panels), and 161P (lowest panels) based on data from JPL database. Left-side panels: $a-e$ plane. Right-side panels: $a-i$.}	\label{fig:MEGNO12JPL}
\end{figure*}

\begin{figure*} 
	\centering
	\includegraphics[width=8.7cm]{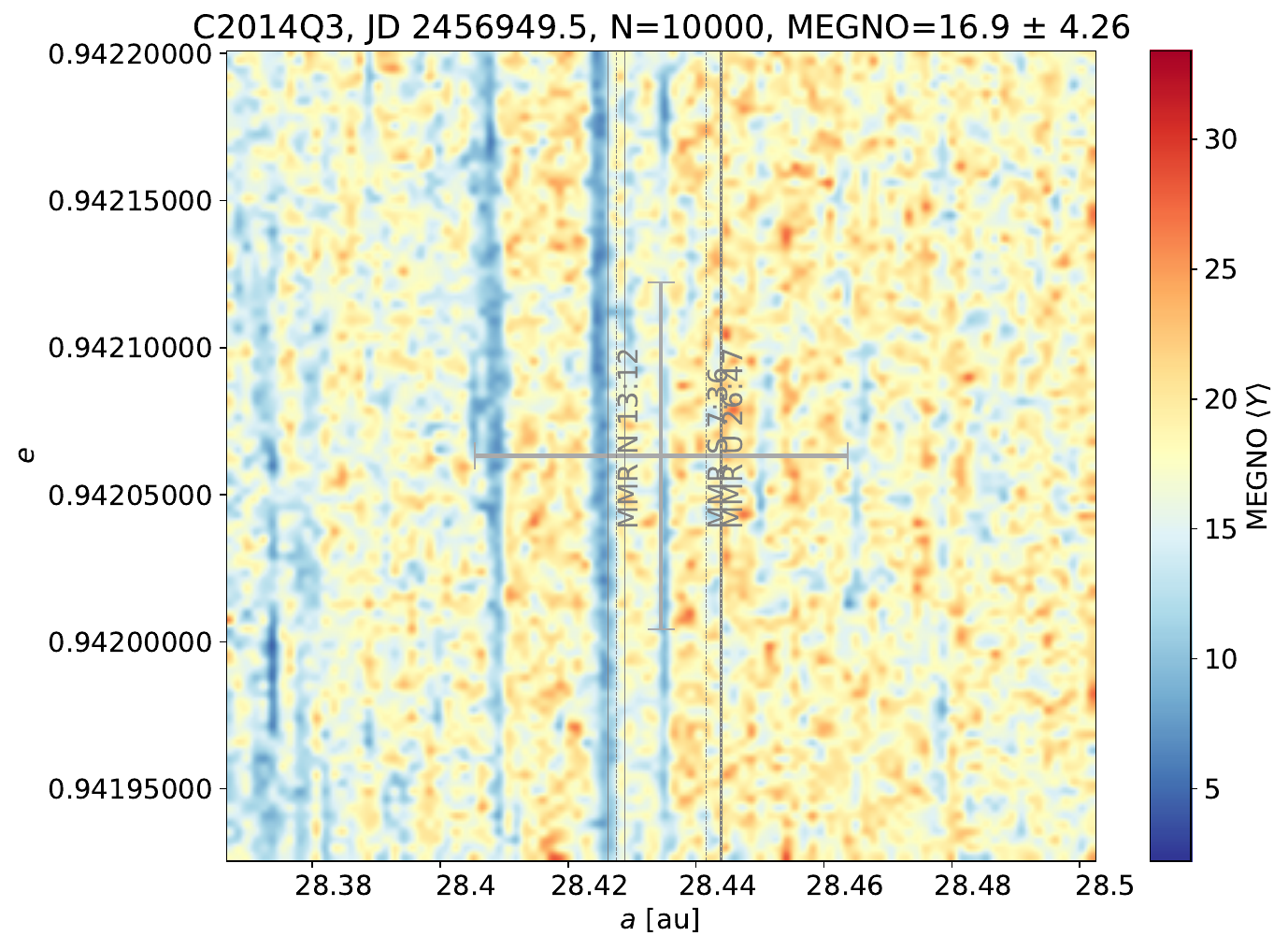}
	\includegraphics[width=8.7cm]{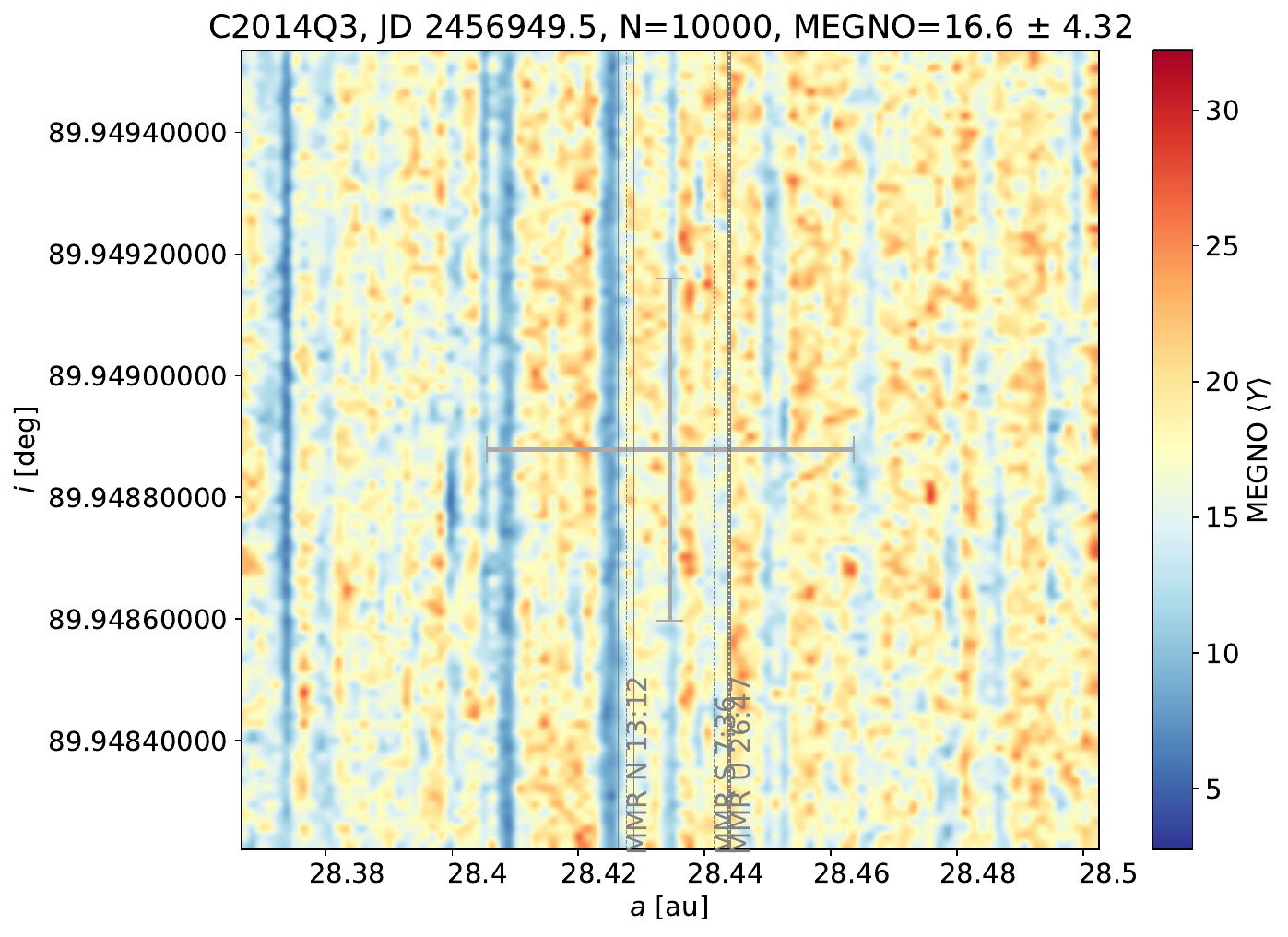}
	\includegraphics[width=8.7cm]{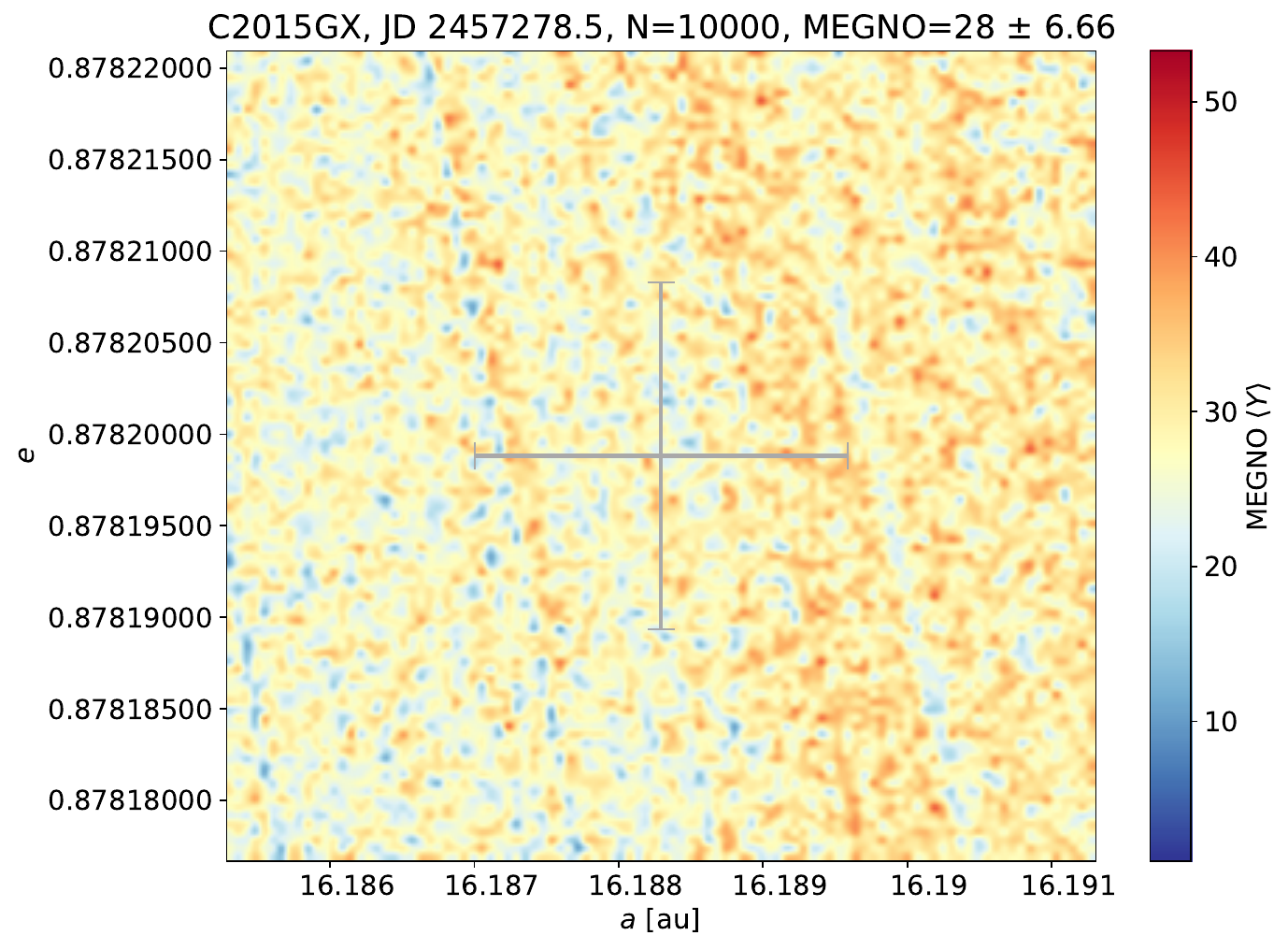}
	\includegraphics[width=8.7cm]{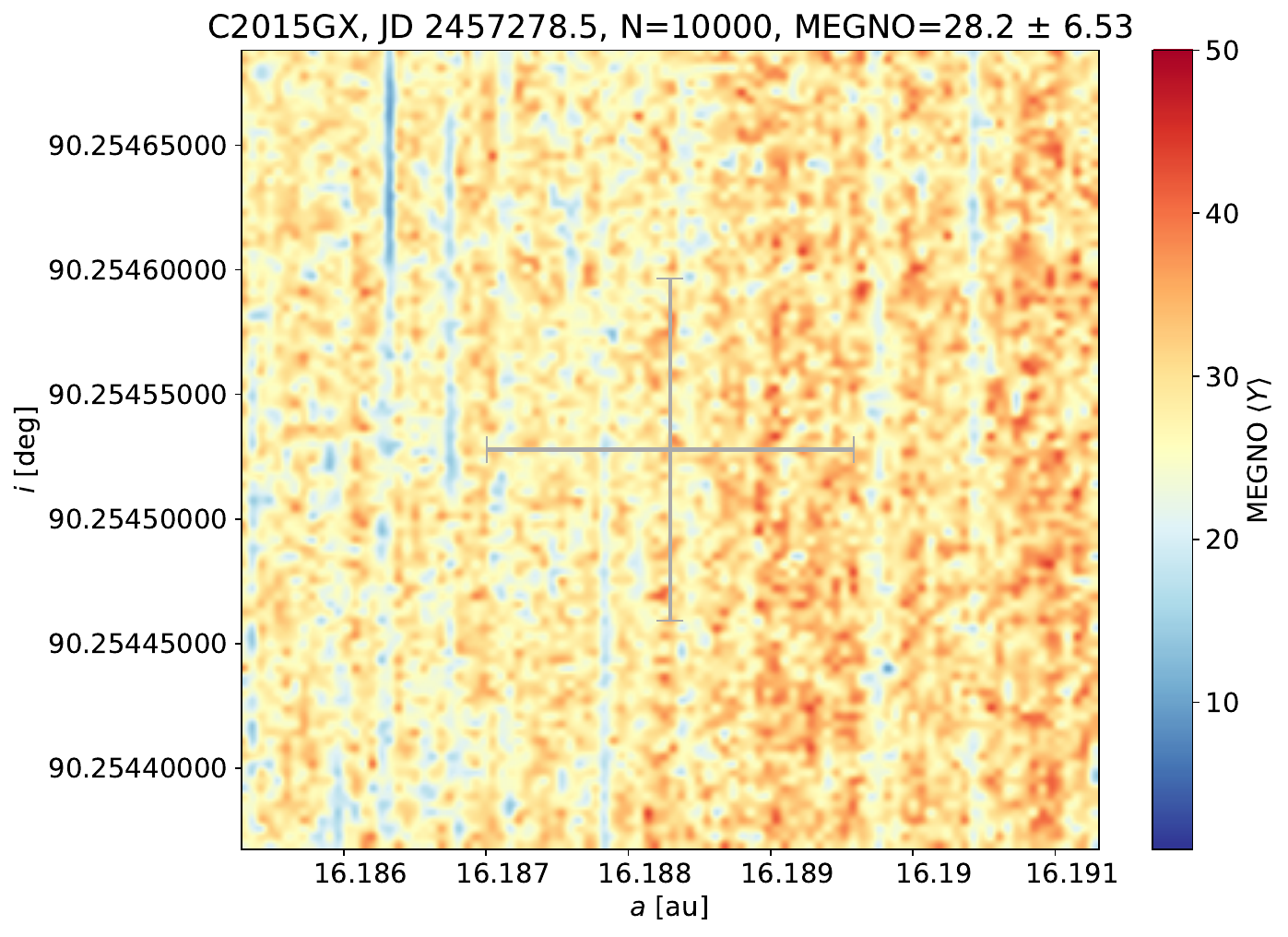}
	\includegraphics[width=8.7cm]{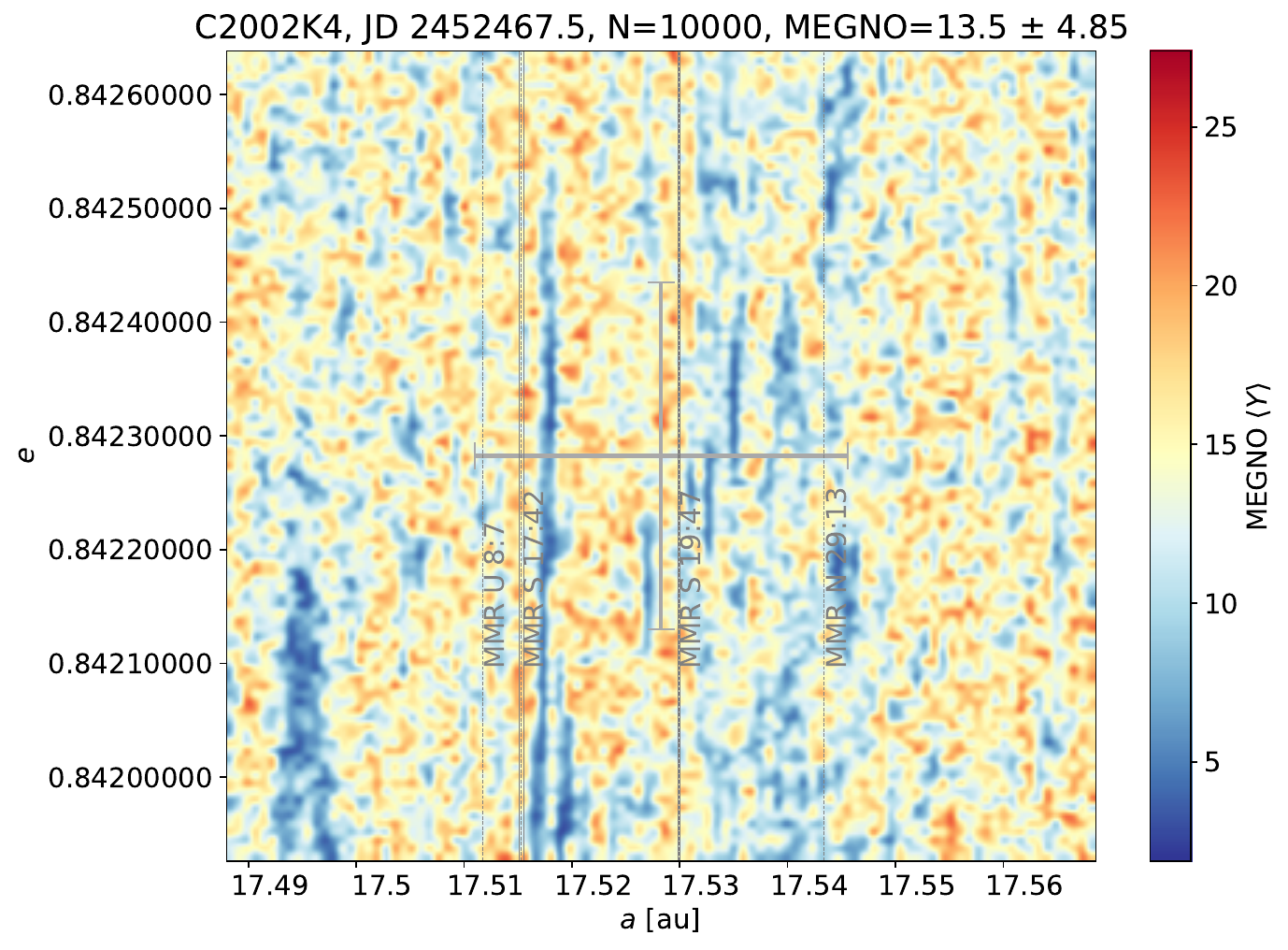}
	\includegraphics[width=8.7cm]{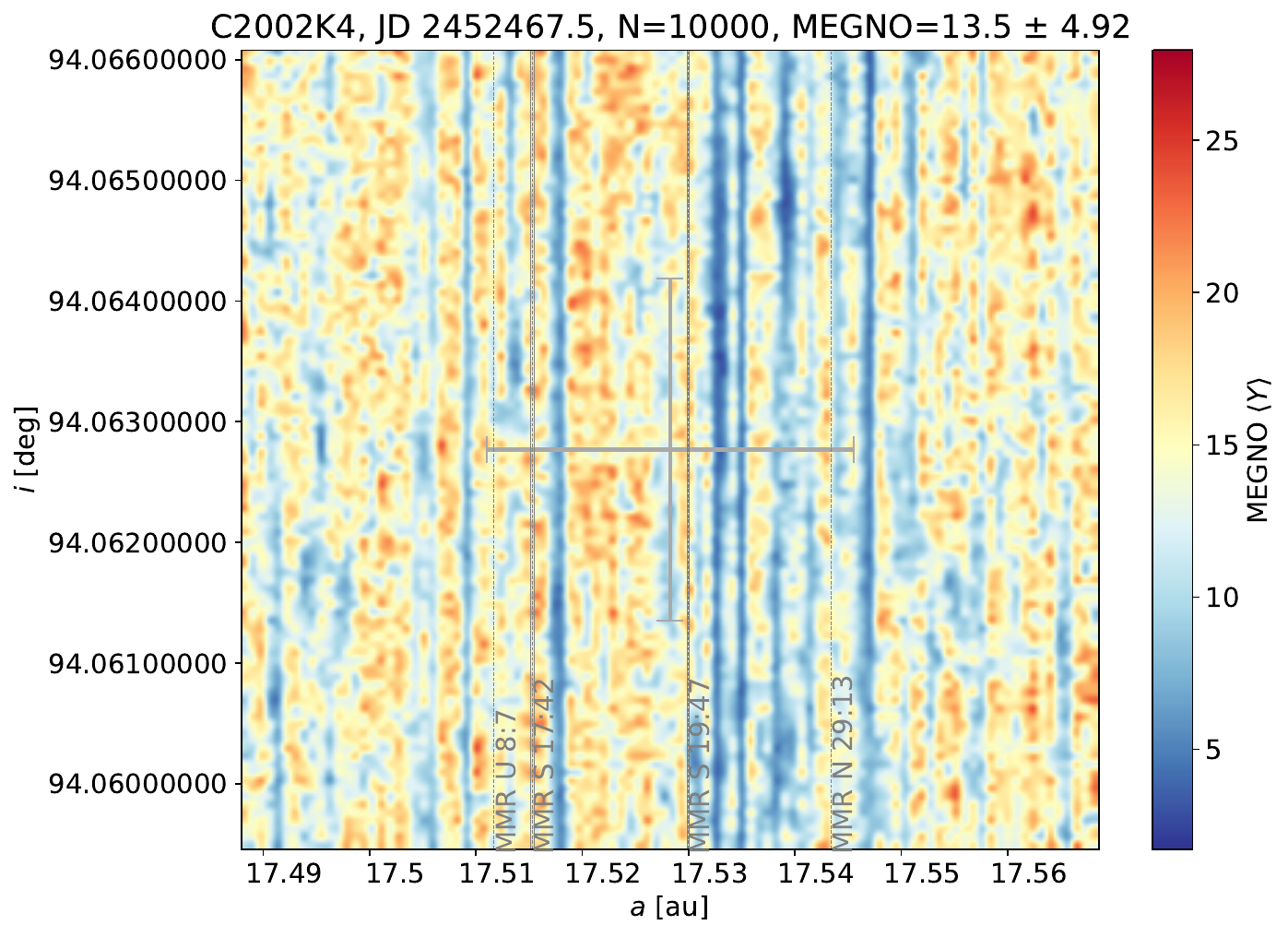}
	\caption{MEGNO maps of comets  C/2014 Q3 (upper panels), C/2015 GX (middle panels), and  C/2002 K4 (lowest panels) based on data from JPL database. Left-side panels: $a-e$ plane. Right-side panels: $a-i$. The strongest MMR resonances are indicated.}	\label{fig:MEGNOc2014q3JPL}
\end{figure*}

}}

\bsp	
\label{lastpage}
\end{document}